\documentclass[twocolumn]{IEEEtran} 
\usepackage[T1]{fontenc}
\usepackage{color}
\usepackage{mathrsfs}
\usepackage{amsmath}
\usepackage{amsthm} 
\usepackage{amssymb}
\usepackage{graphicx}
\usepackage{float} 
\usepackage{cite}
\usepackage{CJK}
\usepackage{indentfirst}
\usepackage{makecell}
\usepackage{algorithm}
\usepackage{algorithmic}
\usepackage{fancyhdr}
\usepackage{diagbox}
\usepackage{bm}
\usepackage{amsmath}
\usepackage{cases}
\usepackage{setspace}
\usepackage[unicode=true,
 bookmarks=true,bookmarksnumbered=true,bookmarksopen=true,bookmarksopenlevel=1,
 breaklinks=false,pdfborder={0 0 0},pdfborderstyle={},backref=false,colorlinks=false]
 {hyperref}
\hypersetup{
 pdftitle={Your Title},
 pdfauthor={Your Name},
 pdfpagelayout=OneColumn, pdfnewwindow=true, pdfstartview=XYZ, plainpages=false}

\usepackage{multirow} 
\usepackage{xcolor}

\allowdisplaybreaks[4]

\usepackage[caption=false,font=footnotesize]{subfig}

\IEEEoverridecommandlockouts

\usepackage{lettrine}

\usepackage{geometry}
\renewcommand{\headrulewidth}{0pt} 
\renewcommand{\arraystretch}{1.2}

\usepackage{subfig}
\begin{document}

\title{\textcolor{black}{
    Blockchain-Enabled Routing for Zero-Trust Low-Altitude Intelligent Networks}}   
\author{
    \IEEEauthorblockN{ Ziye Jia, \IEEEmembership{Member, IEEE}, Sijie He, Ligang Yuan,
    Fuhui Zhou, \IEEEmembership{Senior Member, IEEE},
     Qihui Wu, \IEEEmembership{Fellow, IEEE},\\
     Zhu Han, \IEEEmembership{Fellow, IEEE},
     and Dusit Niyato, \IEEEmembership{Fellow, IEEE}}
    \thanks{This work was supported in part by National Natural Science Foundation of China under Grant 62301251, in part by the Natural Science Foundation on Frontier Leading Technology Basic Research Project of Jiangsu under Grant BK20222001, in part by the Young Elite Scientists Sponsorship Program by CAST 2023QNRC001, and in part by NSF ECCS-2302469, and Japan Science and Technology Agency (JST) Adopting Sustainable Partnerships for Innovative Research Ecosystem (ASPIRE) JPMJAP2326.}
    \thanks{Ziye Jia, Sijie He, and Qihui Wu are with the College of Electronic 
    and Information Engineering, Nanjing University of Aeronautics and Astronautics, 
    Nanjing 211106, China (e-mail: jiaziye@nuaa.edu.cn; hesijie@nuaa.edu.cn; wuqihui@nuaa.edu.cn).}
    \thanks{Ligang Yuan is with the College of Civil Aviation, Nanjing University of Aeronautics 
    and Astronautics, Nanjing 211106, China (e-mail: yuanligang@nuaa.edu.cn).}
    \thanks{Fuhui Zhou is with the College of Artificial Intelligence, Nanjing University of Aeronautics and Astronautics, Nanjing 211106, China (e-mail: zhoufuhui@nuaa.edu.cn).}
    
    \thanks{Zhu Han is with the Department of Electrical and Computer Engineering at the University of Houston, Houston, TX 77004 USA (e-mail: hanzhu22@gmail.com).}
    \thanks{Dusit Niyato is with the College of Computing and Data Science, Nanyang Technological University, Singapore 639798 (e-mail: dniyato@ntu.edu.sg).}
    }
\maketitle
\pagestyle{fancy}
\fancyhf{}
\fancyhead[R]{\fontsize{7}{9}\selectfont \thepage}
\renewcommand{\headrulewidth}{0pt} 
\renewcommand{\footrulewidth}{0pt}

\begin{abstract}
    Due to the scalability and portability, low-altitude intelligent networks (LAINs)
    are essential in various fields such as surveillance and disaster rescue. 
    However, in LAINs, unmanned aerial vehicles (UAVs) are characterized by 
    the distributed topology and high mobility, thus vulnerable to security threats,
    which may degrade routing performances for data transmissions. 
    Hence, how to ensure the routing stability and security of LAINs is challenging.
    In this paper, we focus on the routing with multiple UAV clusters in LAINs.
    To minimize the damage caused by potential threats, 
    we present the zero-trust architecture with the software-defined perimeter and
    blockchain techniques to manage the identify and mobility of UAVs.
    Besides, we formulate the routing problem to optimize the end-to-end (E2E)
    delay and transmission success ratio (TSR) simultaneously, 
    which is an integer nonlinear programming problem and intractable to solve.
    Therefore, we reformulate the problem into a decentralized partially 
    observable Markov decision process.
    We design the multi-agent double deep Q-network-based routing algorithms 
    to solve the problem, empowered by the soft-hierarchical experience replay buffer and 
    prioritized experience replay mechanisms. \textcolor{black}{Finally, extensive simulations are conducted and  the numerical results demonstrate that the proposed framework reduces the average E2E delay by 59\% and improves the TSR by 29\% on average compared to benchmarks, while simultaneously enabling faster and more robust identification of low-trust UAVs.}

\end{abstract}

\newcommand{\CLASSINPUTtoptextmargin}{0.8in}

\newcommand{\CLASSINPUTbottomtextmargin}{1in}

\section{Introduction}
\lettrine[lines=2]{A}{s} key components of the six generation communication networks, 
the low-altitude intelligent network (LAIN) is widely applied to multiple tasks, 
such as disaster rescue and real-time monitoring {\cite{10418158, 11105407, mao2024survey}}. 
In the applications, unmanned aerial vehicles (UAVs) can 
collect the data from sensor devices (SDs), and then transmit them to the ground base stations (BSs).
Besides, in LAINs, the UAV networking can cooperatively accomplish complex tasks, 
thereby providing low-cost, flexible, and versatile services {\cite{Distributionally-Jia,10430396}}.
While transmitting the demanded data, routing is a significant issue {\cite{Routing_UAV_Survey}}. 

However, since UAVs are characterized by the complex application environment, high dynamic, and distributed topology, 
they are vulnerable and unreliable against security threats, such as attacks and node failures {\cite{He_Routing}}. 
\textcolor{black}{
To mitigate the impact of potential attacks, a dynamic identity authentication mechanism is essential. Traditional strategies based on static network boundaries are widely studied, such as firewalls and virtual private networks (VPNs) {\cite{attkan2022cyber}}. 
Nevertheless, when a node passes the initial authentication, it remains trusted and its subsequent malicious behaviors are often ignored due to a lack of dynamic monitoring.
This static nature makes traditional strategies unsuitable for highly dynamic LAINs.}
On the contrary, a zero-trust network security architecture is  based on the dynamic verification of nodes and behaviors, 
adhering to the principle of ``never trust, always verify'' {\cite{annabi2024towards}}.
\textcolor{black}{In zero-trust architectures (ZTAs), the software-defined perimeter (SDP) plays a significant role.}
It achieves secure access control through verifying identities, dynamic authorization, and detailed management, 
etc., reducing the illegal access to network resources and ensuring the security of data transmissions {\cite{9791053_zero}}.

 \begin{table*}[!t]
    \renewcommand\arraystretch{1}
    \color{black}
    \begin{center}
       \caption{\textcolor{black}{ Comparisons of Our Adaptive Framework over Existing Secure Routing Schemes for UAV Networks}}
       {\label{table-related-work}}
        \fontsize{8}{10}\selectfont{
        \begin{tabular}{|c|c|c|c|c|}
            \hline
            Reference 
& Trust Management & Blockchain & Algorithm & Key Characteristic\\
            \hline  
\multirow{2}{*}{Our Work} & \multirow{2}{*}{\makecell{Decentralized\\(Adaptive weights)}} & 
\multirow{2}{*}{Yes} & \multirow{2}{*}{SP-MADDQN} & Provides a comprehensive, adaptive, and 
\\ & & & & Byzantine-resistant secure routing framework. \\      
            \hline
            \multirow{2}{*}{\cite{Trusted_Blockchain}} &\multirow{2}{*}{\makecell{Decentralized  \\ (Fixed weights)}} & \multirow{2}{*}{Yes} & \multirow{2}{*}{Security-aware distributed routing protocol} & Static trust model lacks dynamic adjustability \\
            & & & & and the routing method is non-adaptive. \\            
            \hline
            \multirow{2}{*}{\cite{NIVEDITA2025103385}} & \multirow{2}{*}{\makecell{Decentralized  \\ (Fixed weights)}} & \multirow{2}{*}{No} & \multirow{2}{*}{Adaptive snow geese algorithm} & Static trust models lack adaptability, incur high  \\
            & & & &  audit overhead, and are vulnerable to tampering. \\ 
            \hline
            \multirow{2}{*}{\cite{8417971}} & \multirow{2}{*}{Centralized} & \multirow{2}{*}{No} & \multirow{2}{*}{Centralized trust based secure routing} & Trust management suffers from both single point of  \\ 
            & & & & failure and the lack of scalability. \\
            \hline
            \multirow{2}{*}{\cite{commi_energy}} & \multirow{2}{*}{Without trust} & \multirow{2}{*}{No} & \multirow{2}{*}{Multi-agent reinforcement learning (MARL)} & The lack of effective solutions against malicious  \\
            & & & &  nodes/behaviors results in insecurity.\\
            \hline
        \end{tabular}
        }
    \end{center}
    \vspace{-0.3cm}
\end{table*}
    \color{black}

Moreover, due to the malice and failure of UAVs, 
the reliability of LAINs and the availability of communication links are variational, 
which may degrade  the routing performance. 
Therefore, it is significant to efficiently evaluate the trust of UAVs and manage the mobility of UAVs in LAINs.
Specifically, to manage the trust of UAVs, a couple of works introduce the ground control station (GCS) as a center manager, 
which is not resilient to fault tolerance and suffered from tampering, long distances, and interferences {\cite{wei2024survey}}.
Hence, how to effectively manage the node trust and guarantee the routing security remains challenging
in the distributed and unreliable networks. 
In particular, the blockchain is a distributed ledger that provides  verifiable and traceable records of interactions, 
in which transaction details and records are stored and 
hard to be tempered with {\cite{blockchai_security,Trusted_Blockchain}}.
Therefore, the blockchain can be applied to recording opinions of UAVs and
building the decentralized trust management mechanism {\cite{Blockchain9120287,10415005}}. 
Meanwhile, several works related to the node trust evaluation calculate the values through fixed weights, 
which lack adaptabilities and flexibilities {\cite{tang2022blockchain}}. 
Therefore, the evaluation method based on adaptive weights is required to improve the timeliness.

Besides, the performance of the end-to-end (E2E) delay and transmission success ratios (TSRs) are the key basises for optimizing routing in LAINs,
since the low delay and high TSR can significantly enhance the reliability and timeliness of various emergency applications {\cite{ Wang_security,Review2-4}}.
Consequently, it is necessary  to propose dynamic routing algorithms
to facilitate the  efficient, reliable, and timely data transmission in the varying network topology.
However, most traditional routing algorithms (e.g., A*, Floyd-Warshall, and Dijkstra) 
are designed for static environments or fixed rules, which cannot directly adapt to moving UAVs and varying topologies {\cite{zhang20213d}}.
Hence, to tackle the issue, the multi-agent deep reinforcement learning (MADRL) 
is an effective  algorithm that dynamically interacts with the environment in real-time 
and can achieve satisfactory results with limited information {\cite{9738819}}.
\textcolor{black}{To more clearly illustrate the significance of the article, Table {\ref{table-related-work}} is presented.}


To deal with the above challenges, in this paper, the UAVs in LAINs is firstly divided  
into multiple clusters to collaboratively transmit the demands from SDs to BSs
with security threats and dynamic distributed topologies. 
Considering the joining and exiting of UAVs, 
we present the ZTA with the SDP controller
for the identity authentication and dynamic management of UAVs.
Then, we design the adaptive weight trust model to update the credit values of UAVs, 
including the direct trust via the forwarding rate of demands, 
trust interaction degree, and the reception rate of probe packets,  
as well as the indirect trust via aggregating the opinions from neighboring nodes.
Furthermore, in order to mitigate the impact of security threats, 
we propose the lightweight consortium blockchain to store the immutable 
transaction record and credit values of the trust evaluation.
In the lightweight blockchain, the cluster head UAVs (CHUs) are designated 
as full UAVs and deployed as decentralized managers, 
thereby participating in the consensus for blocks, 
while other UAVs act as light UAVs and only store the block.
Particularly, when credit values are lower than the given safety threshold, 
UAVs are isolated from the network by the consensus among CHUs.
According to the constructed zero-trust LAIN, 
we formulate the routing problem to optimize the E2E delay and TSR simultaneously, 
which is an integer nonlinear programming (INLP) problem and intractable to solve. 
Besides, since it is challenging to obtain the global information in distributed LAINs,
  we reformulate the routing problem into a decentralized partially observable 
Markov decision process (Dec-POMDP).
Furthermore, the improved multi-agent double deep Q-network (MADDQN) 
based algorithm is designed and empowered by the proposed soft-hierarchical 
experience replay buffer (SHERB) and prioritized experience replay (PER) mechanisms. 
Finally, numerical simulations are conducted to verify the performance of proposed algorithms.

In short, the major contributions of the work are summarized as follows.
\begin{itemize}
    \item To enhance the transmission efficiency of demands,
    UAVs are divided into multiple clusters in LAINs.
    Besides, we design the ZTA based network model to minimize the damage of 
    the potential security threats through continuous verifications. 
    The SDP controller is applied to the identity management and authentication of UAVs,
    addressing the security threats and the dynamic joining and exiting of UAVs.
    \item The adaptive weight trust model is designed to comprehensively evaluate 
    the credit values of UAVs combining the direct and indirect factors, 
    thereby mitigating the impact of low-trust UAVs on network performances.
    Furthermore, the blockchain technique is introduced to record 
    the transactions and credit values of UAVs 
    and support the SDP controller for the mobility management of UAVs. 
    Meanwhile, we propose the lightweight consortium blockchain to reduce 
    the consensus delay cost and improve the efficiency in LAINs.
    Furthermore, we present an improved practical Byzantine fault tolerance 
    (IPBFT) to enhance the network security, 
    in which the primary UAV is selected via considering the credit value 
    and resources of the storage and computation power.
    \item We formulate the routing problem to optimize both the total E2E delay and TSR 
    of transmitting demands, 
    which is an INLP problem and tricky to solve by traditional optimization schemes.
    Then, the routing problem is reformulated into a Dec-POMDP form,
    dealing with the challenge of obtaining global information in the decentralized LAIN. 
    Furthermore, the SHERB and PER empowered MADDQN (SP-MADDQN) algorithm 
    is proposed to learn the dynamically varying network topology with the mobility of UAVs
    and make optimized decisions.
    \item Extensive simulations are conducted to evaluate the SP-MADDQN algorithms with the mobility and malicious UAVs. 
    The adaptive weight trust method is demonstrated to identify the malicious UAV more quickly 
    than the average and random evaluation methods.
    Numerical results show that the proposed SP-MADDQN algorithm decreases the E2E delay and increases the TSR,  
    compared to the multi-agent deep Q-network (MADQN) based methods and the MADDQN without the SHERB and PER mechanisms.
\end{itemize}

    The rest of this paper is organized as follows.  In Section {\ref{Related-works}}, we introduce the related works. The system model and problem formulation are illustrated in Section {\ref{System-model}}. 
    Then, we design the algorithms in Section {\ref{sec:Algorithm}}. 
    Simulations are conducted and analyzed  in Section {\ref{sec:Simulation Results}}. 
    Finally, the conclusions are future works are presented in Section {\ref{sec:Conclusions}}.

\vspace{0.06cm}
\section{Related Work\label{Related-works}}
There exist some works related to the network security mechanism.
For example, in {\cite{LI2025110964}}, the authors introduced the VPN technique 
for network boundaries and indicated the existing challenges posed by the partial protection and privacy concerns.
In {\cite{10537758}}, to protect the security, the authors combined the next generation firewall 
with the self authentication due to the inability to deal with targeted and data-focused attacks.
The authors in {\cite{JUMA2020102598}} integrated the  Internet protocol security (i.e., IPSec/IPv6) 
and secure socket layer to solve the security issue of VPNs.
Through VPNs or firewalls as boundaries, these works can enhance  network securities.
However, they are easily bypassed by attackers. When the boundary defenses are bypassed, 
the security protection of the internal network is compromised. 
To mitigate the harm, the ZTA is applied to secure networks.
For instance, in {\cite{abdelhay2024toward}}, 
to enhance the resilience of the traditional static protection against various attacks through VPNs,
the authors introduced the authentication mechanism based on the SDP controller 
to create a dynamic and secure zero-trust network environment.  
Authors in {\cite{Han_GLOBECOM}} proposed a zero-trust strategy to establish continuous 
trust while remaining skeptical of potential betrayal attacks, 
and a Dirichlet-based trust evaluation method was developed 
to select trustworthy participants.
In {\cite{Zero-Trust-2}}, to address the potential security risks, 
the authors proposed a ZTA with {SDP} for 5G core networks, providing secure communications  
through an authorization-based method. 
The above works illustrate that it is possible to employ the ZTA and SDP technique to ensure the routing security in LAINs.

Besides, there exist a couple of works related to node management mechanisms.
In {\cite{GCS_1}}, according to the status information sent by UAVs, 
the GCS played a role in managing the connections among nodes, which were limited by the control range.
The authors in {\cite{GCS_3}} presented that the GCS remotely controlled UAVs by sending the controlling information, 
in which UAVs were limited within the communication range of the GCS, 
restricting the scope of operations.
{\cite{GCS_2}} stated that as the center controller, 
the GCS received the data transmitted by UAVs, 
which suffered from long distances, 
undulating terrains or other interferences.
The above works rely on a center controller, which is not resilient 
to fault tolerance and may be susceptible to tampering.
Therefore, how to guarantee the reliability and security of routing remains challenging
in the distributed LAINs.

\textcolor{black}{
With the decentralized and tamper-proof characteristics,  
the blockchain technique is introduced to build decentralized trust mechanisms and secure networks \cite{Review2-2}. 
For example, in {\cite{DRRS-BC}}, 
a blockchain-based decentralized registration framework 
was proposed to solve the problem of misconfiguration and malicious attacks. 
Authors in \cite{Review2-1} proposed a blockchain-based data storage solution for autonomous vehicle networks, aiming to address the significant challenges posed by the frequent data transmission to the network efficiency and security.
In \cite{Review2-3}, a blockchain-based spectrum sharing framework for UAV networks was designed to maintain decentralization, ensuring the fairness and reliability in transactions.} The authors in {\cite{8951253}} proposed a blockchain-based multi-wireless sensor network authentication scheme, in which nodes were classified into BSs, cluster heads and ordinary nodes according to different capabilities. In {\cite{KUANG2025103981}}, the authors proposed a novel lightweight authentication scheme for Internet of things devices. To study the issue of the identity privacy leakage caused by malicious attacks, the authors utilized the blockchain to enhance the reliability through the hyperledger and ring signature method {\cite{YANHUI202210365}}.
The above schemes demonstrate that based on the blockchain, a decentralized identity management method enables to eliminate the vulnerable central point and enhance the reliability.

Specifically, there exist some works focusing on MARL-based routing problems in dynamic scenarios. 
For example, the authors in {\cite{MARL-o-Manage-4}} utilized 
a value decomposition network-based MARL algorithm to 
minimize the E2E delay of routing in dynamic aerial-terrestrial hybrid networks.
In {\cite{MARL-o-Manage-5}}, 
the routing problem was formulated as a max-min problem 
via the Lagrange method, 
and a constrained MARL dynamic routing algorithm was proposed for the objective-constraint balance.
The authors in {\cite{MARL-o-Manage-6}} developed 
the mean-field enhanced heterogeneous MARL framework
to optimize the routing communication energy efficiency.
The above studies  illustrate that the MARL can solve the routing problem effectively.
However, these works do not consider the credit evaluation and mobility management of UAVs in the vulnerable environment.

In this work, we consider designing the trust evaluation method based on the adaptive weights. 
In the zero-trust LAIN, the SDP controller is supplemented by the blockchain technology to achieve the identity authentication and permission management.
Additionally, we propose the SP-MADDQN algorithm  to tackle the adaptive dynamic routing issue.
 \begin{figure*}[t]
    \centering
    \includegraphics[width=0.9\linewidth]{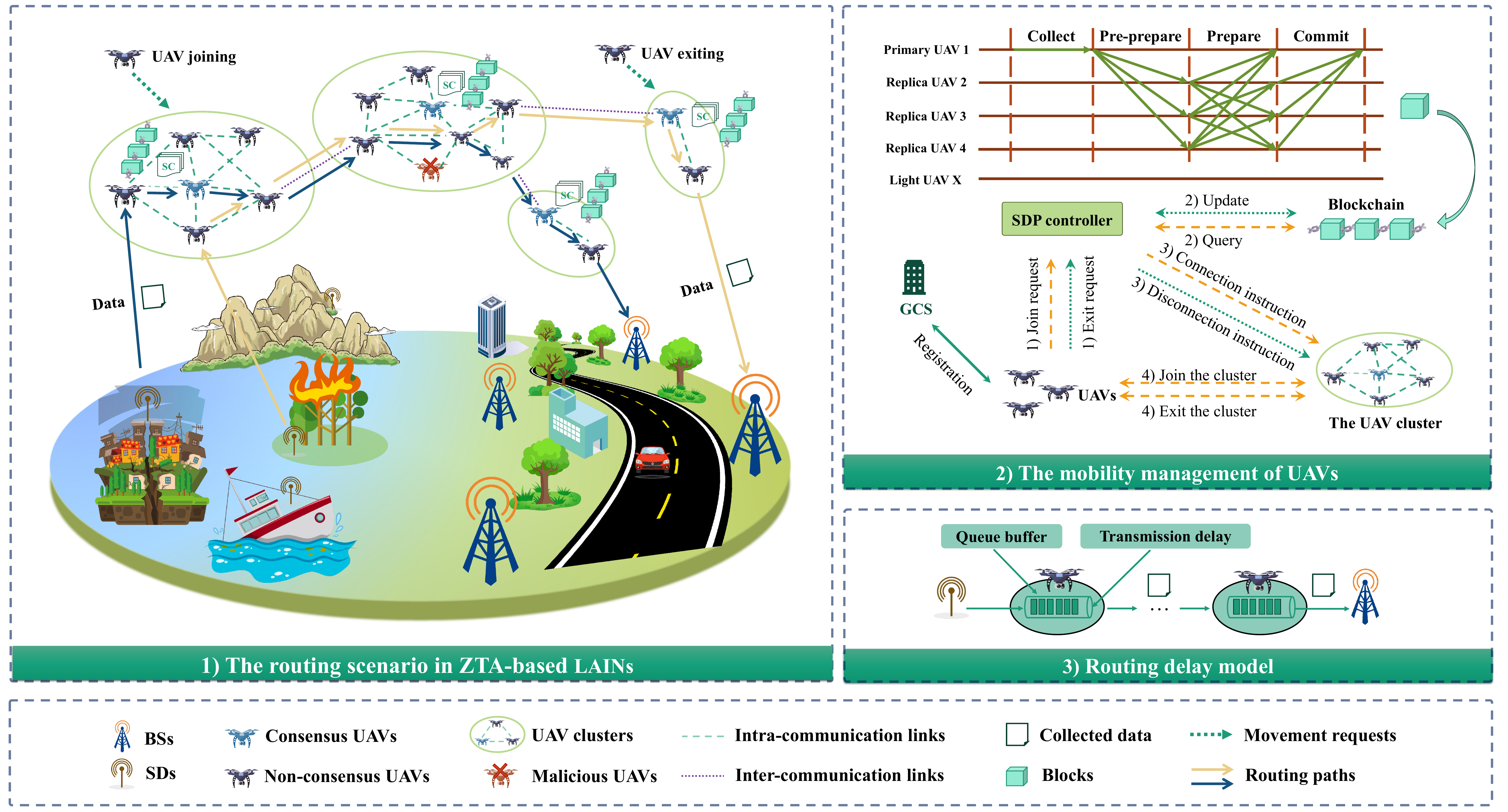}
    \caption{\label{fig:Network_of_UAV} 
    An illustration of blockchain-enabled routing for zero-trust LAINs. 
\textit{ Part  1) } presents the routing scenario in ZTA-based LAINs with the security threats. 
\textit{ Part  2) } introduces the mobility management of UAVs via the blockchain and SDP techniques.
\textit{ Part  3) } illustrates the model of the E2E delay during routing.
    } 
\end{figure*}
 
\section{System Model and Problem Formulation  \label{System-model}}
\subsection{Zero-Trust Architecture Based Network Model}
\subsubsection{Network Model}
The routing process in zero-trust LAINs is shown in Fig. \ref{fig:Network_of_UAV}, 
including  $I$ SDs, $U$ UAVs, and $B$ BSs. 
Thus, in the network, the total number of nodes is $I+U+B$, 
and each node holds a unique identity. 
$\mathcal{B}\! =\!\{1,...,b,...,B\}$, $\mathcal{I}\! =\!\{1,...,i,...,I\}$, 
and $\mathcal{U}\! =\!\{1,...,u,...,U\}$ denote the sets of 
BSs, SDs, and UAVs, respectively. 
\textcolor{black}{ 
Additionally, according to different abilities, the UAV set $\mathcal{U}$ can be classified as three types of
 $\mathcal{U}_\mathtt{s}$, $\mathcal{U}_\mathtt{r}$, and $\mathcal{U}_\mathtt{d}$
for the data collection, relay forwarding, and downlinking, respectively.
Specifically, each type of UAV sets is divided into multiple clusters, 
and in each cluster, the UAV with the most energy and computing power is selected as the cluster head.}
$\mathcal{E}\!=\!\mathcal{E}_{iu}\!\cup \mathcal{E}_{uu}\!\cup \mathcal{E}_{ub}$ indicates the set of communication 
link statuses among all nodes, where $e_{iu}\!\in\!\mathcal{E}_{iu}$, $e_{ub}\!\in\!\mathcal{E}_{ub}$, 
and $e_{u u }\!\in\!\mathcal{E}_{uu}$ represent the connection links
between the SD and UAV, BS and UAV, as well as UAV and UAV, respectively.
$e_{nm}\in \mathcal{E}$ indicates whether there exists a communicable link 
between nodes $n\in \mathcal{I} \cup \mathcal{U}$ and $m\in \mathcal{U} \cup \mathcal{B}$. 
Specifically, $e_{nm}=1$ indicates the link is effective, 
and $e_{nm}=0$ denotes that there exist no direct links. 

In particular, the time period is divided into $\mathsf{T}$  
slots and the length of each time step is $\tau$. 
$T \!=\! \{1,...,t,...,\mathsf{T}\}$ represents the set of time steps.      
At the start of time step $t$, 
demand $r \in R$ is transmitted from SD $i\in \mathcal{I}$ to 
destination BS $b\in \mathcal{B}$, and the information is denoted by 
$\varUpsilon^r\!=\!\{\mathtt{s}^r,\mathcal{L}^r,\mathtt{d}^r,\mathcal{T}^r_{max}\}$,  
in which the source is $\mathtt{s}^r\!=\!i$, and the destination is $\mathtt{d}^r\!=\!b$.
Here, the size of demand $r$ is $\mathcal{L}^r$ (in bit), 
and $\mathcal{T}^r_{max}$ is the maximum delay tolerance of the demand transmission.
Moreover, the demands are uploaded from SD $i \!\in\! \mathcal{I}$ to UAV $u\!\in\! \mathcal{U}_\mathtt{s}$,  
relayed by UAV $u\! \in\! \mathcal{U}_\mathtt{r}$, and downloaded 
from UAV $u\!\in\!\mathcal{U}_\mathtt{d}$ to BS $b \!\in\! \mathcal{B}$. 
$\mathcal{P}_{ib}^{r}\!=\!{(e_{iu},\ldots,e_{ub})}$ 
indicates a completed routing path for transmitting demand $r$
 from source SD $i\!\in\!\mathcal{I}$ to 
corresponding destination BS $b\! \in\!\mathcal{B}$.

The coordinates of SD $i$ 
and BS  $b$ remain fixed and are denoted as  ${\varTheta}_{i}\!=\!{(x_{i},y_{i},0)}$ and 
${\varTheta}_{b}\!=\!{(x_{b},y_{b},0)}$, respectively. 
The location of UAV $u$  is indicated by 
${\varTheta}_{u}(t)={(x_{u}(t),y_{u}(t),z_{u}(t))}$ in three-dimensional Cartesian coordinates at time step $t$.

\vspace{-0.05cm}
\textcolor{black}{
At time step $t$, the velocity of UAV $u$ is defined by the magnitude ${\nu}_{u}(t) \in [{\nu}_{\min}, {\nu}_{\max}]$, along with a horizontal azimuth $\phi_{u}(t)$ and a vertical elevation $\delta_{u}(t)$, based on the random walk mobility model \cite{RWMM-1}. Thus, the corresponding velocity vector in 3D Cartesian coordinates is expressed as $\bm{\nu}_{u}(t) = \left( \nu ^x_{u}(t),{\nu}^y_{u}(t),{\nu}^z_{u}(t)\right)$,
where $\nu ^x_{u}(t)={\nu}_{u}(t)\cos\delta_{u}(t)\cos\phi_{u}(t)$, $\nu ^y_{u}(t)={\nu}_{u}(t)\cos\delta_{u}(t)\sin\phi_{u}(t)$, and $\nu ^z_{u}(t)={\nu}_{u}(t)\sin\delta_{u}(t)$.
The collective velocities of all $U$ UAVs are represented by $\bm{V}(t) = \{\bm{\nu}_{1}(t), \bm{\nu}_{2}(t), \dots, \bm{\nu}_{U}(t)\}$.
At the start of each slot $t$, each UAV can instantaneously update its speed and orientation, then maintain them constant throughout the slot. From a kinematic perspective, the position $\Theta_{u}(t)$ evolves with the velocity $\boldsymbol{\nu}_{u}(t)$. Specifically, if UAV $u$ begins slot $t$ with velocity $\boldsymbol{\nu}_{u}(t)$ and remains at uniform speed, its displacement over the slot is approximated by $\Delta\Theta_{u}(t) \approx \boldsymbol{\nu}_{u}(t) \tau$, where $\tau$ is the slot duration. Thus, the position at the next slot is updated as 
$\Theta_{u}(t+1) = \Theta_{u}(t) + \boldsymbol{\nu}_{u}(t) \tau.$
}

Additionally, the Euclidean distance between nodes $n \in \mathcal{I}\cup\mathcal{U}$ and $m \in \mathcal{U}\cup\mathcal{B}$
is indicated by $d_{n  m}(t)$ at time step $t$, i.e., 
\vspace{-0.1cm}
\begin{equation}{\label{distance}}
    \begin{aligned}
    &d_{n  m}(t) \!= \!\\
    &\sqrt{\!(x_{n}(t)\!-\! x_{m}(t))^2\!+\!(y_{n}(t)\!-\!y_{m}(t))^2\!+\!(z_{n}(t)\!-\!z_{m}(t))^2}.
    \end{aligned}
\end{equation}
\vspace{-0.1cm}
In particular, the distance between UAVs must satisfy
\vspace{-0.1cm}
\begin{equation}{\label{distance_min}}
    d_{min} \leqslant d_{n  m}(t), \forall n ,m \in\mathcal{U}, n\neq m, t\in T.
    \vspace{-0.1cm}
\end{equation}
Here, $d_{min}$ represents the safe distance to avoid collisions among UAVs.
At time step $t$, the set of connected UAVs for UAV $u$ is denoted as $\Gamma_{u}(t)$,   
and distance $d_{u\kappa }(t)$ between UAV $u$ and UAV $\kappa \in \Gamma_{u}(t) $ satisfies 
\vspace{-0.1cm}
\begin{equation}{\label{distance_max}}
    d_{u \kappa}(t)\leqslant d_{u,max},\forall \kappa \in \Gamma_{u}(t), u\in \mathcal{U}, t \in T,
    \vspace{-0.1cm}
\end{equation}
in which $d_{u,max}$ is the maximum communication distance of UAV $u$.

\subsubsection{Zero-Trust Architecture Scheme}
In the distributed LAIN, malicious UAVs are more possible 
to survive and undermine network performances.
To ensure the security of LAINs, a ZTA scheme is proposed to achieve 
the certification and authorization of joining or exiting requests,
combining blockchain and SDP techniques. 
In the ZTA, there are four essential units:
the GCS, UAV, SDP controller, and blockchain, shown in Fig. {\ref{fig:Network_of_UAV}}.

\begin{itemize}
    \item UAVs are the clients for joining or exiting LAINs, which need to obtain the legal registration information from GCS. 
    \item \textcolor{black}{The GCS serves as the root certificate authority within the public key infrastructure that supports the system security. Its primary responsibility is designed to register UAVs and provide them with foundational digital certificates at the start of tasks. }
    \item \textcolor{black}{The SDP controller is logically centralized for the unified access control and physically distributed among trust UAVs for the high-availability, thereby avoiding the failure of single point. Furthermore, it dynamically manages and controls access to network resources, based on the identity information of UAVs.}  Particularly, only the authenticated and authorized UAV has permissions to join LAINs.  
    \item \textcolor{black}{The blockchain stores immutable and trustworthy records for the trust management and identity states, including a certificate revocation list, to enforce  the secure authentication by SDP in the zero-trust environment. In detail, the consensus UAVs periodically record credit values $\mathbb{T}_u(t)$ onto the chain, establishing the reliable basis for trust certifications. The SDP controller monitors these updates and automatically triggers policy reevaluation. If the trust of a UAV falls below the threshold $\mathbb{T}_{thr}$, its token is revoked and the revocation is broadcast via the blockchain. It ensures that all peers can independently deny connections to revoked UAVs, enabling distributed and timely policy enforcement}. 
\end{itemize}
\subsubsection{Joining and Exiting of UAVs}
Before applying to join a cluster, UAVs must obtain the unique identity information 
via registering at the GCS. 
Then, UAVs pack the identity information, signature of the GCS, 
and identity document of the joined cluster into a transaction, 
which is broadcast to LAINs for verifications by the selected CHUs. 
After the consensus approval, the UAV is allowed to join the cluster network. 
When a UAV exits, the CHU of the exiting UAV is required to send 
the corresponding transaction to  LAINs for dynamically 
updating the statuses and changes of link availabilities.

\textcolor{black}{The designed architecture is independent of the underlying cryptographic primitives. The analytical models rely on abstract costs, allowing the system to adopt future standards. Besides, the security is anchored in the continuous verification of immutable on-chain records, not in any specific algorithm. Besides, the system derives its resilience from an integrated adaptive trust and consensus loop, which identifies and isolates the malicious activity regardless of its specific form, eliminating the need for predefined threat models. In addition, we design a trust model to assess the impact of malicious behaviors, ignoring specific attack types.}

\subsection{Trust Model}
In LAINs, to enhance the  routing efficiency, 
UAVs are assigned with credit values representing the trustworthiness, 
which are updated via comprehensively considering direct and indirect factors, 
avoiding the single assessment factor and insufficiently reasonable calculation.
Additionally, $\bm{\mathbb{\textbf{T}}(t)}=\{\mathbb{T}_{1}(t), \ldots,\mathbb{T}_{U}(t)\}$ is the set of 
credit values, and for UAV $u$, the value of $\mathbb{T}_{u}(t) \in \bm{\mathbb{\textbf{T}}(t)}$ is within the range of $[0,1]$.  
Based on the given safety threshold $\mathbb{T}_{thr}$ determined by security requirements, 
UAVs are divided into two states of trust and malice {\cite{trust_parameters}}.
$\xi_u(t)$ denotes whether UAV $u$ is trusted, i.e.,
\vspace{-0.1cm}
   \begin{equation}
    \xi_u(t)=\left\{
        \begin{aligned}
            &1, \mathbb{T}_{u}(t)>\mathbb{T}_{thr},\\
            &0, \mathbb{T}_{u}(t)\leq \mathbb{T}_{thr}.
        \end{aligned}
    \right.
    \vspace{-0.1cm}
   \end{equation}
Therefore, the set and number of malicious UAVs are denoted as 
$\mathcal{U}_{\mathtt{f}}(t)\!=\!\{u \!\mid\! \xi_u(t)\!=\!1,\forall u\!\in \! \mathcal{U}\}$, 
and $U_{\mathtt{f}}(t)\!=\!\left\lvert \mathcal{U}_{\mathtt{f}}(t)\right\rvert $, respectively.
Furthermore, UAVs may reject to forward the received demands and frequently interact with 
the low trust UAVs, which are related to the trust evaluation, overviewed for the following.
\subsubsection{Direct Trust Calculation} The direct trust combines the following factors.
\begin{itemize}
    \item {\em Demand forwarding rate $\mathbb{T}_{u}^{\text{\tiny{D1}}}(t)$:} 
    $\mathbb{T}_{u}^{\text{\tiny{D1}}}(t)$ represents  the ratio of the total transmitted demand 
    $\sum_{k=1}^{t}\varepsilon^{tr}_{u}(k)$ to the total received demand 
    $\sum_{k=1}^{t}\varepsilon^{re}_{u}(k)$ 
    by UAV $u$ from time step 1 to $t$, 
    to evaluate the malicious behaviors rejecting to transmit received demands, i.e.,   
    \vspace{-0.1cm}
    \begin{equation}{\label{equ:D1}}
        \mathbb{T}_{u}^{\text{\tiny{D1}}}(t)\!=\!\frac{\sum\limits_{k=1}^{t}\varepsilon^{tr}_{u}(k)}{\sum\limits_{k=1}^{t}\varepsilon^{re}_{u}(k)}, \forall u \in \mathcal{U},\sum_{k=1}^{t} \!\varepsilon^{re}_{u}(k) \!> \!0, t \in T.
    \vspace{-0.1cm}
    \end{equation} 
    Besides, $\sum_{k=1}^{t} \!\varepsilon^{re}_{u}(k)\!=\!0$ denotes that UAV $u$ does not receive demands, 
    and the credit remains the initial value.
    \item {\em Trusted interaction degree  $\mathbb{T}_{u}^{\text{\tiny{D2}}}(t)$:}  
    For each node, a higher interaction frequency with high-trust UAVs yields a higher credit value, and vice versa.
    Thus, the degree of trust interaction is defined as
    \vspace{-0.1cm}
    \begin{equation}
        \begin{aligned}
        \mathbb{T}_{u}^{\text{\tiny{D2}}}(t)\!=\! \frac{ \rho^{\text{\tiny{H}}}_{u}(t)}{\rho_{u}(t)}, \forall u \in \mathcal{U}, t \in T.   
        \end{aligned}  
        \vspace{-0.05cm}
    \end{equation}  
    $\rho^{\text{\tiny{H}}}_{u}(t)$ and $\rho_{u}(t)$ denote the  numbers  of UAV $u\in\mathcal{U}$ interacting
    with high-trust UAVs and all UAVs, respectively.
    \item {\em Probe packet reception rate $\mathbb{T}_{u}^{\text{\tiny{D3}}}(t)$:} 
    To capture the conditions of surroundings, 
    UAVs periodically probe the links of neighboring UAVs via \textit{Hello} messages.
    In particular, the reception rate is  the probability that probe messages successfully reach the receiver, i.e.,
    \vspace{-0.1cm}
    \begin{equation}
        \mathbb{T}_{u}^{\text{\tiny{D3}}}(t)=\frac{\varTheta_{u}(t-\Im,t)}{\Im/\ell},
        \vspace{-0.1cm}
    \end{equation}  
    
    where $\varTheta_{u}(t\!-\!\Im,t)$ is the number of probe messages actually received by the evaluated UAV ${u}$ during 
    time window $\Im_u $, and $\ell_u $ is the probing period. Hence, $\Im_u/\ell_u$ is the theoretical number of 
    the received probe messages.
\end{itemize}

Considering the above three factors, the comprehensive direct credit value of UAV 
$u$ at time step $t$ is 
\vspace{-0.1cm}
\begin{equation}{\label{direct_trust}}
    \mathbb{T}^{\text{\tiny{D}}}_{u}(t) =\backepsilon_{1} \! \mathbb{T}^{\text{\tiny{D1}}}_{u}(t) +\backepsilon_{2}\! 
    \mathbb{T}^{\text{\tiny{D2}}}_{u}(t)+ \backepsilon_{3} \! \mathbb{T}_{u}^{\text{\tiny{D3}}}(t), \forall  u \in \mathcal{U}, t\in T,
    \vspace{-0.1cm}
\end{equation}
where $\backepsilon_{1}$, $\backepsilon_{2}$, and $\backepsilon_{3}$ indicate the weights of 
$\mathbb{T}^{\text{\tiny{D1}}}_{u}(t)$, $\mathbb{T}^{\text{\tiny{D2}}}_{u}(t)$, and $\mathbb{T}_{u}^{\text{\tiny{D3}}}(t)$, 
respectively. Additionally, $\backepsilon_{1}+\backepsilon_{2} +\backepsilon_{3}=1$.
\subsubsection{Indirect Trust Calculation}The direct trust evaluation mainly derives from the experience 
of UAVs and is subjective and limited. 
Nevertheless, the indirect trust evaluation can mitigate the limitation through fully utilizing the knowledge of networks, 
which is obtained from the trusted neighboring UAVs and calculated as 
\vspace{-0.1cm}
\begin{equation}
    \mathbb{T}^{\text{\tiny{I}}}_{u}(t)\!=\!\sum\limits_{\kappa \in \Gamma_{u}(t) }\! 
    \frac{\mathbb{P} _{\kappa}(t)}{\mathbb{P}_{\kappa}(t)\!+\!\mathbb{N}_{\kappa}(t)}, \mathbb{T}_{\kappa}(t)\!\geqslant\!\mathbb{T}_{thr} 
    ,\forall u \!\in\! \mathcal{U}, t\in T.
    \vspace{-0.1cm}
\end{equation}
Here, for UAV $u$, $\mathbb{P}_{\kappa}(t)$ and $\mathbb{N} _{\kappa}(t)$ are the positive 
and negative recommendations from UAV $\kappa$, respectively.

 Therefore, the credit value $\mathbb{T}_{u}(t+1)$ of UAV $u$ at time step $t+1$ is calculated as
 \vspace{-0.1cm}
\begin{equation}
    \begin{aligned}
         \mathbb{T}_{u}(t+1) =\psi_{u}^{0}(t)\mathbb{T}_{u}(t)+ \psi_{u}^{1}(t) \mathbb{T}^{\text{\tiny{D}}}_{u}(t) &+\psi_{u}^{2}(t) \mathbb{T}^{\text{\tiny{I}}}_{u}(t), \\
         &\forall  u \in \mathcal{U}, t\in T,
    \end{aligned}
    \vspace{-0.1cm}
\end{equation}
where $\mathbb{T}_{u}(0)$ indicates the initial credit value for UAV $u$.
$\psi_{u}^{0}(t)+\psi_{u}^{1}(t)+\psi_{u}^{2}(t)\!=\!1$. 
In detail, $\psi_{u}^{t}(t)$, $\psi_{u}^{1}(t)$, and $\psi_{u}^{2}(t)$ denote the weights of $\mathbb{T}_{u}(0)$,
$\mathbb{T}^{\text{\tiny{D}}}_{u}(t)$, and $\mathbb{T}^{\text{\tiny{I}}}_{u}(t)$, respectively.  
Moreover, we propose the adaptive model to dynamically adjust the  weights, detailed as
\vspace{-0.1cm} 
\begin{align}
    \left\{\begin{aligned}{\label{Adaptive_weights}}
        \psi _{u}^0(t)&\!=\! \beta \times \frac{\mathbb{T}_{thr}}{\mathbb{T}_{u}(t)}, \\
        \psi _{u}^1(t) &\!=\! \frac{(1\!-\!\psi _{u}^0(t)) (1\!-\!\mathbb{T}^{\text{\tiny{D}}}_{u}(t))}{2\! -\! \mathbb{T}^{\text{\tiny{D}}}_{u}(t)\! +\! \mathbb{T}^{\text{\tiny{I}}}_{u}(t)}, 0\leq\! \mathbb{T}^{\text{\tiny{D}}}_{u}(t) \!+ \!\mathbb{T}^{\text{\tiny{I}}}_{u}(t)\!\leq 2,\\
        \psi _{u}^2(t) &\!=\! \frac{(1\!-\!\psi _{u}^0(t)) (1\!-\!\mathbb{T}^{\text{\tiny{I}}}_{u}(t))}{2 \!-\! \mathbb{T}^{\text{\tiny{D}}}_{u}(t)\! + \!\mathbb{T}^{\text{\tiny{I}}}_{u}(t)}.
    \end{aligned}\right.
    \vspace{-0.1cm}
\end{align}

\textcolor{black}{
The design of \( \psi^0_u(t) = \beta \times \mathbb{T}_{thr} / \mathbb{T}_u(t) \) is motivated by the principle of adaptive historical inertia. The weight of the historical credit value \( \mathbb{T}_u(t) \) is inversely proportional to the current magnitude $\mathbb{T}_{u}(t+1)$. When a UAV consistently behaves well (i.e., the \(\mathbb{T}_u(t)\) value is relatively high), the formula assigns a lower weight to history, making the update process focus more on recent direct and indirect evidences (i.e., \(\mathbb{T}^{\text{\tiny{D}}}_u(t)\) and \(\mathbb{T}^{\text{\tiny{I}}}_u(t)\)). The mechanism enables the model to respond quickly to the latest behaviors of UAVs. On the contrary, when \( \mathbb{T}_u(t) \) is low, the weight of history increases. The system pays more attentions to long-term behavioral records, thereby avoiding a sharp drop in credit values due to a single or a few negative reports, and enhancing the robustness of the model. The parameter \( \beta \in (0,1]\) controls the overall sensitivity of the historical behaviors and limits the range of $\psi_{u}^0(t)$. \( \mathbb{T}_{thr} \) serves as a normalization benchmark to evaluate if the UAV needs to be isolated from the LAIN.
As $\mathbb{T}_{u}(t)$ increases, $\psi_{u}^0(t)$ decreases, and thus $\psi_{u}^1(t)$ and $\psi_{u}^2(t)$ increases, enlarging the impact of malicious behaviors on the evaluation of credit values. When $\mathbb{T}^{\text{\tiny{D}}}_{u}(t)\! +\! \mathbb{T}^{\text{\tiny{I}}}_{u}(t)\!=\!2$, $\psi _{u}^1(t) \!=\!\psi _{u}^2(t)\!=\! \frac{1}{2}(1\!-\!\psi _{u}^0(t))$.}

Meanwhile, the centralized trust manager relies on the third-party authentic center
and is susceptible to diverse attacks, causing the entire network to break down {\cite{GCS_1}}. 
To overcome the challenge posed by the above security threats, the scheme
based on the blockchain technology is introduced to provide the distributed 
and reliable managers for credit values, ensuring the secure routing in LAINs{\cite{DRRS-BC}}.

\subsection{Blockchain Model}
Due to decentralization and the distributed ledger maintained by multiple nodes,
the consortium blockchain can be applied for LAINs to create the secure and reliable communication network.
\textcolor{black}{In detail, we design the lightweight consortium blockchain, in which only UAVs 
with the high credit value, storage, and computation power are assigned 
as full UAVs (i.e., CHU) to participate in the consensus {\cite{li2021lightweight, LightTrust_9434372}}.} 
Particularly, since only $U_f$ full UAVs are required for the consensus 
instead of $U$ UAVs, the resource consumption of the consensus is reduced to $U_f/U$, 
benefiting for the sustainability and reliability.
Meanwhile, as the major consensus mechanism of the consortium blockchain, 
the PBFT is applied to the consensus process which can tolerate 1/3 malicious UAVs {\cite{11201899}}. 
However, within the PBFT, the random selection of the 
primary UAV may lead to the entire consensus process to fail 
when the nominated primary UAV is malicious {\cite{channel_model_PL_2}}.
Therefore, to enhance the security of LAINs and the reliability of the PBFT, 
we designate the full UAV with the largest credit value and strongest ability as the primary UAV that is updated periodically. The details are as follows.

\subsubsection{Role Selection}
In the lightweight blockchain, 
the set of UAVs $\mathcal{U}= \mathcal{U}_f \cup \mathcal{U}_l$ 
is divided into the full UAV set 
$\mathcal{U}_f= \{1, 2, 3, \ldots, U_f \}$ and light UAV set
$\mathcal{U}_l= \{1, 2, 3, \ldots, U_l \}$.
\paragraph{Full UAV}Full UAVs have the full blockchain ledger and are mainly responsible 
for the blockchain consensus including the broadcasting and verification of transactions. 
\paragraph{Light UAV}In LAINs, apart from the full UAVs, the other UAVs act as light UAVs   
that can only store the header of blocks and are mainly responsible for generating 
local transactions and forwarding transactions.
\paragraph{Primary UAV}
The primary UAV serves as the client of the blockchain. 
It gathers all transactions, verifies the signature and message authentication code (MAC) of each transaction, 
and then compiles the collected transactions into a block \cite{Blockchain_delay}.
\paragraph{Replica UAV} Except for the primary UAV, the other full UAVs operate as non-primary (replica) UAVs.

\textcolor{black}{The hierarchical role mechanism effectively reduces the computational and storage overhead of the entire network, thereby enhancing the scalability of the network. By concentrating security and management responsibilities on highly reliable full UAVs, the probability of single-point failures in the network is reduced. Therefore, the balance between the security and resource consumption centers on the scales of full UAVs. In addition, based on the hierarchical role method, the IPBFT mechanism is detailly introduced in the following.}

\subsubsection{IPBFT based Consensus Process} 
At time step $t$, the numbers of total and malicious consensus UAVs 
are indicated as $K(t)$ and $U^c_{\mathbf{f}}(t)$, respectively. 
To achieve the consensus, every node needs to aggregate a minimum of $2\lceil F(t)\rceil+1$ consistent prepare messages from the diverse replica, 
where $ F(t) =(K(t)-1)/3$.
If $U^c_{\mathbf{f}}(t)\!>\!F(t)$, 
the IPBFT based consensus may fail, leading to untimely information updates.
In Fig. {\ref{fig:Network_of_UAV}}, the consensus process is detailed, 
in which the time cost of the cryptographic operations is considered, 
including generating signatures, verifying signatures, and generating/verifying MACs, 
with costs of $\epsilon_s$, $\epsilon_v$, and $\epsilon_m$ cycles, respectively.
Thus, the consensus process and corresponding delay are discussed as follows {\cite{channel_model_PL_2}}.
\paragraph{Collection}
In this  phase, the primary UAV collects routing information from UAVs and then verifies.
Specifically, the light UAV sends its own information to the nearest surrounding full UAVs after transmitting demands.
Then, each replica UAV generates a transaction according to the received information from surroundings.
Particularly, the transactions are signed by the private key of replica UAVs 
and forwarded to the primary UAV with an MAC. 

At this step, 
 the delay is mainly caused by the primary UAV in verifying all the packaged transactions. 
Specifically, to verify the signatures and MACs of the transactions from the $K(t)-1$ UAVs, 
$(K(t)-1)(\epsilon_v+\epsilon_m)$ CPU cycles are required. 
Additionally, $\epsilon_v+\epsilon_m$ CPU cycles are needed to verify the request messages. 
Hence, the delay of this step is 
\vspace{-0.1cm}
\begin{equation}
    \mathcal{T}_1=\frac{K(t)(\epsilon_v+\epsilon_m)}{\text{$\partial^{p}_{ }(t)$}},\vspace{-0.1cm}
\end{equation}
where ${\partial^{p}(t)}$ is the computation resource, e.g., CPU speed, assigned by the primary UAV for the block consensus.

\paragraph{Pre-prepare}
In this stage, the primary UAV first packages the validated transactions into a block and then generates a signature for the block 
and $K(t)-1$ MACs for $K(t)-1$ replica UAVs individually. Hence, the delay of the primary UAV isolate
\vspace{-0.1cm}
\begin{equation}
    \mathcal{T}_{2,1}=\frac{\epsilon_s+(K(t)-1)\epsilon_m}{\text{$\partial^{p}(t)$}}.\vspace{-0.1cm}
\end{equation}

Subsequently, each replica UAV consumes $\epsilon_v + \epsilon_m$ CPU cycles for verifying the pre-prepare message, 
and $K(t)(\epsilon_v + \epsilon_m)$ CPU cycles for verifying the signatures and MACs of $K(t)$ transactions. 
Therefore, the delay is
\vspace{-0.1cm}
\begin{equation}
    \mathcal{T}_{2,2}=\max_{c\in\mathcal{U}_c}\frac{(K(t)+1)(\epsilon_\nu+\epsilon_m)}{\partial^{np}_{c}(t)}.\vspace{-0.1cm}
\end{equation}
$\partial^{np}_{c}({t})$ is the allocated CPU cycles of replica UAV $c$ for the blockchain consensus.
Therefore, the total delay of the pre-prepare step is $\mathcal{T}_{2} =\mathcal{T}_{2,1}+\mathcal{T}_{2,2}$.

\paragraph{Prepare}
In the phase, 
since the primary UAV only needs to verify $2\lceil F(t)\rceil+1$ MACs and signatures of the received prepare messages, 
the delay of the primary UAV is
\begin{equation}
    \mathcal{T}_3^p=\frac{(2\lceil F(t)\rceil+1)(\epsilon_\nu+\epsilon_m)}{\text{$\partial^{p}_{ }(t)$}}.
\end{equation}
Compared with the primary UAV, the replica UAVs consume extra $\epsilon_s + (K(t)- 1)\epsilon_m$ CPU cycles 
to generate a signature for the prepare message, and $K(t)-1$ MACs for other $K(t)-1$ full UAVs.
Hence, the delay of each replica UAV $c$ is
\begin{equation}
    \mathcal{T}_{3,c}^{np}=\frac{(2\lceil F(t)\rceil+1)(\epsilon_\nu+\epsilon_m)+\epsilon_s+(K(t)-1)\epsilon_m}{\partial^{np}_{c}(t)}.
\end{equation}
Based on the above analysis, the delay of this stage is 
\begin{equation}
    \mathcal{T}_3=\max\{\mathcal{T}_3^p,\mathcal{T}_{3,c}^{np}\},\forall c\in\mathcal{U}_c,
\end{equation}
   where $\mathcal{U}_c$ is the set of all replica UAVs.            

\paragraph{Commit}
After receiving $2\lceil F(t) \rceil+1$ consistent prepare messages, each UAV broadcasts a commit message to all other consensus UAVs  
and must accumulate at least $2\lceil F(t) \rceil+1$ consistent commit messages.
At this step, each UAV generates one signature for the commit message and $K(t)-1$ MACs 
for $K(t)-1$ full UAVs. Besides, each full UAV verifies $2\lceil F(t)\rceil+1$ signatures and MACs. 
Therefore, the total delay of this step is
\begin{equation}
    \mathcal{T}_4\!=\!\frac{\epsilon_s \!+\!(K(t)\!-\!1)\epsilon_m\! +\! (2\lceil F(t)\rceil\!+\!1)(\epsilon_v \!+\! \epsilon_m)}
    {\min\left\{\text{$\partial^{p}(t)$}, \partial _c^{np}(t)\right\}},\!\mathrm{~} \forall c\!\in\!\mathcal{U}_{c}.
\end{equation}
Therefore, the total consensus delay $\mathcal{T}_b(t)$ is
\begin{equation}
    \mathcal{T}_b(t)=\mathcal{T}_1(t)+\mathcal{T}_2(t)+\mathcal{T}_3(t)+\mathcal{T}_4(t).\vspace{-0.2cm}
\end{equation}

\subsubsection{Transaction Generation and Broadcasting}
After routing, the transaction waiting for the consensus is generated periodically 
by UAVs and includes the status of UAVs, 
i.e., the queue buffer, location, and information of neighboring UAVs.
Then, all generated transactions are broadcast to the entire network via the Gossip protocol for verifying {\cite{saldamli2022improved}}. 
After verification, the transactions are added to the transaction pool of the blockchain and then synchronized to the 
entire blockchain. Once a consensus is reached, the data transaction is recorded in the blockchain with tamper-proof.

\textcolor{black}{It is noted that message complexity $\mathcal{O}(K^2)$  of IPBFT  is controlled via a small  fixed $K$, and its radio airtime is a predictable overhead in per slot. This validates that the design of the lightweight blockchain does not affect real-time performance.}

\subsection{Routing Delay Model }
In LAINs, the E2E routing delay is the total transmission delay of 
multiple demands from  different source  UAVs to different destination UAVs, 
shown in Fig. {\ref{fig:Network_of_UAV}}.
Binary variable $\eta_{n m}^{r}(t) \!\in\! \{0,1\}$ denotes whether demand $r$ 
passes link $e_{n m}$ ($n \! \in\! \mathcal{I}\cup \mathcal{U},m \! \in \!\mathcal{U}\cup \mathcal{B}$), i.e.,
\vspace{-0.1cm}
\begin{equation}{\label{eta-link}}
    \eta_{n m}^{r}(t)\!=\!
    \left\{
        \begin{aligned}
            &1, \!\text{ if }\!  r \! \text { is transmitted via }\! e_{n m}\! \text { at time step }\! t,\\
            &0, \!\text{ otherwise}.
        \end{aligned}
    \right. \vspace{-0.1cm}
\end{equation}
Binary variable $\zeta_{n}^{r}(t) \!\in\! \{0,1\}$ indicates the UAV where the demand is located, i.e.,
\vspace{-0.1cm}
\begin{equation}{\label{zeta-demands}}
    \zeta_{n}^{r}(t)\!=\!
    \left\{
        \begin{aligned}
            &1, \!\text{ if }\!  r \! \text { is located on }  n  \in \mathcal{I}\cup \mathcal{U}\text { at time step }\! t,\\
            &0, \!\text{ otherwise}.
        \end{aligned}
    \right. \vspace{-0.1cm}
\end{equation}

Since the transmission delay is constrained by channel transmission rate $G$, 
we analyze the channel characteristics of both ground-to-air and air-to-air wireless communication links {\cite{channel_model_PL, zheng2025rotatable, 11222668}}.
Specifically, the path loss between nodes $n \in \mathcal{I} \cup \mathcal{U}$ and $m \in \mathcal{U} \cup \mathcal{B}$ is denoted as $L_{n m}(t)$, 
which remains constant within time step $t$, i.e.,
\vspace{-0.1cm}
\begin{equation}{\label{PLAG}}
    \begin{aligned}
        &L_{n m}(t)=20\log \left(\frac{4\pi  \lambda   }c d_{n m}(t)\right)\\
        &+\omega\left[{\Pr}_{n m}(t) \left(\eta_{n m}^\mathrm{LoS}(t)-\eta_{nm}^\mathrm{NLoS}(t)\right) +\eta_{n m}^\mathrm{NLoS}(t)\right],
    \end{aligned}
    \vspace{-0.1cm}
\end{equation} 
where $\lambda$ denotes the carrier frequency, and $c$ is the speed of light.
$\eta_{n m}^\mathrm{LoS}(t)$ and $\eta_{nm}^\mathrm{NLoS}(t)$  
represent the additional path losses under line of sight (LoS) and non-LoS (NLoS) propagations, respectively. 
When $\omega\!=\!0$, $L_{n m}(t)$ indicates the path loss between UAVs, 
and otherwise if $\omega\!=\!1$, $L_{n m}(t)$ denotes the path loss from $n \!\in\! \mathcal{I}$ to $m\!\in\!\mathcal{U}$, or from $n\!\in\!\mathcal{U}$ to $m\!\in\!\mathcal{B}$.
Besides, ${\Pr}_{n m}(t)$ is the probability of an LoS link between SDs/BSs and UAVs, i.e.,
\vspace{-0.1cm}
\begin{equation}{\label{P}}
    {\Pr}_{n m}(t)\!=\!\frac1{1\!+\!\varrho _1\!\exp\left\{\!-\!\varrho_2 \!
    \left[ \! \frac{180}{\pi}\!\arctan \! \left( \! \frac{h_{n m}(t)}{\chi _{n  m}(t)} \! \right)\!-\!\varrho _1 \! \right] \!\right\}},
\end{equation} 
where ${h_{n m}(t)}$ and ${\chi_{n m}(t)}$ are the differences of altitudes and horizontal  distances, respectively.
 $\varrho_1$ and $\varrho_2$ are the constant parameters {\cite{channel_model_PL}}.
\textcolor{black}{In particular, the antenna gains for UAVs, BSs, and SDs are modeled as constants incorporated into the link budget, thus not affecting the relative performance comparison of the algorithms.}

Due to high mobility and unstable data traffic fluctuations of LAINs,
the queue buffer is introduced for each node $n \!\in\! \mathcal{I}  \cup  \mathcal{U}$ to mitigate the congestion of networks.
Specifically,  at time step $t\!-\!1$, the amount of demands received by node $n$ is $\varepsilon_{n}^{re}(t\!-\!1)$. 
Additionally, at time step $t$, $C_{n}(t)$, $C_{n}^{max}$,
$\mathcal{C}_{n}(t)=\{r , \zeta_{n}^{r}(t)\!=\!1, \forall r \!\in\! R\}$, and $\varepsilon_{n}^{tr}(t)$
denote the queue length, maximum queue capacity, queued demand set, 
and the number of transmitted demands of node $n\!\in\! \mathcal{I}  \cup  \mathcal{U}$, respectively, i.e.,
\begin{align}{\label{re_tr}}
    \left\{\begin{aligned}
    &\varepsilon_{n}^{re}(t\!-\!1)\!=\!\sum_{r\in R} \sum_{{m}  \in \mathcal{I}  \cup \mathcal{U} } \eta^r_{{m}n}(t\!-\!1), 0\!\leq\! \varepsilon_{n}^{re}(t\!-\!1)\!\leq\!C_{n}^{max},\\
    &C_{n}(t)=\sum_{r\in R}\zeta_{n}^{r}(t), 0\leq C_{n}(t)\!\leq \!C_{n}^{max}, C_{n}(t)\!=\!\left\lvert \mathcal{C}_{n}(t)\right\rvert, \\
    &\varepsilon_{n}^{tr}(t)\!=\!\sum_{r\in \mathcal{C}_{n}(t)} \sum_{{m}  \in  \mathcal{U}\cup \mathcal{B}  } \eta^r_{n {m}}(t),\varepsilon_{n}^{tr}(t)\!=\! C_{n}(t).
\end{aligned}\right.
\end{align}
Here, UAVs transmit demands simultaneously,
which indicates that all demands received from time step $t\!-\!1$ are forwarded at  time step $t$.
Besides, to narrow the transmission delay gap of different demands in one time step,
we design an adaptive channel bandwidth allocation scheme, based on the bit length of demand 
$r \!\in\! \mathcal{C}_{n}(t)$ and total bandwidth $ B_n(t)$ at time step $t$.
Hence, the allocated bandwidth $B^r_n(t)$ is
\begin{equation}{\label{equ:BW}}
    B^r_n(t)\!= \!\frac{\mathcal{L}^{r} B_n(t)}{\sum\limits_{k  \in \mathcal{C}_{n}\!(t)}\!\mathcal{L}^{k}}, 
    \forall n \!\in\! \mathcal{I} \! \cup \!\mathcal{U},r \!\in \!\mathcal{C}_{n}(t), t \!\in \!T.
\end{equation}
Based on Shannon theory, at time step $t$, the transmission rate $G^r_{n m}\!(t)$ for demand $r$ from nodes $n\!\in\! \mathcal{I} \! \cup \!\mathcal{U}$ 
to $m\!\in\! \mathcal{U} \! \cup \!\mathcal{B}$ is
\begin{equation}{\label{Channel-Rate}}  
    G_{n m}^r(t)\!=\!B^r_n(t)\log_{2}\!\left(\!1\!+\!
    \frac{P^{tr}_{nm}(t) \!\cdot\! 10^\frac{\!-\!{L_{n m}(t)}}{10}}{\sigma^{2}_{n m}(t)}\!\right)\!, \forall r \!\in\! \mathcal{C}_{n}(t),
\end{equation}
where $P^{tr}_{n m}(t)$ and $\sigma^{2}_{n m}(t)$
are the transmission and noise powers between nodes $n$ and $m$, respectively. 
Furthermore, the delay for transmitting demands from node $n$ to node $m$ at time step $t$ is
\begin{equation}{\label{trans-delay1}}
    \begin{aligned}
         \mathcal{T}^{tr}_{n  m}(t)  \!=  \!\max_{r \in R} \mathcal{T}^{tr,r}_{n  m}(t),
       \forall n\!\in\!\mathcal{I}\cup \mathcal{U}, m  \! \in \!  \mathcal{U}\cup\mathcal{B}, t  \!\in \!  T,
    \end{aligned}
\end{equation}
where $\mathcal{T}^{tr,r}_{n  m}(t)\!=\!\frac{\mathcal{L}^{r}}{G^r_{n  m}(t)} \eta^{r}_{n  m}(t)$.
When routing path $\mathcal{P}^{r}_{ib}$ for transmitting demand $r$ 
from source SD $i$ to destination BS $b$ is determined, the  E2E delay $\mathcal{T}^{r} (t)$ is
\begin{equation}{\label{trans_delay}}
    \mathcal{T}^{r}(t)\!=\!\underset{ t \in T}{\sum}\underset{{e_{n m}\in \mathcal{P}_{ib}^r}}{\sum}\mathcal{T}^{tr}_{n  m}(t), 
    \forall r\!\in\! R, n \!\in \!\mathcal{I}\cup \mathcal{U}, m\! \in\! \mathcal{U}\cup\mathcal{B}.
\end{equation}

\subsection{Problem Formulation}
The objective is designed to minimize the total E2E delay $\mathcal{T}^{r}$ in (\ref{trans_delay})
and improve TSR $\mathbb{R}$ during the transmision in LAINs,
 with the mobility of UAVs and the malicious UAVs.
Therefore, the optimization problem is formulated as
\begin{equation}{\label{optimal}}
    \begin{aligned}
    \mathscr{P}0:\;&\underset{{\boldsymbol{\eta,\zeta}}}{\textrm{max}}\;
    \frac{\mathbb{R}}{\mathcal{T}^{r}} \\
       \textrm{s.t.}\;
        &\text{(\ref{distance_min}), (\ref{distance_max}),  (\ref{eta-link}), (\ref{zeta-demands}), (\ref{re_tr})},\\
        &\sum_{m  \in \mathcal{U} \cup \mathcal{B} }\!\eta^{r}_{n m}(t)\!=\!1, \forall r \!\in\! R, t \!\in\! T,e_{n m}\! \in \mathcal{E},\\
        &\forall  n\in \mathcal{I} \cup\mathcal{U}, m \in \mathcal{U}\cup \mathcal{B}, 
    \end{aligned}
\end{equation}
where $\boldsymbol{\eta}=\{\eta_{n m}^{r}(t),\forall t\in T, e_{n m}\in \mathcal{E}, 
n \in \mathcal{I} \cup \mathcal{U},m \in \mathcal{U}\cup \mathcal{B}, r \in R\}$, 
and $\eta_{n m}^{r}(t)$ denotes whether demand $r$ 
passes link $e_{n m}$ between $n \! \in\! \mathcal{I}\cup \mathcal{U}$ and $m \! \in \!\mathcal{U}\cup \mathcal{B}$.
The set $\boldsymbol{\zeta}=\{\zeta_{n}^{r}(t),\forall t\in T, 
n \in \mathcal{I} \cup \mathcal{U}, r \in R\}$, 
and $\zeta_{n}^{r}(t)$ indicates whether demand $r$ is located on $n \in \mathcal{I} \cup \mathcal{U}$.
It is noted that demand $r$ from node $n \!\in\! \mathcal{I} \cup \mathcal{U}$ can 
only be received by one another node $m \!\in \! \mathcal{U} \cup \mathcal{B}$.

Specifically, $\mathbb{R}$ is calculated through the number $N_{\text{fail}}$ of  
unsuccessfully transmitted demands  and the number $N_{\text{total}}\!=\!|R|$ of 
totally transmitted demands, detailed as $\mathbb{R}\!=\!1-[N_{\text{fail}} / {N_{\text{total}}}]$.
In particular, 
a failed transmission may be caused by the following situations: \textit{{a)}} $\mathcal{T}^{r}$  is longer than   $\mathcal{T}^{r}_{max}$.
\textit{{b)}} $\mathcal{T}^{tr}_{n  m}(t)$ or $\mathcal{T}_b(t)$ is longer than $\tau$.
\textit{{c)}} $\mathcal{T}^{r}+\mathcal{T}_b(t)$ is longer than $\tau$.
\textit{{d)}} The demands are not transmitted to the corresponding destination BS, denoted as  $arrive=0$.

Based on the above definition, up to time step $t$, the total number of failures is
\begin{equation}
    \begin{aligned}
       &N_{\text{fail}}(t) = N_{\text{fail}}(t - 1) + \\
&\begin{cases}
1, & \text{if }   \mathcal{T}^{r} \geq \mathcal{T}^{r}_{max} 
\text{ or } \mathcal{T}^{tr}_{n  m}(t) \geq \tau \text{ or } \mathcal{T}_{\text{b}}(t) \geq \tau\\
& \text{ or } \mathcal{T}^{r} + \mathcal{T}_{\text{b}}(t) > \tau  \text{ or } arrive\!=\!0, \\
0, & \text{otherwise,}
\end{cases} 
\forall t\in T.
    \end{aligned}
\end{equation}
Note that $N_{\text{fail}}(t)$ is initialized as 0 at time step $1$, 
and when $t=T$, $N_{\text{fail}}=N_{\text{fail}}(t)$.
Additionally, it is observed that $\mathscr{P}0$ is in the form of INLP and 
intractable to deal with. 
Therefore, in the following, a feasible solution is proposed via the MDP-based reformulation.

\subsection{Dec-POMDP based Problem Reformulation}
To adapt to the dynamically changing network environment 
and the local observations of each UAV, $\mathscr{P}0$ is reformulated as a Dec-POMDP form.
In the Dec-POMDP, each UAV $u$ is an agent and makes the routing decision to 
send demands to next alternative UAVs. 
Therefore, the agent set is the UAV set $\mathcal{U}$.
At each time step $t$, agent $u\in \mathcal{U}$ observes local state $o_{u}(t)$ 
from the environment and performs action $a_{u}(t)$ according to $o_{u}(t)$, 
with an immediate reward $\mathcal{R}_{u}(t+1)$, and 
the environment transferring to next observation state $o_{u}({t+1})$.
Specifically, $\left\langle o_{u}(t),a_{u}(t),\mathcal{R}_{u}(t+1),o_{u}({t+1}),f_{u}(t)\right\rangle$ 
indicates the transition experience and is detailed  as follows.
   \subsubsection{State Space} 
   At time step $t$, state $o_{u}(t)$ is the available information of agent $u$ and neighboring UAV $\kappa \in \Gamma_{u}(t)$, i.e.,
    \begin{equation}{\label{equ:observation}}
        \begin{aligned}
        o_{u}(t) = \{\Theta_{n}(t), \mathcal{C}_{n}(t),\mathbb{T}_n(t)\}, \forall n \in \{{u}\} \cup \Gamma_{u}(t).
        \end{aligned}
    \end{equation}
    $\Theta_{n}(t)$, $\mathcal{C}_{n}(t)$, and $\mathbb{T}_n(t)$ indicate the locations, 
   the sets of queued demands, and the credits of UAV $n$, respectively.
   At time step $t$, the  joint state $\bm{s}(t)$ aggregates observations of all agents and is 
   denoted as $\bm{s}(t)=\{o_{u}(t), u\in \mathcal{U}\}$.
   Thus, the state space is indicated as $\mathcal{S} = \{\bm{s}(t)| t\in T\}$.
    
    \subsubsection{Action Space} Agent $u$ makes decisions for each carried demand $r \!\in \!\mathcal{C}_{u}(t)$ independently, 
    and   $a_{u}(t)= \{a_{u}^r(t), r \!\in \!\mathcal{C}_{u}(t)\}$ denotes actions 
    for demands of $\mathcal{C}_{u}(t)$ at time step $t$.
    The sub-action $a_{u}^r(t)$ represents the next-hop neighboring node 
    selected by UAV $u$ for relaying demand $r$, i.e.,
    \begin{equation}
        \begin{cases}
            a_{u}^r(t) \in \{\Gamma_{u}(t)\}, \forall u \in \mathcal{U}_\mathtt{s} \cup \mathcal{U}_\mathtt{r},\\
            a_{u}^r(t) \in \{ \Gamma_{u}(t), Z _u(t)\},  \forall u \in \mathcal{U}_\mathtt{d},
        \end{cases}    
    \end{equation}
    where $Z _u(t)$ is the set of BSs that are connected with UAVs of $\mathcal{U}_\mathtt{d}$ at time step $t$. 
    If UAV $u \!\in\!\mathcal{U}_\mathtt{d}$ connects to the destination BS $b$ of demand $r \!\in\! \mathcal{C}_u(t)$, 
    it is directly transmitted to BS $b$ by UAV $u$.
    On the contrary, the demand is required to be relayed by neighboring UAVs.
    In time step $t$, $\bm{a}(t)=\{a_{u}(t), u\in \mathcal{U}\}$ represents the action set of all agents,
    and  $\mathcal{A} = \{\bm{a}(t)| t\in T\}$ denotes the action space.
    \subsubsection{Reward}  $\mathcal{R}_u(t)\!=\!\{\mathcal{R}_{u}^r(t)|r \!\in\! \mathcal{C}_{u}(t)\}$, 
    where $\mathcal{R}_{u}^r(t)$ is the reward that agent $u$ obtains after 
    transmitting demand $r$ at time step $t$. 
    It is noted that the routing paths of  demands are determined hop by hop.
    Hence, we calculate the reward with one hop as the minimum period.
    Considering $\mathscr{P}0$, for transmitting demand $r$ from UAV $u$ to UAV $\kappa$,
    we propose the reward value to be inversely correlated with $\mathcal{T}^{tr,r}_{u\kappa}(t)$ 
    and positively correlated with $\mathbb{T}_{u}(t)\!\cdot\! \mathbb{T}_{\kappa}(t)$.
    Particularly, $\mathbb{T}_{u}(t)\!\cdot\! \mathbb{T}_{\kappa}(t)$  represents
    the transmission success probability of link $e_{u\kappa}$. 
    It is rational since small $\mathbb{T}_{u}(t)$ and $\mathbb{T}_{\kappa}(t)$ lead to 
    the incorrect transmission and random loss of demands, directly decreasing the TSR. 
    Hence, the TSR $\mathbb{R}$ of one hop can 
    be reasonably represented by the value of $\mathbb{T}_{u}(t)\!\cdot\! \mathbb{T}_{\kappa}(t)$.
Hence, the reward is 
\begin{equation}{\label{equ:reward}}
    \mathcal{R}^{r}_{u}(t)=\frac{\mathbb{T}_{u}\!\cdot\! \mathbb{T}_{\kappa}}{\mathcal{T}^{tr}_{u\kappa}(t)\cdot \iota  +\varsigma}
    \cdot \eta_{u \kappa}^r(t), \forall r \in \mathcal{C}_u(t), \kappa \in \Gamma_{u}(t),
\end{equation}
where $\iota$ is set to ensure that $\mathcal{T}^{tr}_{u\kappa}(t)$ and $\mathbb{T}_{u} \cdot \mathbb{T}_{\kappa}$ 
are of the same order of magnitude. $\varsigma$ is the hyperparameter balancing the effect 
of the delay and credit values.
Besides, at time step $t$, the total reward of all demands is  
\begin{equation}{\label{equ:reward-total}}
    \mathcal{R}(t)=\sum_{r \in \mathcal{C}_u(t)} \sum_{u\in \mathcal{U}} \mathcal{R}^{r}_{u}(t), \forall t \in T.
\end{equation}

    \subsubsection{Transition Flag} Flag $f^r(t)$ indicates whether demand $r$ arrives at destination BS $b \!\in\! \mathcal{B}$, defined as
    \begin{equation}
        f^r{(t)}\!=\!
        \begin{cases}
        \!1,  \text{if }a_u^r{(t)} \text{ is destination BS }b, \\
        \!0,  \text{otherwise,} 
        \end{cases}
    \end{equation}
    and $\bm{f}(t)\!=\!\{f^r(t)| r \!\in\! R\}$ is the flag set for all demands.
    \subsubsection{Discount Factor} $\gamma_u$ is the cumulative reward of UAV $u$, 
    and the set of all UAVs is $\bm{\gamma}=\{\gamma_u| u\!\in\! \mathcal{U}\}$.
    A larger $\gamma$ indicates decisions focusing on the long-term reward.

At time step $t$, policy $\pi_{u}^r(t)$ drives agent $u$ to take action $a_{u}^r(t)$ for demand $r$
under observation $o_{u}(t)$.  The joint policy of demands on agent $u$ is denoted as 
$\bm \pi_u(t)=\{\pi_{u}^r(t), \forall r \!\in\! \mathcal{C}_u(t)\}$, 
and the policy of all agents is $\bm \Pi(t)=\{\bm\pi_{u}(t), \forall u \in \mathcal{U}\}$. 
After determining the Dec-POMDP and specific policy $\bm \Pi(t)$, 
the routing path for all demands can be obtained. 
Hence, the objective is transformed to obtain the optimal 
policy $\bm \Pi^{\star}(t)=\{\bm\pi_{u}^{\star}(t), \forall u\!\in\! \mathcal{U}\}$, 
and then the SP-MADDQN-based routing method is designed to 
obtain $\bm \Pi^{\star} =\{\bm \Pi^{\star} (t)| t \in T\}$  
for optimizing the total E2E delay and TSR.

\section{Algorithm Design}{\label{sec:Algorithm}}

\color{black}
\subsection{Trust-Aware CHU Selection and Update Algorithm}

We propose a trust-aware CHU selection algorithm that dynamically elects leaders based on a composite metric balancing the security (i.e., credit value) and processing capability, while preventing the leadership thrashing.

\subsubsection{Composite Score Design}
The selection is driven by a lightweight composite score $S_u(t)$ for each UAV $u$, which is designed to prioritize the security while balancing computational resources, i.e.,  
\begin{equation} 
S_u(t) = \mathbb{T}_u(t) \cdot (w_1 \cdot \tilde{\partial}_u(t)+ w_2 \cdot \tilde{C}_{u}(t) ), \label{eq:composite_score}
\end{equation}
where $\mathbb{T}_u(t)$ is the current credit value of UAV $u$. $\tilde{\partial}_u(t) = \partial_u(t) / \partial_{\text{max}}(t)$ and $\tilde{C}_{u}(t)={C}_{u}(t)/{C}_{\text{max}}(t)$  are the normalized computational and storage capabilities, relative to the highest capability in the network, respectively. $w_1$ and $w_2$ are the weights, and $w_1+w_2=1$. 

\begin{algorithm}[t]
\color{black}
\caption{\textcolor{black}{Trust-Aware CHU Selection and Update Algorithm}} 
\label{Alg:chu_selection}
\begin{algorithmic}[1]
\REQUIRE  \textcolor{black} {\(\mathcal{U}\),  \(\mathbb{T}_{thr}\), and stability period \(\mathcal{M}\).}
\ENSURE Full UAV set \(\mathcal{U}_f\) and primary UAV.
\STATE \textbf{Initialization:} \(\mathcal{U}_f \gets \emptyset\), counter \(\text{count}_u \gets 0, \forall u \in \mathcal{U}\). \label{line:init}
\FOR{each UAV \(u \in \mathcal{U}\)} \label{line:loop_start}
    \IF{\(\mathbb{T}_{u}(t) > \mathbb{T}_{thr}\) \AND \(\partial_u(t) > \partial_{thr}\)} \label{line:threshold_check}
        \STATE \(\mathcal{U}_f \gets \mathcal{U}_f \cup \{u\}\). \label{line:add_to_set}
    \ENDIF
\ENDFOR \label{line:loop_end}
\STATE Calculate \(S_u(t)\) for all UAVs, and select the highest-scoring UAV as the primarily via ({\ref{eq:composite_score}}). \label{line:calculate_score}
\FOR{each time step \(t\)} \label{line:time_loop_start}
    \STATE Perform consensus and routing tasks. \label{line:perform_tasks}
    \IF {\(t \mod \mathcal{M} == 0\)} \label{line:update_trigger}
        \STATE Update \(\mathcal{U}_f\) based on current states (lines \ref{line:loop_start}-\ref{line:loop_end}). \label{line:update_set}
        \FOR{each \(u \in \mathcal{U}_f\)} \label{line:inner_loop_start}
            \IF{\(\mathbb{T}_{u}(t) < \mathbb{T}_{thr}\) \OR \(\partial_u(t) < \partial_{thr}\)} \label{line:remove_check}
                \STATE Remove \(u\) from \(\mathcal{U}_f\), and reset \(\text{count}_{u}\). \label{line:remove_node}
            \ELSIF {\(\text{count}_{u} < \mathcal{M}\)} \label{line:hysteresis_check}
                \STATE \(\text{count}_{u} \gets \text{count}_{u} + 1\). \label{line:increment_counter}
            \ENDIF
        \ENDFOR \label{line:inner_loop_end}
        \STATE Update the primary UAV to highest \(S_u(t)\) in \(\mathcal{U}_f\). \label{line:update_primary}
        \IF {\(\max(\text{count}_u) \geq 2\mathcal{M}\) for \(u \in \mathcal{U}_f\)} \label{line:rotation_check}
            \STATE Remove longest-serving UAV \(u^*\) from \(\mathcal{U}_f\), and reset \(\text{count}_{u^*}\). \label{line:rotation_remove}
        \ENDIF \label{line:rotation_end}
    \ENDIF \label{line:update_end}
\ENDFOR \label{line:time_loop_end}
\end{algorithmic}
\end{algorithm}


The proposed trust-aware CHU (i.e., the full UAV)  selection and update algorithm is detailed in Algorithm~\ref{Alg:chu_selection}. First, the algorithm initializes the full UAV set $\mathcal{U}_f$ and the stability counters (line \ref{line:init}). By evaluating whether the credit value and computational capability of each UAV exceed their respective thresholds, it constructs the initial $\mathcal{U}_f$ (lines \ref{line:loop_start}-\ref{line:loop_end}). Subsequently, based on (\ref{eq:composite_score}), the composite score $S_u(t)$ is calculated for all candidate UAVs, and the UAV with the highest score is selected as the primary node (line \ref{line:calculate_score}).
At each time step, the algorithm performs consensus and routing tasks (line \ref{line:perform_tasks}). Every $\mathcal{M}$ time steps, an update to $\mathcal{U}_f$ is triggered (line \ref{line:update_trigger}). During the update, any UAV whose trust or capability falls below the required thresholds is removed from $\mathcal{U}_f$, and its stability counter is reset (lines \ref{line:inner_loop_start}-\ref{line:inner_loop_end}). For UAVs that remain qualified, the stability counter is incremented, providing a hysteresis period of $\mathcal{M}$ slots to shield them from removal due to minor score fluctuations (lines \ref{line:hysteresis_check}-\ref{line:increment_counter}). The primary UAV is then updated to the node with the highest $S_u(t)$ within the revised $\mathcal{U}_f$ (line \ref{line:update_primary}). Finally, we perform a fairness-driven rotation to balance the long-term load (lines \ref{line:rotation_check}-\ref{line:rotation_remove}). This mechanism removes a UAV from $\mathcal{U}_f$ once it has completed $2\mathcal{M}$ time steps of service. The algorithm employs hysteresis  via an $\mathcal{M}$-slot shield  and rotation via retirement after $2\mathcal{M}$ slots, to ensure the cluster stability and long-term fairness.
\color{black}

\subsection{SP-MADDQN-based Routing Algorithm}
Based on the trust-aware full UAV selection and update mechanism, we propose the MADDQN-based routing algorithm in the multiple sources and destinations scenarios. The routing algorithm improves the learning speed and efficiency in dynamic scenarios, combining the SHERB and PER mechanisms {\cite{11149011,11122503}}.  
Particularly, 
PER allows the agent to sample experiences with higher importance more frequently based on their priorities. 
This is particularly beneficial when dealing with large-scale states and action spaces, 
since it enables the algorithm to focus on the most informative experiences for faster and more effective learning.
SHERB constructs the experience buffer of each agent 
via embedding the distances between nodes into rewards, i.e.,
\begin{figure}[t]
  
    \centering
    \includegraphics[width=0.90\linewidth]{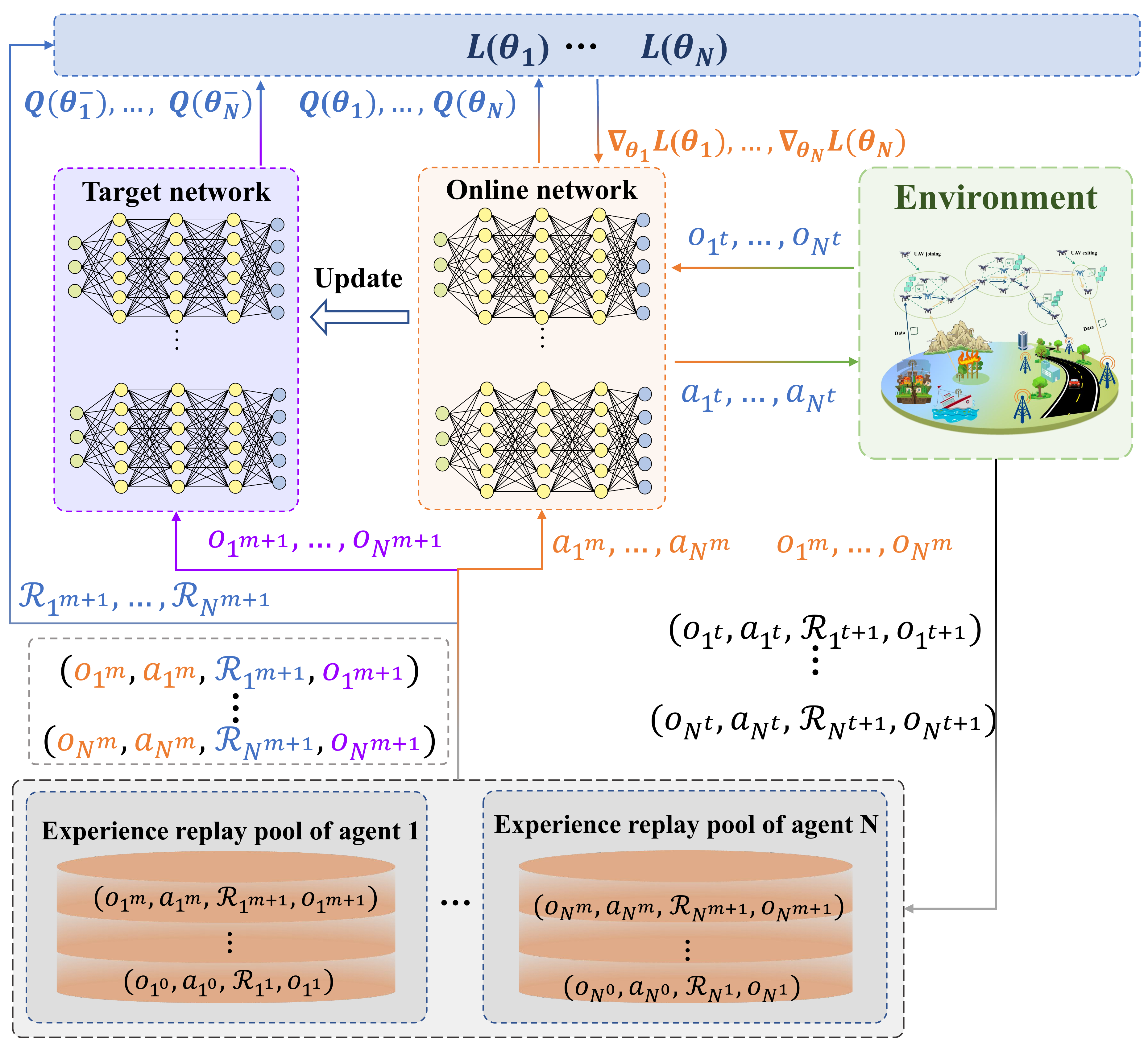}
    
    \textcolor{black}{\caption{\label{fig:consensus_update} 
    SP-MADDQN algorithm framework for trust routing in LAINs.}}
    
\end{figure}

\begin{equation}{\label{equ:Algorithm-reward}}
    \begin{aligned}
    \mathcal{R}_{u}^r(t)=\frac{\mathbb{T}_{u}\cdot \mathbb{T}_{\kappa}}{\mathcal{T}^{tr}_{u\kappa}(t)\cdot \iota  +\varsigma} 
    \cdot &  \frac{d_{u\kappa}(t)}{d_{u\kappa}(t)+d_{\kappa b}(t)} \eta_{u \kappa}^r(t),\\
    &\forall r \in  \mathcal{C}_{u}(t), \kappa \in \Gamma_{u}(t), t\in T,
    \end{aligned}
\end{equation}
where $d_{u\kappa}(t)$ and $d_{\kappa b}(t)$ represent the distances from UAV $u$ to UAV $\kappa$ and from UAV $\kappa$ to destination BS $b$, respectively.
Since a larger $d_{u\kappa}(t)$ requires a fewer hop counts and saves the time cost,
$d_{u\kappa}(t)$ is designed to be positively correlated with rewards.
Meanwhile, ${d_{u\kappa}(t)+d_{\kappa b}(t)}$ is the minimal transmission distance 
from the selected next-hop UAV $\kappa$ to UAV $u$ and BS $b$.
Agent $u$ may obtain the greater value of rewards, 
when $d_{u\kappa}(t)+d_{\kappa b}(t)$ is closer to 
the shortest straight distance $d_{u b}(t)$,  and vice versa.

\begin{algorithm}[t!]
    \caption{{\label{algorithm-MADDQN}}SP-MADDQN-based Routing Algorithm}
    \begin{algorithmic}[1]
    \REQUIRE {$\mathcal{I}$, $\mathcal{U}$, $\mathcal{B}$, $\mathcal{E}$, $R$, ${\bm{\alpha}}$, and ${\bm{\gamma}}$ }.
    \ENSURE Optimal policy ${\bm{\Pi^{*}}}$.
    \STATE\textbf{Initialization:}\label{initialize} Initialize the network environment, hyper-parameters, 
    ERB set $\bm {\mathcal{D}}$, and the online 
    and target network parameters $ \theta_u $  and $ \theta_u^{-} $ for each agent $u \in \mathcal{U}$, respectively.
    \FOR{each episode} {\label{episode}}
    \FOR{$t = 1, \dots, T$}
    \FOR{$u = 1,\dots, U$} 
    \STATE {\label{episode-initi}} The observation of agent $u$ is set as $o_{u}(t)$.
    \STATE {\label{Allocate-BW}} Allocate the bandwidth based on queued demands in $\mathcal{C}_u(t)$ via (\ref{equ:BW}). 
    \FOR{$r = 1, \dots, |\mathcal{C}_u(t)|$}{\label{FOR-demands}} 
        \STATE Select action $a_{u}^r(t)$ for demand $r$ under observation $o_{u}(t)$ 
        using a $\epsilon$-greedy policy.
        \STATE Execute $a_{u}^r(t)$, and obtain reward $\mathcal{R}_{u}^r(t\!+\!1)$.
    \ENDFOR{\label{ENDFOR-demands}}
   
     \IF{$|\mathcal{D}_u| > D$} {\label{sample}}
        \STATE  Select $D$ samples from $\mathcal{D}_u$ according to the importance priority.
        \STATE {\label{target}} Compute Q-target value $y_{u}(t)$ via (\ref{Q-target}).
        \STATE {\label{update-theta}} Update $\theta_u$ via the gradient descent in (\ref{theta-update}). 
        \STATE {\label{update-theta-}} Periodically update target network parameter $\theta_u^{-}$ via (\ref{target-undate}) every $\mathcal{W}$ steps.
    \ENDIF 
    \ENDFOR
    \STATE {\label{update-environment}} Update the environment, 
    and set observation $o_{u}(t)\leftarrow o_{u}({t + 1})$ for all agents in $\mathcal{U}$.
    \STATE {\label{Store-E}}Store transition $( o_{u}(t), a_{u}^r(t), \mathcal{R}_{u}^r(t\!+\!1), o_{u }({t\! + \!1}))$
     of each demand $r$ into corresponding $\mathcal{D}_u$.
    \ENDFOR
    \ENDFOR
    \end{algorithmic}
\end{algorithm}

Furthermore, the DDQN algorithm is presented via combining double Q-learning with DQNs
to approximate Q-value functions via deep neural networks (DNNs) and decouple the action selection 
and calculation of Q-target values. 
Besides, for each agent $u$, the DDQN contains the online network with parameter $\theta_{u}$ 
and the target network with parameter $\theta^{-}_{u}$.
The historical experience is denoted as 
$\mathcal{D}_u\!=\!\{(o_{u}(t), a_{u}^r(t), \mathcal{R}_{u}^r({t+1}), o_{u}({t+1}))\vert r \!\in\! \mathcal{C}_u(t)\}$,
which is stored in the experience replay buffer (ERB) of agent $u$.
According to the importance priority, $D$ historical experience transition tuples 
are sampled from $\mathcal{D}_u$. Then, $L(\theta_{u})$ is calculated as the mean squared 
error between Q-value function $Q(o_{u}(t), a_{u}(t);  \theta_{u})$ and Q-target $y_{u}(t)$, i.e.,
\begin{equation}{\label{loss}}
    L(\theta_{u}) = \mathbb{E}_{\nu  \thicksim \mathcal{D}_{u}} \left[ \left( y_{u}(t)  
    - Q(o_{u}(t), a_{u}(t);  \theta_{u}) \right)^2 \right],
\end{equation}  
where tuple $\nu $ is a transition data of $\mathcal{D}_{u}$, and $y_{u}(t)$ is
\begin{equation}{\label{Q-target}}
    \begin{aligned}
         &y_{u}(t)  =  \mathcal{R}_{u}({t+1}) +  \\
          &\gamma_u Q(o_{u}({t+1}), \arg\max_{a^{\prime}_{u}(t)} Q(o_{u}({t+1}), a^{\prime}_{u}(t);  \theta_{u}^b)|\theta^{-}_{u}).
    \end{aligned}
\end{equation}
Additionally, $L(\theta_{u})$ is minimized via constantly training $\theta_{u}$. 
$\nabla_{\theta_{u}}L(\theta_{u})$ denotes the gradient of $L(\theta_{u})$ 
and is leveraged to update $\theta_{u}$ via the gradient descent, i.e., 
\begin{equation}{\label{theta-update}}
    \theta_{u} \leftarrow \theta_{u} - \alpha_{u} \nabla_{\theta_{u}} L(\theta_{u}).
\end{equation}
Here, $\alpha_{u}$ is the learning rate.
To stabilize training and improve the convergence,
parameter $\theta^{-}_{u}$ is periodically updated in every $\mathcal{W}$ steps 
to match the online network parameter $\theta_{u}$, 
with a soft update coefficient $\mu$, i.e.,
\begin{equation}{\label{target-undate}}
    \theta_{u}^{-}\leftarrow \mu   \theta_{u} +(1-\mu)\theta_{u}^{-}.
\end{equation}

The detail of the proposed SP-MADDQN-based routing method is shown in Algorithm {\ref{algorithm-MADDQN}}.
Firstly, the network environment, parameters, ERB set $\bm{\mathcal{D}}$ are initialized ({line \ref{initialize}}).
At  each episode,  $o_{u}(t)$ is the observation of agent $u$  (lines {\ref{episode}}-{\ref{episode-initi}}).
Further, the bandwidth of agent $u$ is allocated according to the queued demands in $\mathcal{C}_u(t)$ (line \ref{Allocate-BW}).
In the light of  observation $o_{u}(t)$, sub-action $a_{u}^r(t)$  is selected for demand $r$ via a $\epsilon$-greedy policy
and is executed to obtain reward $\mathcal{R}_{u}^r({t+1})$ (lines {\ref{FOR-demands}}-{\ref{ENDFOR-demands}}). 
If the mini-batch $D$ is satisfied, $D$ samples are selected from $\mathcal{D}_{u}$ based on the importance priority,
to calculate Q-target value $y_{u}(t)$ via (\ref{Q-target}) (lines {\ref{sample}}-{\ref{target}}). 
Besides, the gradient descent steps are performed to update $\theta_{u}$ (line {\ref{update-theta}}).
Every $\mathcal{W}$ steps,
the target network parameter $\theta_{u}^{-}$ is periodically updated (line {\ref{update-theta-}}).
After finishing the actions of all demands, environments are updated.
Then, next observation $o_{u}({t \!+ \!1})$ is obtained to update $o_{u}(t)$ (line \ref{update-environment}).
Additionally, $( o_{u}(t), a_{u}^r(t), \mathcal{R}_{u}^r(t\!+\!1), o_{u }({t\! + \!1}))$
is stored into corresponding $\mathcal{D}_u$ for training (line \ref{Store-E}).

 \color{black}
\subsection{Computational Complexity and Convergence Analysis}
\subsubsection{Computational Complexity Analysis}In Algorithm {\ref{Alg:chu_selection}}, the per-update cost within period $\mathcal{M}$ is dominated by sorting $U$ candidates, i.e., $\mathcal{O}(U \log U)$, which defines the overall worst-case complexity. This ensures scalability in networks with hundreds of UAVs.

In Algorithm {\ref{algorithm-MADDQN}}, assume that $W_i$ is the width of $i$-th layer of neural networks, and $M$ denotes the total number of layers. For the entire forward propagation process, the computational complexity of the agent is \(\mathcal{O} (S\cdot A \cdot \sum_{i = 1}^{M - 1}W_i \cdot W_{i + 1})\), since the input and output layers are the dimensions of state space $S$ and action space $A$, respectively.
Suppose that \(h\) hops are needed for a demand during routing, and the complexity becomes $\mathcal{O} (h \cdot S \cdot A \cdot \sum_{i=1}^{M-1} W_i \cdot W_{i+1})$. When there exist \(n\) demands, the computational complexity is \(\mathcal{O} (n \cdot h \cdot S\cdot A\cdot\sum_{i = 1}^{M - 1}W_i \cdot W_{i + 1})\). 
 \subsubsection{Convergence Analysis}The core of SP-MADDQN is based on DDQNs whose convergence is theoretically guaranteed under the experience replay, target network, and appropriate learning rate \cite{van2016deep, mnih2015human}. The SHERB and PER proposed as sampling strategies for the replay area enhance learning efficiency and stability by focusing on high-value samples. This mechanism does not change the convergence basis of DDQN but accelerates the training process and reduces variance \cite{schaul2015prioritized}. 

The result indicates that the computational complexity  $\mathcal{O} (S\cdot A \cdot \sum_{i = 1}^{M - 1}W_i \cdot W_{i + 1})$ of algorithms linearly scales with the number of demands \( n \) and the average hop count \(h\). Furthermore, the online computational load of each UAV is independent of the total network size, owing to the decentralized multi-agent design in which each agent only relies on its local observation \( o_u(t) \) in (\ref{equ:observation}) \cite{4445757,7989385}. The relative low computational load and independence distribute the workload across all UAVs and avoids the centralized bottleneck, which is a key feature for scalability and makes it feasible for practical deployment in UAV networks. 

 Moreover, the algorithm framework follows the centralized training and distributed execution paradigm, which is well established for deploying DRL {\cite{10233034, 10547350,11045994}}. The computationally intensive training is performed offline, while online execution requires only a lightweight forward pass through a pre-trained network. Thus, the operational overhead of the approach remains manageable even in large-scale networks. 
\color{black}
\section{Simulation Results and Analyses\label{sec:Simulation Results}}
In this section, we conduct a couple of simulations via Python $3.12$ and PyTorch $2.5.1$. Table I lists the specific parameters \cite{Block_Size-1, 11284890}. 
In LAINs, nodes are distributed within a range of 15 km$\times $5 km,
and the altitude range of UAVs is within $[0.2,0.4]$ km.
Besides, the size of demanded data is randomly set within $[400,600]$ kbits.
\textcolor{black}{The hyperparameters for the SP-MADDQN algorithm are detailed in the following. Specifically, the training is conducted over 5,000 episodes, with each episode comprising  1,000 steps. An $\epsilon$-greedy policy is employed for exploration, with $\epsilon$ decaying linearly from 1.0 to 0.01 over the first 80\% of training. The policy and target networks are identical  multi-layer perceptrons with two 256-unit ReLU hidden layers. The optimization uses Adam with $\alpha=0.005$. The target network is softly updated every 100 steps with a parameter of $\mu = 0.01$. The experience replay buffer holds $10^6$ transitions. With PER, the priority exponent $\alpha=0.6$, and the importance-sampling exponent $\curlyvee$ is annealed from 0.4 to 1.0.}

\begin{table}[!t]
    \renewcommand\arraystretch{1}
    \begin{center}
        \color{black}
       \caption{Parameter Notation}
       {\label{table1}}
        \fontsize{8}{10}\selectfont{
        \begin{tabular}{|c|c||c|c|}
            \hline
            Notation & Value & Notation & Value \\  
            \hline
            $\tau$ & 0.5 s& $d_{min}$ & 10 m \\            
            \hline            
            $\beta$ & 0.5 & $\varsigma$ & 20 \\            
            \hline
            $P^{tr}_{n m}(t)$ & 40 dBm  & $\sigma^{2}_{n m}(t)$ & -110 dBm \\ 
            \hline
            $f_c$ & 2.4 Ghz& $c$ & 3 $\times 10^8$ m/s  \\ 
            \hline
            $\eta_{n^t m^t}^\mathrm{LoS}$& 0.1 dB & $\epsilon_v $ & 1 M CPU cycles\\
            \hline
            $\varrho_1$ & 5.0188 & $\partial $ & (2,4) G CPU cycles/s \\
            \hline
            $\varrho_2$ & 0.3511 & $\epsilon_s$ & 1 M CPU cycles \\
            \hline
            $\eta_{n^t m^t}^\mathrm{NLoS}$ & 21 dB & $\epsilon_m$ & 1 M CPU cycles \\   
            \hline            
            $B$ &  2.4 MHz  & $ S_b$ & 15KB  \\
            \hline
            ${\nu}_{\min}$ & 3 m/s    & ${\nu}_{\max}$ &   5 m/s \\
            \hline
            $ \phi_{u}$ & $[0, 2\pi)$  & $ \delta  _{u}$ & $[-\frac{\pi}{2}, \frac{\pi}{2}]$  \\
            \hline
            $w_1 $ &  0.5    & $w_2$ & 0.5 \\
            \hline
        \end{tabular}
        }
    \end{center}
\end{table}
 \color{black}

To assess the performance of the proposed adaptive weight mechanism, 
two benchmark methods are introduced, in which the calculation of 
\(\psi _{u}^0(t)\) follows the same formula in ({\ref{Adaptive_weights}}), 
described as following.
\subsubsection{Average weight methods} 
    $\psi_{u}^1(t)$ and $\psi_{u}^2(t)$ are the average value of $1-\psi_{u}^0(t)$, expressed as 
    \begin{equation}
       \left\{  
        \begin{aligned}    
        \psi _{u}^0(t)  &= \frac{1}{2}\times \frac{\mathbb{T}_{thr}}{\mathbb{T}_{u}(t)}, \\
        \psi _{u}^1(t) &= \psi _{u}^2(t) = 0.5 \times (1-\psi _{u}^0(t)).
        \end{aligned}
        \right.
    \end{equation}
\subsubsection{Random weight methods}  $\psi _{u}^1(t)$ is modeled as a random variable
     uniformly distributed within the range $[0.2(1-\psi_{u}^0(t)), 0.8(1-\psi_{u}^0(t))]$, i.e.,
    \begin{equation}
        \left\{
        \begin{aligned}
            \psi _{u}^0(t) &= \frac{1}{2} \times \frac{\mathbb{T}_{thr}}{\mathbb{T}_{u}(t)}, \\
            \psi _{u}^1(t) &\sim  {\mathcal{N}}(0.2(1-\psi _{u}^0(t)), 0.8(1-\psi _{u}^0(t))),  \\
            \psi _{u}^2(t) &= 1-\psi _{u}^0(t) - \psi _{u}^1(t).
    \end{aligned} 
    \right.
    \end{equation}

The experimental setup is configured as follows, to simulate the effectiveness of the trust evaluation methods. 

  \textit{1)} The number of UAVs is \(12\), and the number of malicious UAVs  is $2$.
  
  \textit{2)} The credit value is initialized as 1, representing UAVs are completely 
    trust when departing from the GCS.
  
  \textit{3)} Considering the stringent security requirements, $\mathbb{T}_{thr}$ is set as 0.8, 
    and the probability range of malicious behaviors is set to $[0.1, 0.5]$ {\cite{trust_parameters}}. 
    Particularly, in this paper, we select $0.4$ and $0.2$ as 
    the probabilities of all malicious behaviors for simulations. 
    Thus, the demand forwarding probability, the trust interaction probability with high-trust UAVs, 
    and probe packet reception probability of malicious UAVs are characterized by 
    $p_1 \! \in \! \{0.6, 0.8\}$, $p_2 \! \in \! \{0.6, 0.8\}$, and $p_3 \! \in \! \{0.6, 0.8\}$, respectively.
    Furthermore, the simulation parameter space $\Lambda$ of $(p_1,p_2,p_3)$ is
    \begin{equation}{\label{parameter-space}}
        \Lambda \!=\!\begin{bmatrix}
            (0.6,0.6,0.6)   (0.6,0.6,0.8)   (0.6,0.8,0.6)   (0.6,0.8,0.8)\\
            (0.8,0.6,0.6)   (0.8,0.6,0.8)  (0.8,0.8,0.6)   (0.8,0.8,0.8)
        \end{bmatrix}.
    \end{equation}
   
    \textit{4)} The evaluation metric is defined as the minimum time steps that
    can identify all malicious UAVs, i.e.,
    \begin{equation}
        \Upsilon^{\text{detect}}_u \!=\! \min\{t \in \mathbb{N}^+ | \mathbb{T}_{u}(t)\!<\!\mathbb{T}_{thr}\}, \forall i \!\in\! \mathcal{U}, t\!\in\! T.
    \end{equation}

\begin{figure}[t]
    \centering
    \includegraphics[width=1\linewidth]{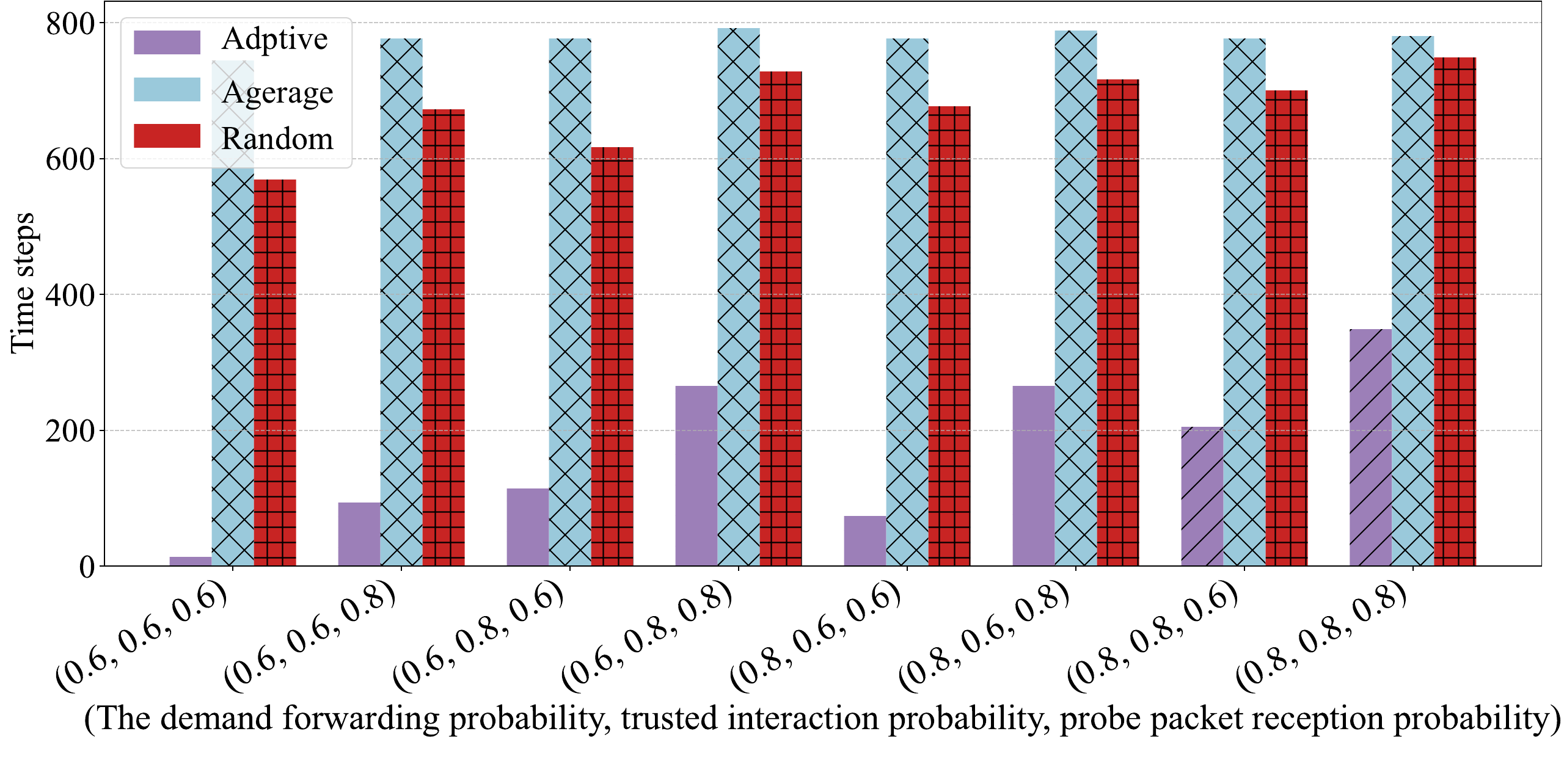}
    \vspace{-0.5cm}
    \textcolor{black}{\caption{ The minimum time step for identifying malicious UAVs with different simulation parameters in space $\Lambda $.}
  \label{fig:trust-value}}
\end{figure}

To evaluate the efficiency of the proposed adaptive weight approach for calculating the credit values, 
results are compared with the average and random weight methods in Fig. {\ref{fig:trust-value}}.
Specifically, we evaluate $\Upsilon^{\text{detect}}_u$ across a diverse range of 
simulation parameters in (\ref{parameter-space}).  For identifying the malicious UAVs,
 it requires the fewest time steps when $(p_1, p_2, p_3)\!=\!(0.6, 0.6, 0.6)$. 
On the contrary, when $(p_1, p_2, p_3)\!=\!(0.8, 0.8,0.8)$, the most time steps are needed. 
It lies that the values of $p_1$, $p_2$, and $p_3$ decrease, 
indicating that the frequencies of incorrect behaviors of malicious UAVs are higher, 
resulting in the increment of the evaluated incorrect behavior ratio 
 (i.e., $1 \!-\!\mathbb{T}^{\text{\tiny{D1}}}_{u}(t)$, $1\!-\!\mathbb{T}^{\text{\tiny{D2}}}_{u}(t)$, and $1\!-\!\mathbb{T}^{\text{\tiny{D3}}}_{u}(t)$). 
Furthermore, compared with other methods, the proposed method consistently obtains 
the minimum time steps in diverse simulation parameters.
The reason lies in that the trust weights in the adaptive weight trust model are proportional to incorrect behavior rates, 
and the credit values of UAVs are dynamically adjusted at a quadratic rate.
Therefore, the penalties for incorrect behaviors become more severe, and  the detection of malicious UAVs is accelerated.
In the same case, the proposed method obtains better performances than both the average and random weight trust methods.

\begin{figure}[t]
    \centering
    \includegraphics[width=0.6\linewidth]{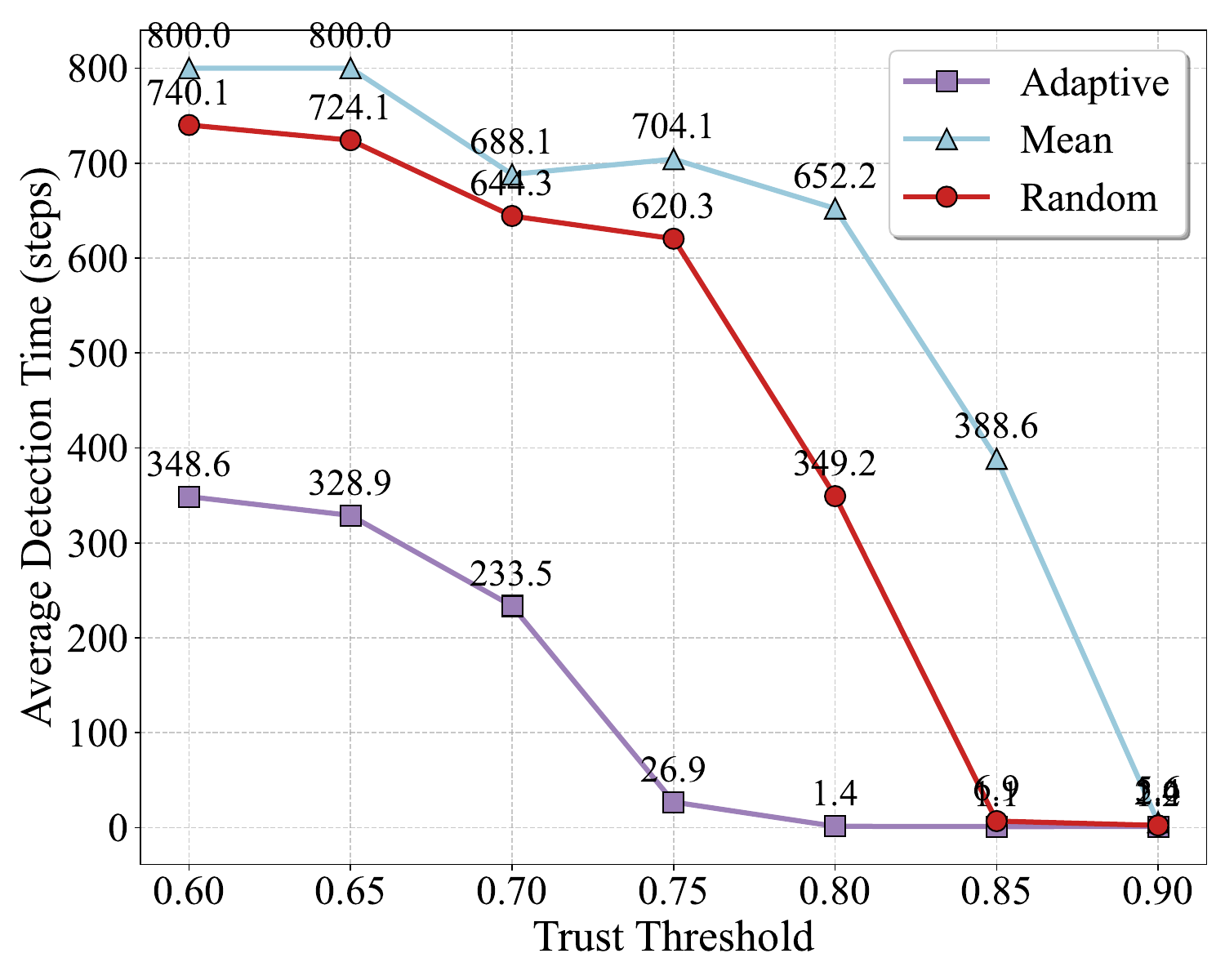}
    \textcolor{black}{\caption{\textcolor{black}{ The minimum time step for identifying malicious UAVs with different trust thresholds.}}
  \label{fig:trust-thr}}
 \end{figure}

\textcolor{black}{ 
In Fig. 4, the sensitivity of trust threshold $\mathbb{T}_{thr}$ is evaluated and the proposed adaptive trust evaluation method in Section III-B are compared with the two benchmark methods. Results demonstrate that the adaptive method consistently requires the minimum time steps to recognize the UAVs with malicious behaviors at the probability tuple $(p_1, p_2, p_3)= (0.5,0.5,0.5)$, compared with other two methods.
In addition, a higher trust threshold requires a smaller time step to isolate malicious UAVs, resulting in a higher TSR and a lower average E2E latency. It is the result that a stricter threshold ($\mathbb{T}_{thr}=0.9$) enables the system to proactively and rapidly isolate potentially malicious or unreliable nodes, thereby protecting the integrity and efficiency of the routing paths. Conversely, a lower threshold ($\mathbb{T}_{thr}=0.6$) retains more time slots for malicious nodes in the network, which may increase the risk of allowing malicious behavior while degrading the overall performance.
}

\begin{figure}[t]
    \vspace{-0.3cm}
    \centering
    \subfloat[\textcolor{black}{ }]{\centering
    \includegraphics[width=0.5\linewidth]{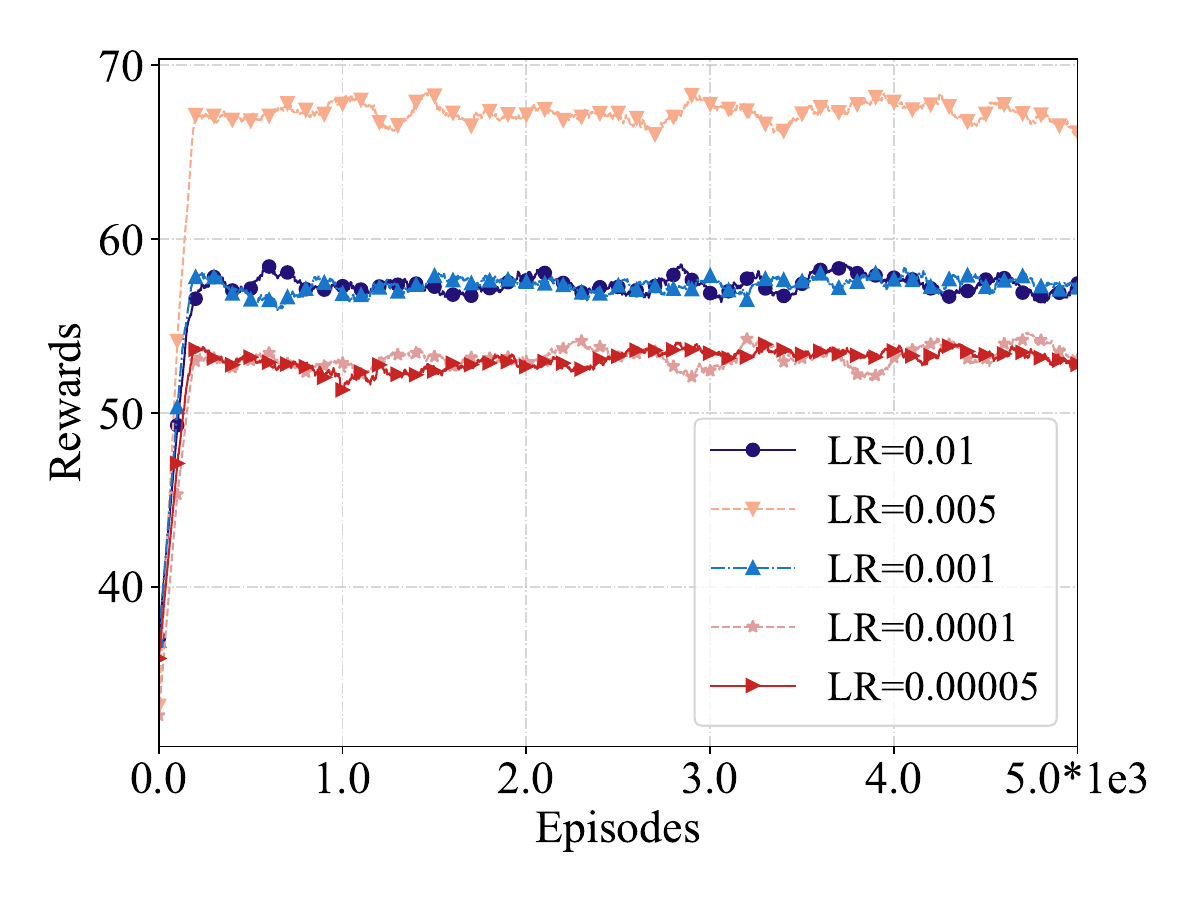}}
    \subfloat[\textcolor{black}{ }]{\centering
    \includegraphics[width=0.5\linewidth]{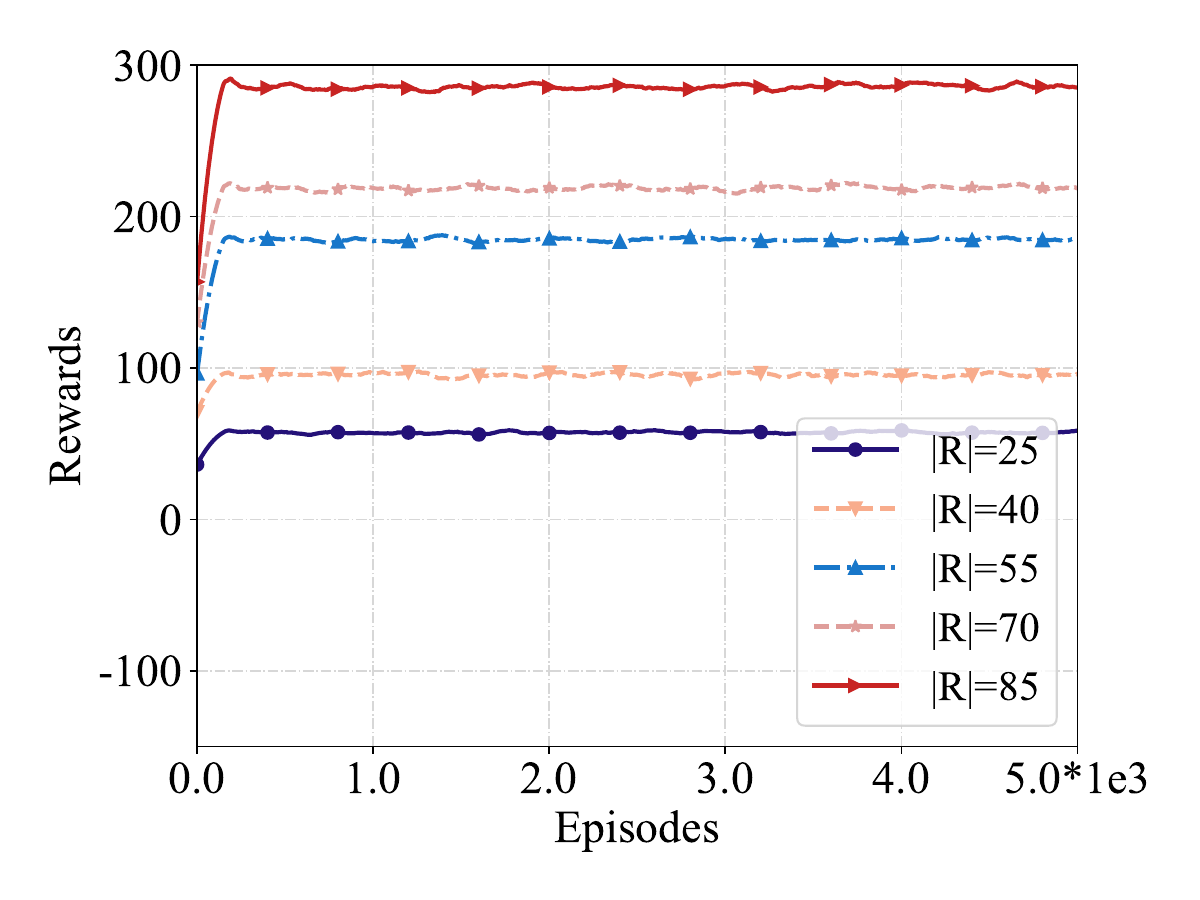}}
    \caption{Comparison of convergence performances for transmitting demands 
    with 8 UAVs and 2 malicious UAVs during training SP-MADDQN.
    (a) Different learning rates 25 demands.
    (b) Different demand numbers. 
    }
    {\label{fig:Rewards-proposed}}
\end{figure}

In Fig. {\ref{fig:Rewards-proposed}}, to evaluate the convergence performance
of the proposed SP-MADDQN algorithm, 
 the relationship between the accumulative rewards 
and episodes is presented with different learning rates (LRs)
and demand numbers. 
Fig. {\ref{fig:Rewards-proposed}}(a) clearly shows that different LRs result in varied convergence behaviors.
In the initial episodes (especially around the first 500), 
rewards change drastically for all LRs, and  tend to stabilize gradually with the increase of episodes.
Notably, LR $\!=\!$ 0.005 achieves the highest and most stable rewards in the long term, compared with other LRs.
When LR $\!=\!$ 0.01 and LR $\!=\!$ 0.001, the convergence value of rewards descends.
However,   LR $\!=\!$ 0.0001 has the worst convergence, with rewards staying at a low level and fluctuating more.
This indicates that an too small or too large learning rate may slow down the convergence process and limit the final reward values. 
Considering the execution effect, LR $\!=\!$ 0.005 is selected for subsequent simulations.
Fig. {\ref{fig:Rewards-proposed}}(b) shows the rewards of the proposed algorithm across 
episodes under diverse numbers of demands. 
It is noticeable that different numbers of demands have different impacts on the convergence and value of rewards. 
As the number of demands grows, the rewards converge rapidly in the early episodes. 
The reason lies in that a higher demand value can meet the batch size requirements for training more quickly.
Besides, more demands obtain high rewards, 
since the total reward of LAINs is defined as the sum of all transmitted demands in ({\ref{equ:reward-total}}).

\begin{figure}[t]
    \vspace{-0.3cm}
    \centering
    \subfloat[\textcolor{black}{ }]{\centering
    \includegraphics[width=0.5\linewidth]{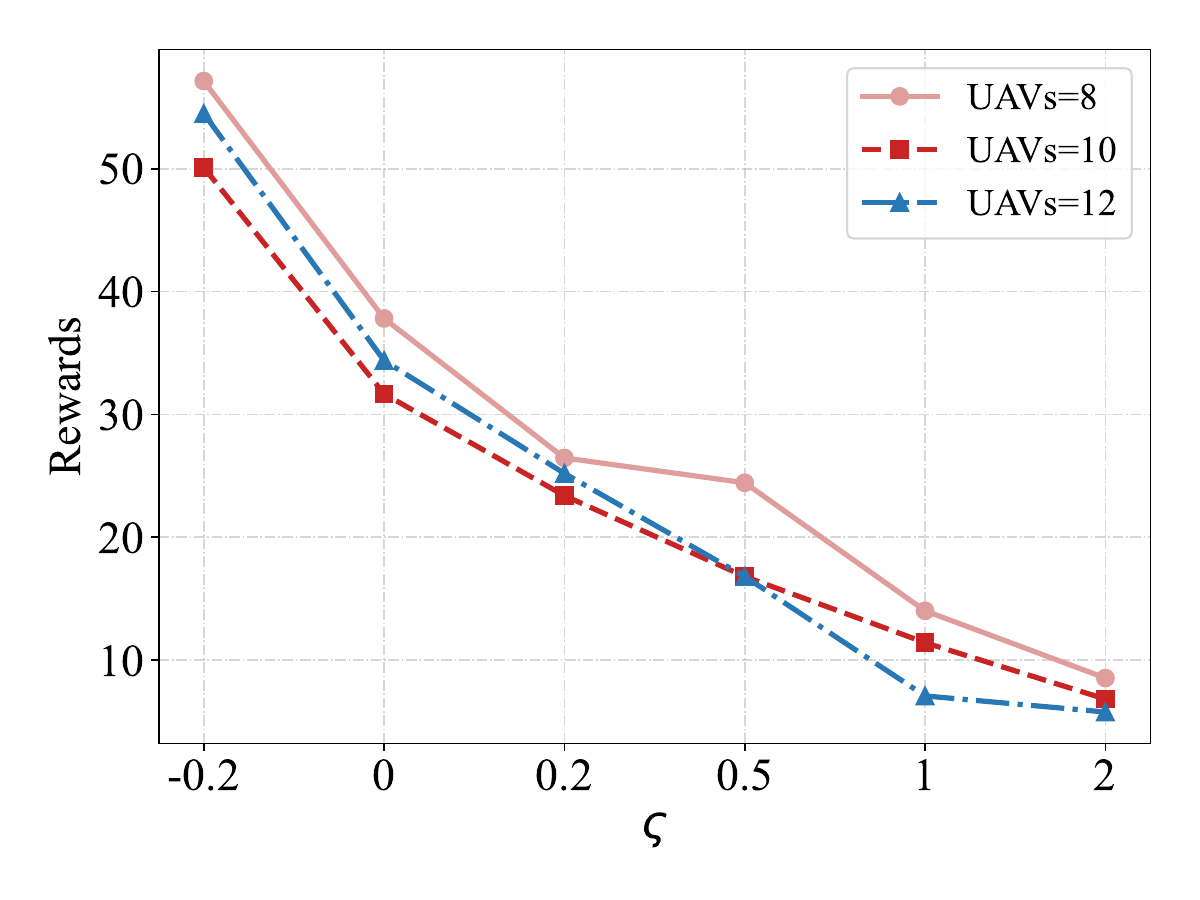}}
    \subfloat[\textcolor{black}{ }]{\centering
    \includegraphics[width=0.5\linewidth]{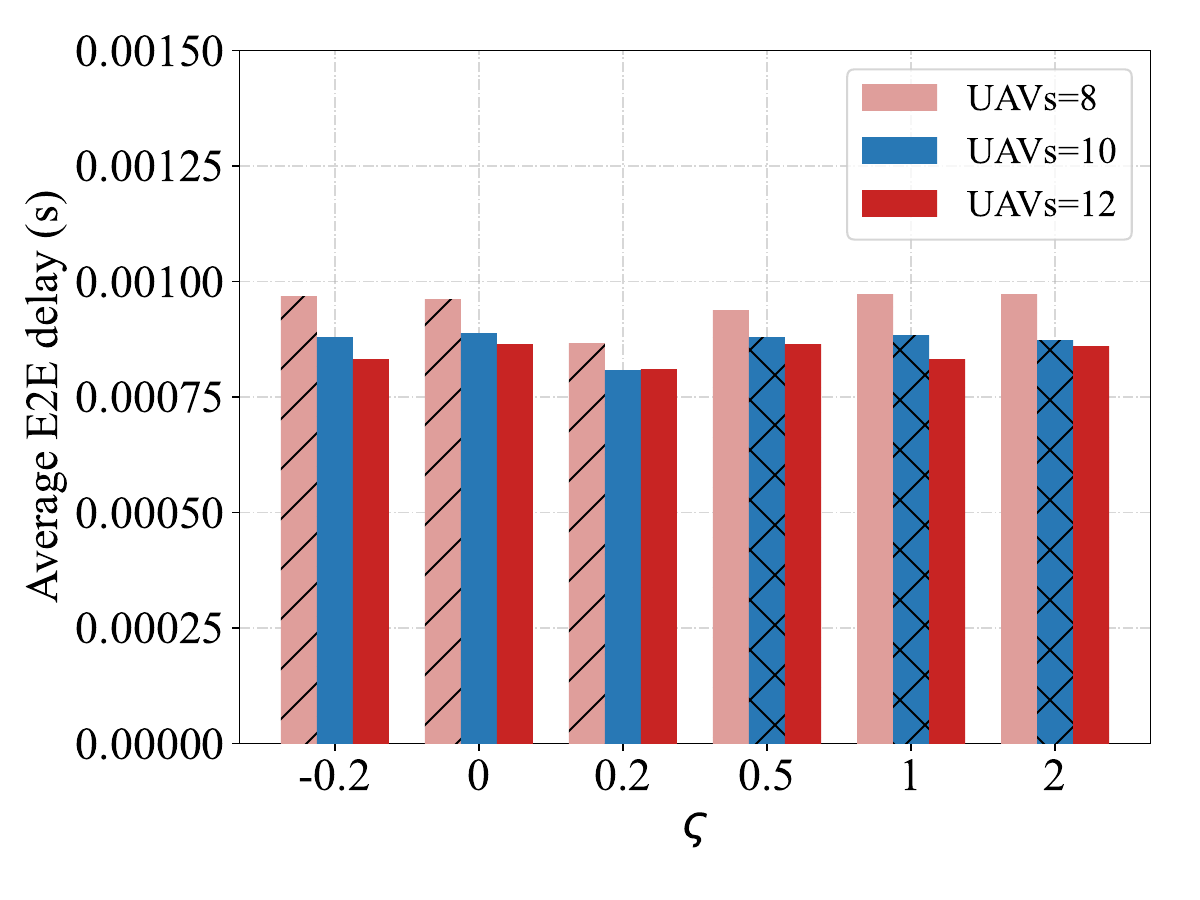}}
    \caption{The evaluation of different $\varsigma$ in terms of the reward 
    and the average E2E delay with 25 demands and 8 UAVs including 2 malicious UAVs.
    }
    {\label{fig:varsigma}}
\end{figure}

Fig. {\ref{fig:varsigma}} evaluates the impact of hyperparameter $\varsigma$ defined in (\ref{equ:reward}) under different network scales.
In detail, Fig. {\ref{fig:varsigma}}(a) shows that as $\varsigma$ increases, the average reward decreases in various network scales, 
since the reward in ({\ref{equ:Algorithm-reward}}) is inversely correlated with $\varsigma$.
Fig. {\ref{fig:varsigma}}(b) indicates the average E2E delay of diverse $\varsigma$. 
It is noted that when $\varsigma$ is 0.2, the delay cost is the lowest among all numbers of UAVs  
and is lower than that without using $\varsigma$ (i.e., $\varsigma=0$).
While the value of $\varsigma$ increases or decreases, the benefit decreases.
Through the sensitivity analysis of $\varsigma$, we apply $\varsigma=0.2$ in the paper.

\begin{figure}[t]
    \centering
    \subfloat[\textcolor{black}{ }]{\centering
    \includegraphics[width=0.7\linewidth]{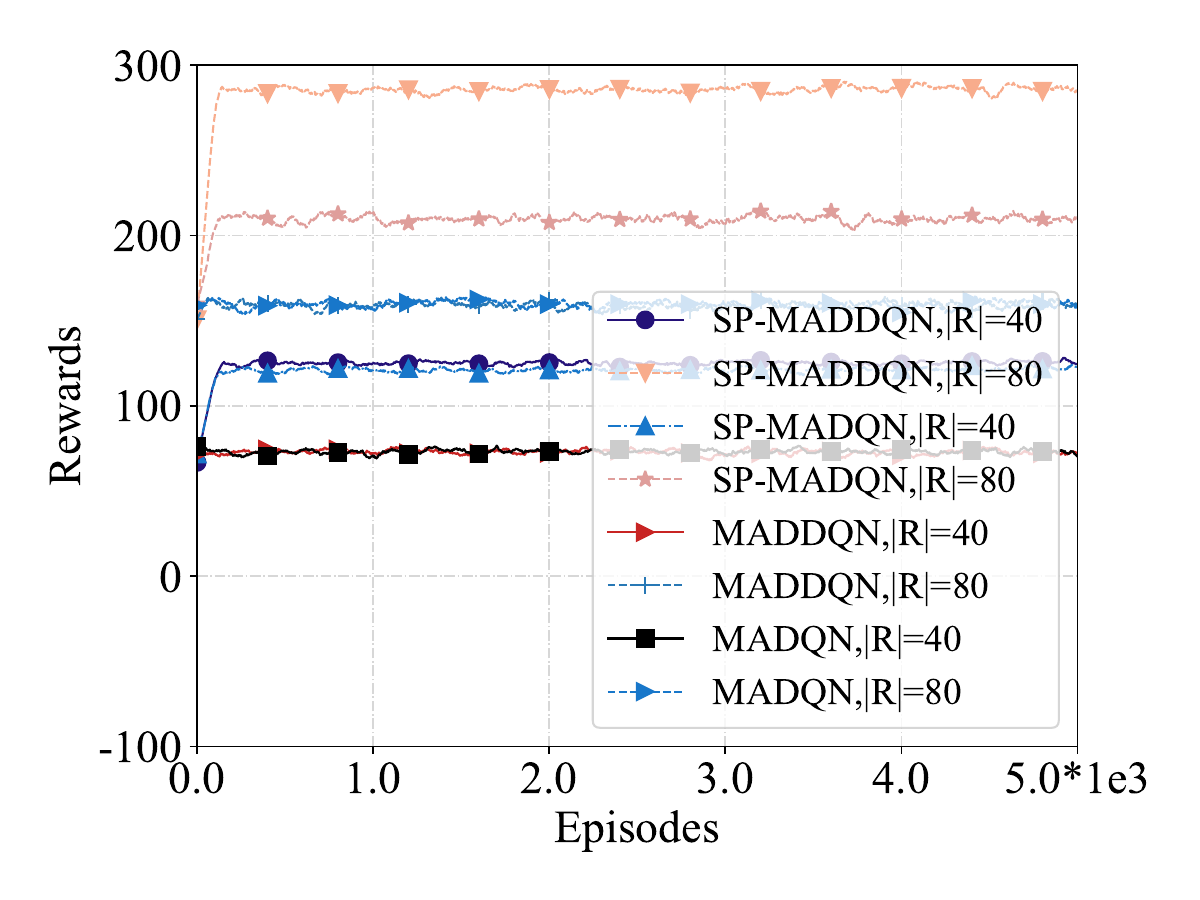}}
    \hspace{0.01\linewidth}
    \subfloat[\textcolor{black}{ }]{\centering
    \includegraphics[width=0.7\linewidth]{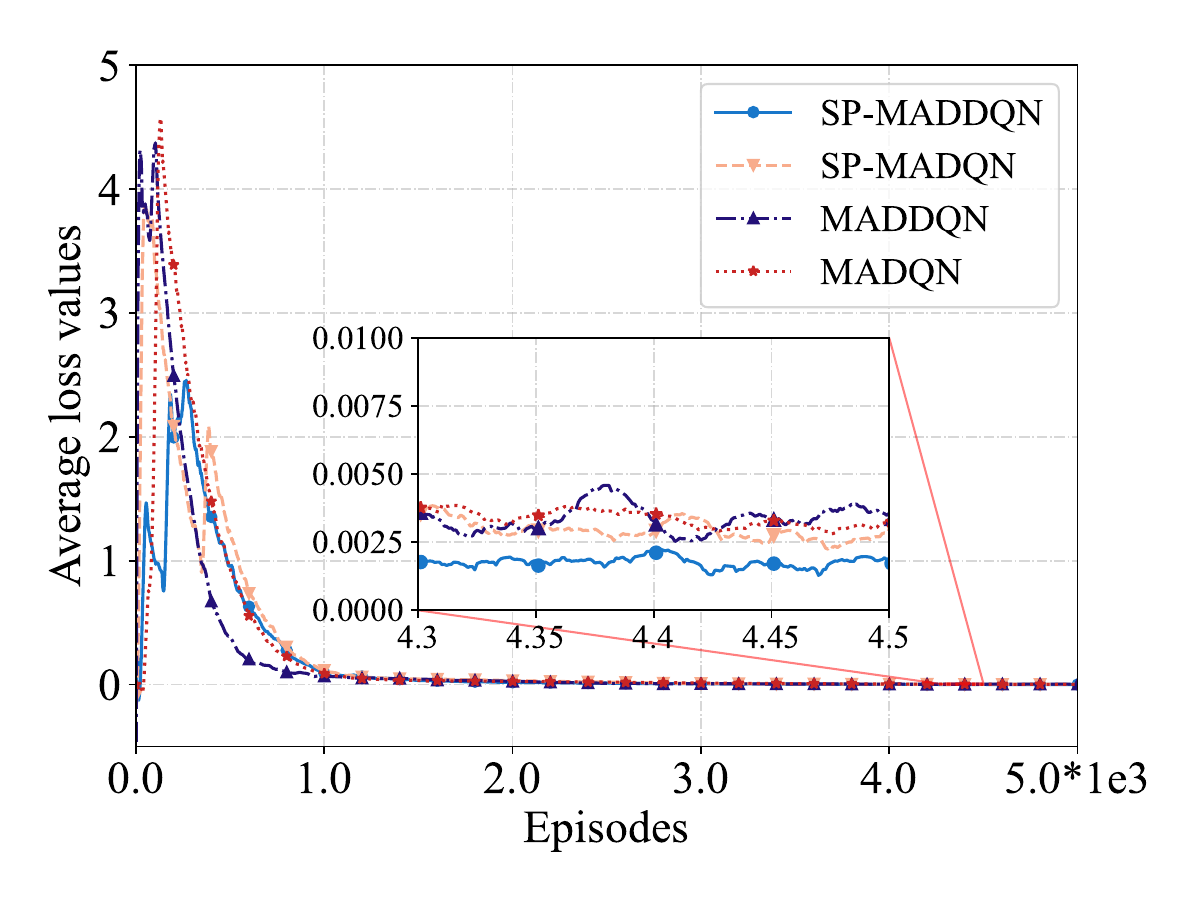}}
    \caption{Comparison of different algorithms in terms of the convergence performance, 
    and loss values with 25 demands and 8 UAVs including 2 malicious UAVs.
     }
    {\label{fig:3Algorithms}}
\end{figure}

\begin{figure}[t]
    \vspace{-0.25cm}
    \centering
    \subfloat[\textcolor{black}{ }]{\centering
    \includegraphics[width=0.5\linewidth]{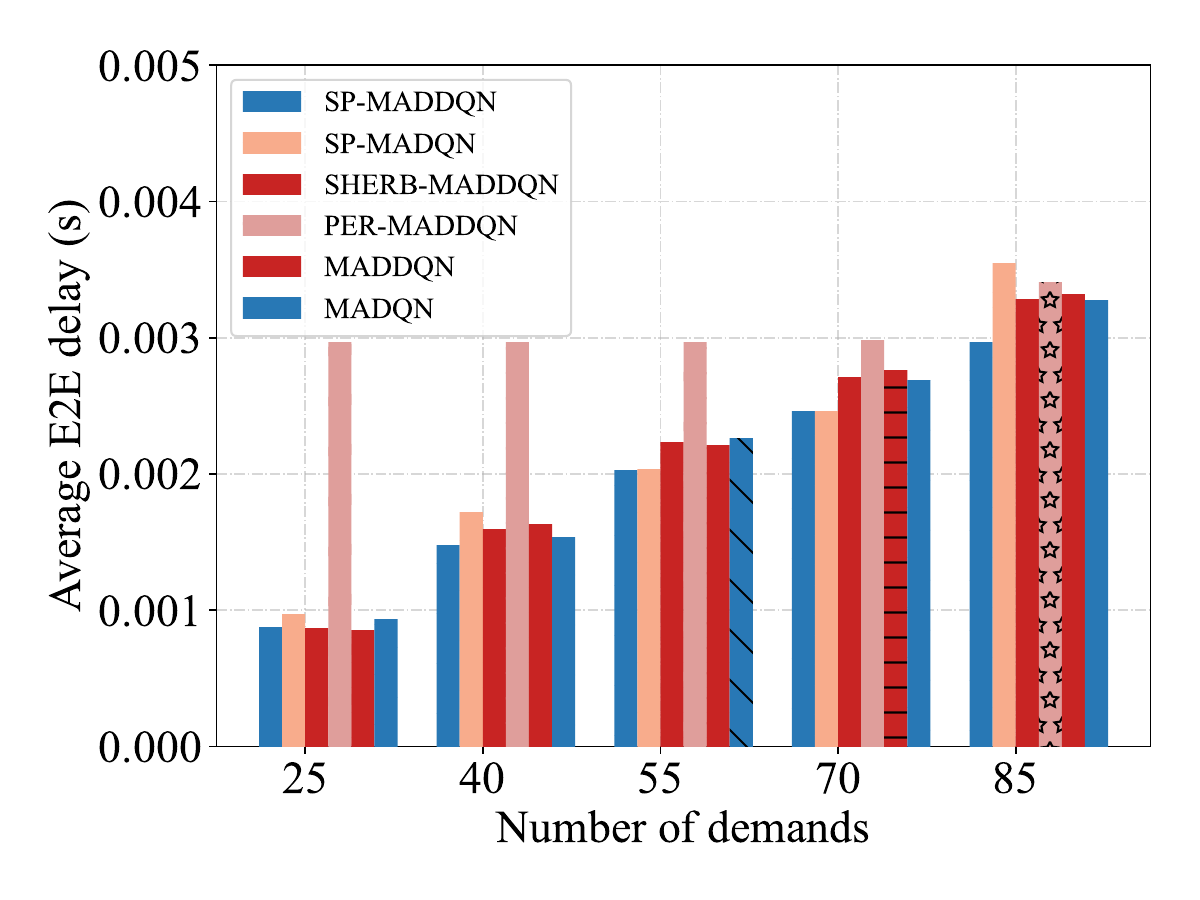}}
    \subfloat[\textcolor{black}{ }]{\centering
    \includegraphics[width=0.5\linewidth]{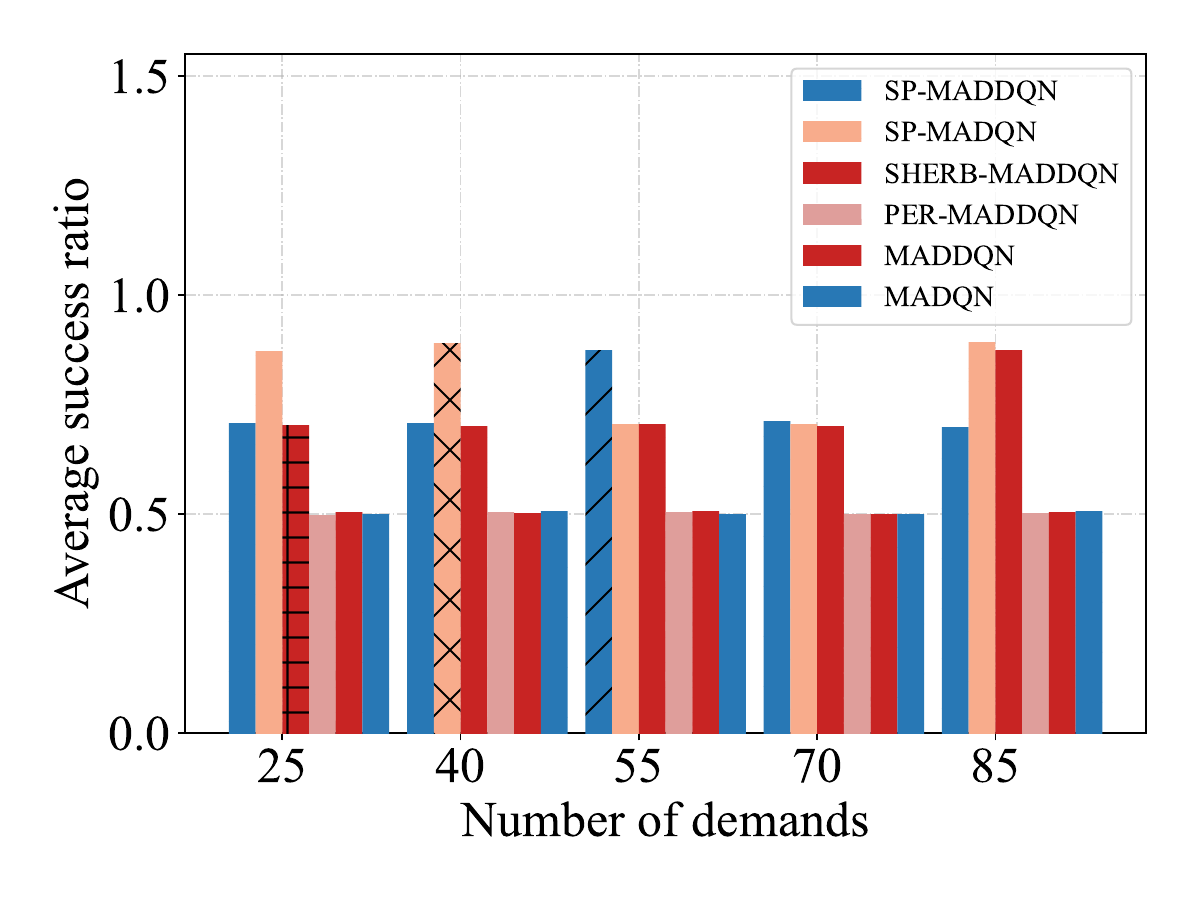}}
    \caption{\textcolor{black}{Comparison of the diverse algorithms in terms of the average E2E delay and TSR with 25 demands and 8 UAVs including 2 malicious UAVs.}
     }
    {\label{fig:Algorithms-DS}}
\end{figure}

In Figs. {\ref{fig:3Algorithms}}-{\ref{fig:Algorithms-DS}}, the SP-MADDQN is compared with SP-MADQN, MADDQN, and MADQN
algorithms with regard to the convergence performance, loss value, average E2E delay, and average TSR, with 2 malicious UAVs.
Specifically, in Fig. {\ref{fig:3Algorithms}}(a), as the episode accumulates, all algorithms show a trend of stable rewards. 
Around episode 1,000, most curves converge. 
However, SP-based algorithms achieve higher stable rewards than non-SP algorithms.
This proves that the introduction of the SHERB and PER mechanisms is conducive to obtaining better performances.
Besides, for the same algorithm, higher demands bring higher rewards, since the reward in (\ref{equ:reward-total}) is calculated via the sum of all demands, generating the larger accumulative rewards.
Fig. {\ref{fig:3Algorithms}}(b) displays that all algorithms present a rapid loss drop in the early 0-1,000 episodes via adjusting parameters to minimize errors. Then, the losses decrease to very small values, indicating that the algorithms already converge. Particularly, the SP-based algorithm achieves a lower loss value earlier and keeps it more stable than the non-SP algorithm. Besides, the enlarged subplot shows the subtle difference that SP-MADDQN has the smoothest loss curve in the later stage, while the non-SP algorithms fluctuate more. This confirms that SP-based mechanisms improve the stability and efficiency of learning, thereby achieving more consistent performances and reducing errors.

\textcolor{black}{Fig. {\ref{fig:Algorithms-DS}} compares six algorithms focusing on average E2E delay and TSR.
Specifically, in Fig. {\ref{fig:Algorithms-DS}}(a), with the increasing number of demands, 
the average E2E delay of all algorithms shows an upward trend, aligned with (\ref{equ:reward-total}). 
Among them, SP-MADDQN obtains the lowest delay cost compared to other five algorithms. 
This indicates that the variants of SP-based methods are more effective in reducing the E2E delay.
It is observed that as the demands grow, the E2E delay increases in all six algorithms. 
Nevertheless, the E2E delay of the proposed SP-MADDQN algorithm decreases 
by 59\% than SP-MADQN, SHERB-MADDQN, PER-MADDQN, MADDQN, and MADQN on average. 
Besides, the E2E delay results indicate that SHERB and PER mechanisms independently enhance routing efficiency compared to baselines, but fall short of the fully integrated both SHERB and PER approach. SHERB-MADDQN consistently shows a marginal advantage over PER-MADDQN, highlighting its superior adaptability to the multiple sources and destinations network conditions.}

\textcolor{black}{
Regarding the TSR in Fig. \ref{fig:Algorithms-DS}(b), the SP-based methods consistently outperform non-SP-based algorithms. Specifically, as the number of demands varies, the TSR of SP-based methods remains above 0.7, whereas non-SP-based algorithms achieve a lower TSR around 0.5. This demonstrates the superior capability of the proposed approach in ensuring successful transmissions, even in environments with malicious UAVs. In particular, SP-MADDQN achieves an average TSR that is 29\% higher than the other five benchmark algorithms. Meanwhile, the ablation variants SHERB-MADDQN and PER-MADDQN exhibit intermediate performance, with TSR values consistently between 0.6 and 0.75 that are higher than non-SP methods but below SP-MADDQN. This indicates that while SHERB and PER mechanisms enhance experience replay and priority scheduling to some extent, they still fall short of the deep integration offered by the SP strategy in adversarial settings. Additionally, SHERB-MADDQN shows a slight advantage over PER-MADDQN, with an approximately 3.2\% higher average TSR, reflecting its better sample utilization and policy stability. In summary, in the presence of 2 malicious UAVs, the proposed SP-MADDQN achieves the best overall performance in both E2E delay and TSR compared to all other considered algorithms}
\begin{figure}[t]
    \centering
    \vspace{-0.3cm}
    \subfloat[\textcolor{black}{ }]{\centering
    \includegraphics[width=0.5\linewidth]{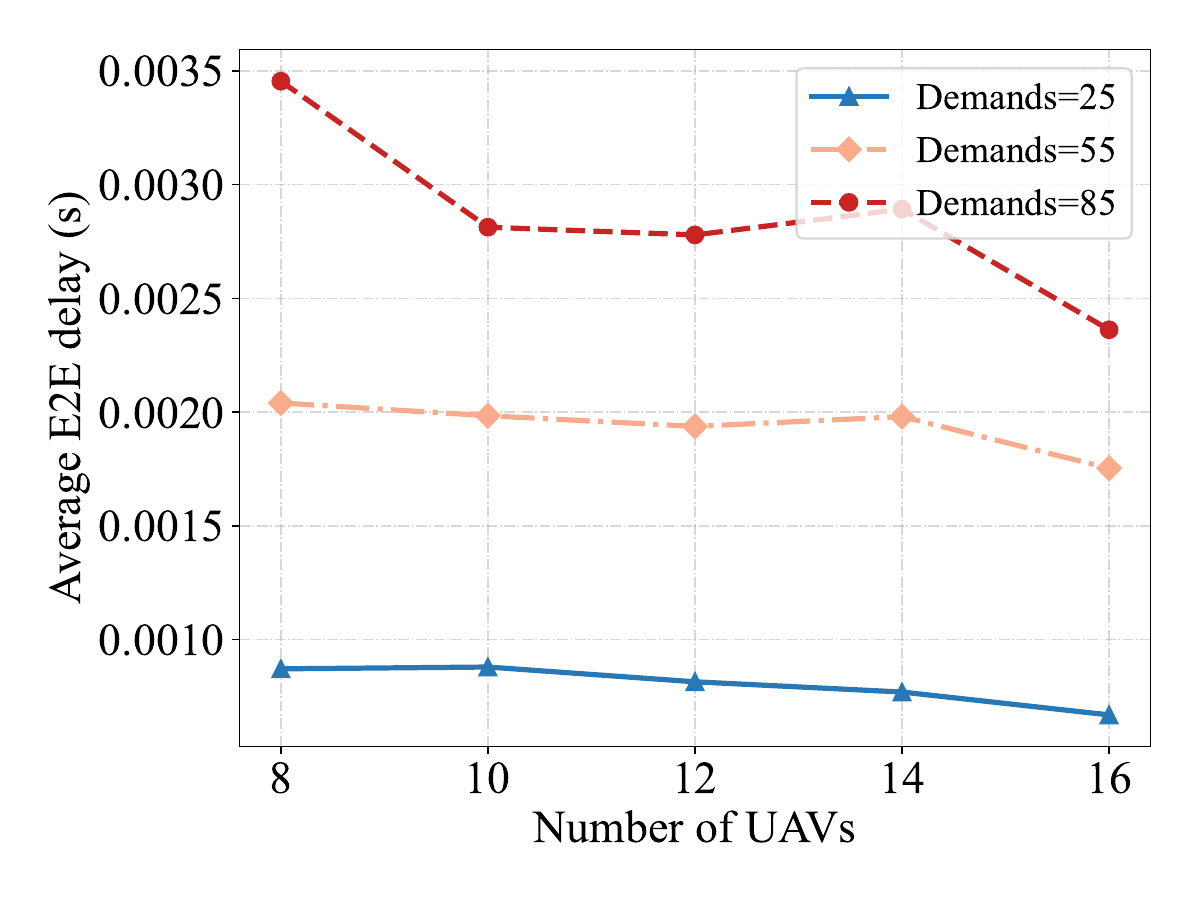}}
    \subfloat[\textcolor{black}{ }]{\centering
    \includegraphics[width=0.5\linewidth]{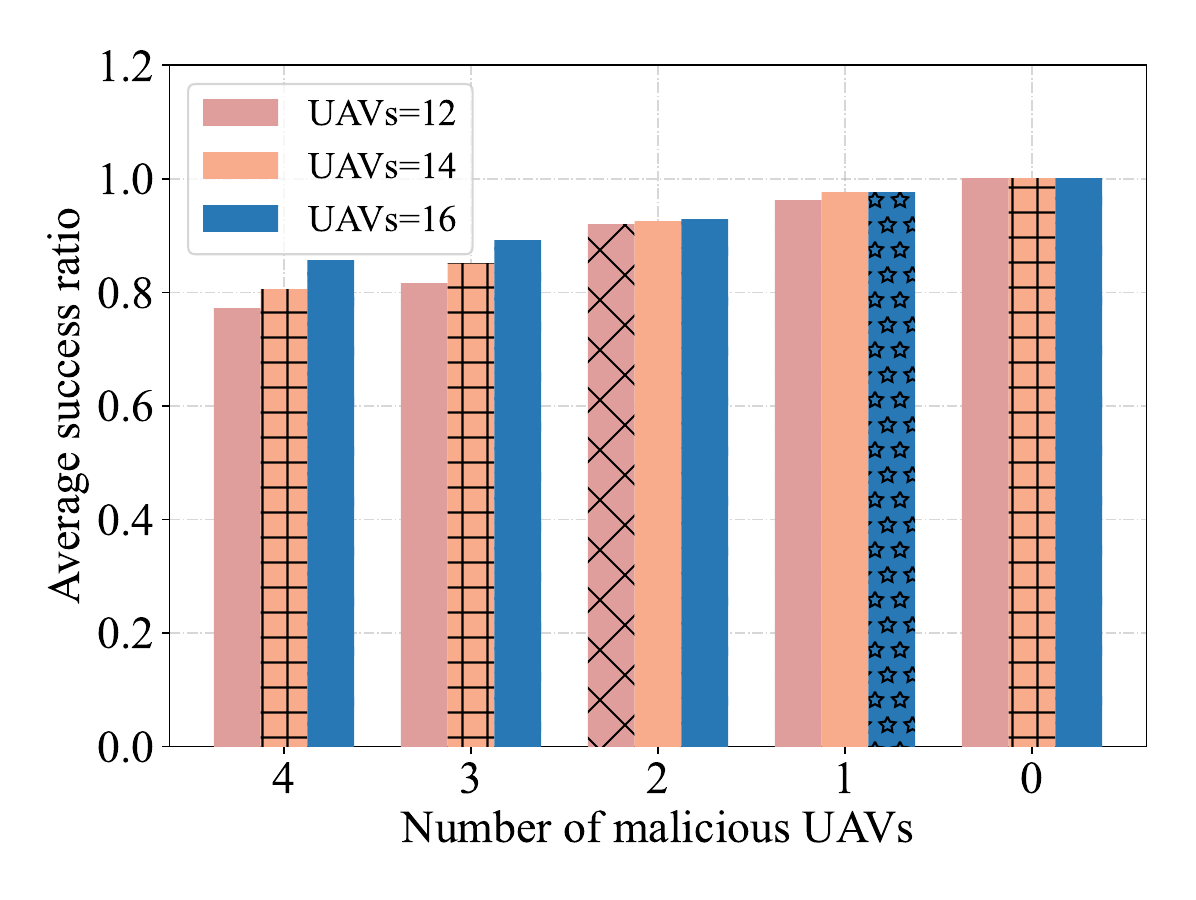}}
    \caption{Comparison of the different network scales on performances. (a) The average E2E delay 
    with 2 malicious UAVs. (b) The TSR under various numbers of malicious UAVs.
{\label{fig:delay-mali-succ}}}
\end{figure}

In Fig. \ref{fig:delay-mali-succ}, the performance of the average E2E delay 
and TSR is compared under the different network scales.
Specifically, Fig. \ref{fig:delay-mali-succ}(a) evaluates 
the average E2E delay of the proposed SP-MADDQN 
algorithm with different scales of LAINs and demand numbers.
With the increment of demands, the average E2E delay rises 
under the same UAV number in LAINs.
When the demand number is identical, 
the average E2E delay decreases as the number of UAVs increases,
since the increased UAVs bring more abundant network resources.
Fig. {\ref{fig:delay-mali-succ}}(b) shows the performance of 
the average TSR when the numbers of malicious UAVs are different 
in various scales of LAINs. 
It can be observed that the average TSR is the highest 
when the number of malicious UAVs is 0. 
While there are malicious UAVs, the TSR decreases significantly at different effects,
denoting that the malicious UAVs decrease the transmission 
performance during routing.
Additionally, the average TSR declines sharply 
as the number of malicious UAVs increases. 
It is because the malicious UAVs may reject to transmit the demands, thereby losing demands.
As the number of UAVs grows, the total network resources are richer, 
such as the channel bandwidth, storage, and computation power, resulting in the higher TSR of routing.

\section{Conclusions and Future Work \label{sec:Conclusions}}
In this work, we have considered that UAVs are divided into multiple clusters 
to collaboratively transmit the demands with security threats and dynamic distributed topologies. 
We have presented the ZTA with the SDP controller for the dynamic management of UAV joining and exiting, 
via the identity authentication and permission. Then, by combining the direct and indirect trust, we have designed the adaptive weight trust model to update the credit value of UAVs. Furthermore, we have proposed the lightweight consortium blockchain to store the transactions and the credit values, which is deployed on the CHUs, mitigating the impact of security threats on routing performances. \textcolor{black}{The proposed framework has been particularly well-suited for real-world scenarios that demand secure and resilient communication in dynamic, open, and potentially adversarial environments.} In the constructed zero-trust LAIN, we have formulated the routing problem to optimize both the average E2E delay and TSR, which is an INLP problem and intractable to solve. To deal with the challenge of obtaining the global information,
we have reformulated the routing problem into a Dec-POMDP form and designed the SP-MADDQN algorithm. 
\textcolor{black}{Furthermore, since the complexity has been independent of the total scale of the network, the SP-MADDQN algorithm can be deployed in practice.}
Simulation results have demonstrated that the adaptive weight trust model
can identify malicious UAVs more quickly than the average and random weight trust methods.
\textcolor{black}{Additionally, the designed SP-MADDQN algorithm has been proven to outperform other benchmarks, reducing the average E2E delay by 59\% and improving the TSR by 29\% on average, also demonstrating superior convergence and lower training loss.}


\textcolor{black}{
To advance the proposed framework toward the practical deployment, several research directions merit further explorations. A crucial step is to refine the cryptographic key management scheme, establishing a more robust and efficient foundation for trust in dynamic networks. Furthermore, the resilience of the IPBFT consensus mechanism under the high mobility requires the in-depth stability analysis. Concurrently, the integrated security model must be extended and stress-tested against a broader and more realistic spectrum of adversarial scenarios to ensure the holistic robustness. Addressing these challenges are pivotal in transitioning the system from a validated prototype to a field deployment solution. }

     \bibliographystyle{IEEEtran}
     \bibliography{ref2}

\begin{thebibliography}{10}
\providecommand{\url}[1]{#1}
\csname url@samestyle\endcsname
\providecommand{\newblock}{\relax}
\providecommand{\bibinfo}[2]{#2}
\providecommand{\BIBentrySTDinterwordspacing}{\spaceskip=0pt\relax}
\providecommand{\BIBentryALTinterwordstretchfactor}{4}
\providecommand{\BIBentryALTinterwordspacing}{\spaceskip=\fontdimen2\font plus
\BIBentryALTinterwordstretchfactor\fontdimen3\font minus \fontdimen4\font\relax}
\providecommand{\BIBforeignlanguage}[2]{{%
\expandafter\ifx\csname l@#1\endcsname\relax
\typeout{** WARNING: IEEEtran.bst: No hyphenation pattern has been}%
\typeout{** loaded for the language `#1'. Using the pattern for}%
\typeout{** the default language instead.}%
\else
\language=\csname l@#1\endcsname
\fi
#2}}
\providecommand{\BIBdecl}{\relax}
\BIBdecl

\bibitem{10418158}
B.~He, X.~Ji, G.~Li, and B.~Cheng, ``Key technologies and applications of {UAVs} in underground space: A review,'' \emph{IEEE Trans. Cognit. Commun. Networking}, vol.~10, no.~3, pp. 1026--1049, Jan. 2024.

\bibitem{11105407}
H.~Luo, G.~Sun, J.~Wang, H.~Yu, D.~Niyato, S.~Dustdar, and Z.~Han, ``Wireless blockchain meets {6G}: The future trustworthy and ubiquitous connectivity,'' \emph{IEEE Commun. Surv. Tutorials}, early access, Jul. 2025.

\bibitem{mao2024survey}
K.~Mao, Q.~Zhu, C.-X. Wang, X.~Ye, J.~Gomez-Ponce, X.~Cai, Y.~Miao, Z.~Cui, Q.~Wu, and W.~Fan, ``A survey on channel sounding technologies and measurements for {UAV}-assisted communications,'' \emph{IEEE Trans. Instrum. Meas.}, vol.~73, pp. 1--24, Aug. 2024.

\bibitem{Distributionally-Jia}
Z.~Jia, C.~Cui, C.~Dong, Q.~Wu, Z.~Ling, D.~Niyato, and Z.~Han, ``Distributionally robust optimization for aerial multi-access edge computing via cooperation of {UAVs} and {HAPs},'' \emph{IEEE Trans. Mob. Comput.}, 2025.

\bibitem{10430396}
S.~Javed, A.~Hassan, R.~Ahmad, W.~Ahmed, R.~Ahmed, A.~Saadat, and M.~Guizani, ``State-of-the-art and future research challenges in {UAV} swarms,'' \emph{IEEE Internet Things J.}, vol.~11, no.~11, pp. 19\,023--19\,045, Jun. 2024.

\bibitem{Routing_UAV_Survey}
D.~Shumeye~Lakew, U.~Sa’ad, N.-N. Dao, W.~Na, and S.~Cho, ``Routing in flying ad hoc networks: A comprehensive survey,'' \emph{IEEE Commun. Surv. Tutorials}, vol.~22, no.~2, pp. 1071--1120, 2nd Quart. 2020.

\bibitem{He_Routing}
S.~He, Z.~Jia, C.~Dong, W.~Wang, Y.~Cao, Y.~Yang, and Q.~Wu, ``Routing recovery for {UAV} networks with deliberate attacks: A reinforcement learning based approach,'' in \emph{Proc. IEEE Glob. Commun. Conf. (GLOBECOM)}, Kuala Lumpur, Malaysia, Dec. 2023, pp. 952--957.

\bibitem{attkan2022cyber}
A.~Attkan and V.~Ranga, ``Cyber-physical security for {IoT} networks: {A} comprehensive review on traditional, blockchain and artificial intelligence based key-security,'' \emph{Complex Intell. Syst.}, vol.~8, no.~4, pp. 3559--3591, Feb. 2022.

\bibitem{annabi2024towards}
M.~Annabi, A.~Zeroual, and N.~Messai, ``Towards zero trust security in connected vehicles: A comprehensive survey,'' \emph{Comput. Secur.}, vol. 145, pp. 104\,018--104\,075, Oct. 2024.

\bibitem{9791053_zero}
Y.~Palmo, S.~Tanimoto, H.~Sato, and A.~Kanai, ``A consideration of scalability for software defined perimeter based on the zero-trust model,'' in \emph{Proc. Int. Congr. Adv. Appl. Inf. (IIAI-AAI)}, Niigata, Japan, Jul. 2021, pp. 717--724.

\bibitem{Trusted_Blockchain}
J.~Yang, X.~Liu, X.~Jiang, Y.~Zhang, S.~Chen, and H.~He, ``Toward trusted unmanned aerial vehicle swarm networks: A blockchain-based approach,'' \emph{IEEE Veh. Technol. Mag.}, vol.~18, no.~2, pp. 98--108, Jun. 2023.

\bibitem{NIVEDITA2025103385}
\BIBentryALTinterwordspacing
V.~Nivedita, C.-S. Shieh, and M.-F. Horng, ``An integrated trust-based secure routing with intrusion detection for mobile ad hoc network using adaptive snow geese optimization algorithm,'' \emph{Ain Shams Eng. J.}, vol.~16, no.~7, p. 103385, Jul. 2025. [Online]. Available: \url{https://www.sciencedirect.com/science/article/pii/S2090447925001261}
\BIBentrySTDinterwordspacing

\bibitem{8417971}
J.~Yun, S.~Seo, and J.-M. Chung, ``Centralized trust-based secure routing in wireless networks,'' \emph{IEEE Wireless Commun. Lett.}, vol.~7, no.~6, pp. 1066--1069, Dec. 2018.

\bibitem{commi_energy}
Z.~Wang, H.~Yao, T.~Mai, Z.~Xiong, X.~Wu, D.~Wu, and S.~Guo, ``Learning to routing in {UAV} swarm network: A multi-agent reinforcement learning approach,'' \emph{IEEE Trans. Veh. Technol.}, vol.~72, no.~5, pp. 6611--6624, May 2023.

\bibitem{wei2024survey}
X.~Wei, J.~Ma, and C.~Sun, ``A survey on security of unmanned aerial vehicle systems: Attacks and countermeasures,'' \emph{IEEE Internet Things J.}, vol.~11, no.~21, pp. 34\,826--34\,847, Nov. 2024.

\bibitem{blockchai_security}
J.~Leng, M.~Zhou, J.~L. Zhao, Y.~Huang, and Y.~Bian, ``Blockchain security: A survey of techniques and research directions,'' \emph{IEEE Trans. Serv. Comput.}, vol.~15, no.~4, pp. 2490--2510, Jul./Aug. 2022.

\bibitem{Blockchain9120287}
D.~E. Kouicem, Y.~Imine, A.~Bouabdallah, and H.~Lakhlef, ``Decentralized blockchain-based trust management protocol for the {I}nternet of things,'' \emph{IEEE Trans. Dependable Secure Comput.}, vol.~19, no.~2, pp. 1292--1306, Mar./Apr. 2022.

\bibitem{10415005}
H.~Luo, Y.~Wu, G.~Sun, H.~Yu, and M.~Guizani, ``{ESCM:} an efficient and secure communication mechanism for {UAV} networks,'' \emph{IEEE Trans. Netw. Serv. Manage.}, vol.~21, no.~3, pp. 3124--3139, Jun. 2024.

\bibitem{tang2022blockchain}
F.~Tang, C.~Wen, L.~Luo, M.~Zhao, and N.~Kato, ``Blockchain-based trusted traffic offloading in space-air-ground integrated networks {(SAGIN)}: {A} federated reinforcement learning approach,'' \emph{IEEE J. Sel. Areas Commun.}, vol.~40, no.~12, pp. 3501--3516, Dec. 2022.

\bibitem{Wang_security}
Y.~Wang, Z.~Su, Q.~Xu, R.~Li, T.~H. Luan, and P.~Wang, ``A secure and intelligent data sharing scheme for {UAV}-assisted disaster rescue,'' \emph{IEEE/ACM Trans. Networking}, vol.~31, no.~6, pp. 2422--2438, Dec. 2023.

\bibitem{Review2-4}
P.~Ganesan and S.~K. Jagatheesaperumal, ``Revolutionizing emergency response: The transformative power of smart wearables through blockchain, federated learning, and beyond {5G/6G} services,'' \emph{IT Prof.}, vol.~25, no.~6, pp. 54--61, Nov.-Dec. 2023.

\bibitem{zhang20213d}
L.~Zhang, F.~Hu, Z.~Chu, E.~Bentley, and S.~Kumar, ``{3D} transformative routing for {UAV} swarming networks: A skeleton-guided, {GPS}-free approach,'' \emph{IEEE Trans. Veh. Technol.}, vol.~70, no.~4, pp. 3685--3701, Apr. 2021.

\bibitem{9738819}
T.~Li, K.~Zhu, N.~C. Luong, D.~Niyato, Q.~Wu, Y.~Zhang, and B.~Chen, ``{Applications of multi-agent reinforcement learning in future {Internet}: A comprehensive survey},'' \emph{IEEE Commun. Surv. Tutorials}, vol.~24, no.~2, pp. 1240--1279, 2nd Quart. 2022.

\bibitem{LI2025110964}
J.~Li, B.~Feng, and H.~Zheng, ``A survey on {VPN}: Taxonomy, roles, trends and future directions,'' \emph{Comput. Networks}, vol. 257, pp. 110\,964--110\,984, Feb. 2025.

\bibitem{10537758}
B.~Singh and S.~S. Cheema, ``Next generation firewall and self authentication for network security,'' in \emph{Proc. Int. Conf. Image Inf. Process. (ICIIP)}, Solan, India, May 2023, pp. 707--713.

\bibitem{JUMA2020102598}
M.~Juma, A.~A. Monem, and K.~Shaalan, ``Hybrid end-to-end {VPN} security approach for smart {IoT} objects,'' \emph{J. Network Comput. Appl.}, vol. 158, pp. 102\,598--102\,611, May 2020.

\bibitem{abdelhay2024toward}
Z.~Abdelhay, Y.~Bello, and A.~Refaey, ``Toward zero-trust {6GC}: A software defined perimeter approach with dynamic moving target defense mechanism,'' \emph{IEEE Wireless Commun.}, vol.~31, no.~2, pp. 74--80, Apr. 2024.

\bibitem{Han_GLOBECOM}
X.~Zhang, D.~Wang, Y.~Zhu, W.~Chen, Z.~Chang, and Z.~Han, ``When zero-trust meets federated learning,'' in \emph{Proc. IEEE Glob. Commun. Conf. (GLOBECOM)}, Cape Town, South Africa, Dec. 2024, pp. 794--799.

\bibitem{Zero-Trust-2}
Z.~Yan, G.~Yu, M.~Zhan, Y.~Zhang, and J.~Hu, ``{5GC-SDP}: Security enhancement of {5G} core networks with zero trust,'' in \emph{Proc. Int. Conf. Comput. Supported Cooperative Work Des. (CSCWD)}, Tianjin, China, May 2024, pp. 1597--1602.

\bibitem{GCS_1}
D.~J.~S. Agron, M.~R. Ramli, J.-M. Lee, and D.-S. Kim, ``Secure ground control station-based routing protocol for {UAV} networks,'' in \emph{Proc. Int. Conf. Inf. Commun. Technol. Converg. (ICTC)}, Jeju, Korea (South) ,Oct. 2019, pp. 794--798.

\bibitem{GCS_3}
Y.~Chen, G.~Liu, Z.~Zhang, L.~He, and S.~He, ``Improving physical layer security for multi-{UAV} systems against hybrid wireless attacks,'' \emph{IEEE Trans. Veh. Technol.}, vol.~73, no.~5, pp. 7034--7048, May 2024.

\bibitem{GCS_2}
D.~Yin, X.~Yang, H.~Yu, S.~Chen, and C.~Wang, ``An air-to-ground relay communication planning method for {UAVs} swarm applications,'' \emph{IEEE Trans. Intell. Veh.}, vol.~8, no.~4, pp. 2983--2997, Apr. 2023.

\bibitem{Review2-2}
S.~Kumar~Jagatheesaperumal, M.~Rahouti, A.~Chehri, K.~Xiong, and J.~Bieniek, ``Blockchain-based security architecture for uncrewed aerial systems in {B5G/6G} services and beyond: A comprehensive approach,'' \emph{IEEE Open J. Commun. Soc.}, vol.~6, pp. 1042--1069, Jan. 2025.

\bibitem{DRRS-BC}
H.~Lu, Y.~Tang, and Y.~Sun, ``{DRRS-BC}: Decentralized routing registration system based on blockchain,'' \emph{IEEE/CAA J. Autom. Sin.}, vol.~8, no.~12, pp. 1868--1876, Dec. 2021.

\bibitem{Review2-1}
T.~Liao, L.~Wei, X.~Hu, J.~Hu, M.~Hu, K.~Peng, C.~Cai, and Z.~Xiong, ``Blockchain-enhanced {UAV} networks: Optimizing data storage for real-time efficiency,'' \emph{IEEE Internet of Things J.}, vol.~12, no.~19, pp. 39\,910--39\,924, Oct. 2025.

\bibitem{Review2-3}
Q.~Wang, C.~Qian, S.~Mia, H.~Zhang, H.~Zhao, Y.~Lu, and H.~Zhu, ``Blockchain-enabled credible multi-operator spectrum sharing in {UAV} communication systems,'' \emph{IEEE Trans. Veh. Technol.}, vol.~74, no.~6, pp. 8989--9001, Jun. 2025.

\bibitem{8951253}
Z.~Cui, F.~XUE, S.~Zhang, X.~Cai, Y.~Cao, W.~Zhang, and J.~Chen, ``A hybrid blockchain-based identity authentication scheme for multi-{WSN},'' \emph{IEEE Trans. Serv. Comput.}, vol.~13, no.~2, pp. 241--251, Mar.-Apr., 2020.

\bibitem{KUANG2025103981}
Y.~Kuang, Q.~Wu, R.~Chen, and X.~Liu, ``Blockchain based lightweight authentication scheme for {Internet} of things using lattice encryption algorithm,'' \emph{Comput. Stand. Interfaces}, vol.~93, pp. 103\,981--103\,990, Apr. 2025.

\bibitem{YANHUI202210365}
Y.~Liu, J.~Zhang, S.~Muhammad, Y.~Yuan, P.~Zhang, M.~Sarah, and N.~Avishek, ``Research on identity authentication system of {Internet} of things based on blockchain technology,'' \emph{J. King Saud Univ. Comput. Inf. Sci.}, vol.~34, no.~10, pp. 10\,365--10\,377, Nov. 2022.

\bibitem{MARL-o-Manage-4}
Z.~Wang, S.~Li, E.~J. Knoblock, H.~Li, and R.~D. Apaza, ``Delay sensitive routing in an aerial and terrestrial hybrid wireless network via multi-agent reinforcement learning,'' in \emph{Proc. IEEE Glob. Commun. Conf. (GLOBECOM)}, Cape Town, South Africa, Dec. 2024, pp. 1497--1502.

\bibitem{MARL-o-Manage-5}
Y.~Lyu, H.~Hu, R.~Fan, Z.~Liu, J.~An, and S.~Mao, ``Dynamic routing for integrated satellite-terrestrial networks: A constrained multi-agent reinforcement learning approach,'' \emph{IEEE J. Sel. Areas Commun.}, vol.~42, no.~5, pp. 1204--1218, May 2024.

\bibitem{MARL-o-Manage-6}
H.~Zhang, H.~Tang, Y.~Hu, X.~Wei, C.~Wu, W.~Ding, and X.-P. Zhang, ``Heterogeneous mean-field multi-agent reinforcement learning for communication routing selection in {SAGI}-net,'' in \emph{Proc. IEEE Veh. Technol. Conf. (VTC-Fall)}, London, U.K., Sep. 2022.

\bibitem{RWMM-1}
N.~Lin, F.~Gao, L.~Zhao, A.~Al-Dubai, and Z.~Tan, ``A {3D} smooth random walk mobility model for {FANETs},'' in \emph{2019 IEEE 21st International Conference on High Performance Computing and Communications; IEEE 17th International Conference on Smart City; IEEE 5th International Conference on Data Science and Systems (HPCC/SmartCity/DSS)}, Zhangjiajie, China, Aug. 2019, pp. 460--467.

\bibitem{trust_parameters}
Z.~Teng, B.~Pang, M.~Sun, L.~Xie, and L.~Guo, ``A malicious node identification strategy with environmental parameters optimization in wireless sensor network,'' \emph{Wireless Pers. Commun.}, vol. 117, pp. 1143--1162, Mar. 2021.

\bibitem{li2021lightweight}
C.~Li, J.~Zhang, X.~Yang, and L.~Youlong, ``Lightweight blockchain consensus mechanism and storage optimization for resource-constrained {IoT} devices,'' \emph{Inf. Process. Manage.}, vol.~58, no.~4, pp. 102\,602--102\,625, 2021.

\bibitem{LightTrust_9434372}
I.~Ud~Din, A.~Bano, K.~A. Awan, A.~Almogren, A.~Altameem, and M.~Guizani, ``Lighttrust: Lightweight trust management for edge devices in industrial {Internet} of things,'' \emph{IEEE Internet of Things J.}, vol.~10, no.~4, pp. 2776--2783, Feb. 2023.

\bibitem{11201899}
H.~Luo, G.~Sun, Y.~Liu, D.~Zhao, D.~Niyato, H.~Yu, and S.~Dustdar, ``A weighted {Byzantine} fault tolerance consensus driven trusted multiple large language models network,'' \emph{IEEE Trans. Cognit. Commun. Networking}, early access, Oct. 2025.

\bibitem{channel_model_PL_2}
J.~Du, J.~Wang, A.~Sun, J.~Qu, J.~Zhang, C.~Wu, and D.~Niyato, ``Joint optimization in blockchain and {MEC}-enabled space-air-ground integrated networks,'' \emph{IEEE Internet Things J.}, vol.~11, no.~19, pp. 31\,862--31\,877, Jul. 2024.

\bibitem{Blockchain_delay}
J.~Du, Z.~Yu, A.~Sun, J.~Jiang, H.~Zhao, N.~Zhang, C.~Wu, and F.~Richard~Yu, ``Secure task offloading in blockchain-enabled {MEC} networks with improved {PBFT} consensus,'' \emph{IEEE Trans. Cognit. Commun. Networking}, vol.~11, no.~2, pp. 1225--1243, Apr. 2025.

\bibitem{saldamli2022improved}
G.~Saldamli, C.~Upadhyay, D.~Jadhav, R.~Shrishrimal, B.~Patil, and L.~Tawalbeh, ``Improved gossip protocol for blockchain applications,'' \emph{Cluster Comput.}, vol.~25, no.~3, pp. 1915--1926, Jan. 2022.

\bibitem{channel_model_PL}
Y.~Chen, N.~Zhao, Z.~Ding, and M.-S. Alouini, ``{Multiple {UAVs} as relays: Multi-hop single link versus multiple dual-hop links},'' \emph{IEEE Trans. Wireless Commun.}, vol.~17, no.~9, pp. 6348--6359, Sep. 2018.

\bibitem{zheng2025rotatable}
B.~Zheng, Q.~Wu, T.~Ma, and R.~Zhang, ``Rotatable antenna enabled wireless communication: Modeling and optimization,'' \emph{arXiv preprint arXiv:2501.02595}, Jan. 2025.

\bibitem{11222668}
B.~Zheng, T.~Ma, C.~You, J.~Tang, R.~Schober, and R.~Zhang, ``Rotatable antenna enabled wireless communication and sensing: Opportunities and challenges,'' \emph{IEEE Wireless Commun.}, early access, Oct. 2025.

\bibitem{11149011}
S.~He, Z.~Jia, Q.~Zhu, F.~Zhou, and Q.~Wu, ``Trusted routing for blockchain-enabled low-altitude intelligent networks,'' in \emph{Proc. IEEE/CIC Int. Conf. Commun. China (ICCC)}, Shanghai, China, Aug. 2025.

\bibitem{11122503}
Z.~Jia, S.~He, Q.~Zhu, W.~Wang, Q.~Wu, and Z.~Han, ``Trusted routing for blockchain-empowered {UAV} networks via multi-agent deep reinforcement learning,'' \emph{IEEE Trans. Commun.}, early access, Aug. 2025.

\bibitem{van2016deep}
H.~Van~Hasselt, A.~Guez, and D.~Silver, ``Deep reinforcement learning with double {Q}-learning,'' in \emph{Proc. 13th AAAI Conf. Artif. Intell.}, vol.~30, no.~1, Phoenix, AZ, Feb. 2016, pp. 2094--2100.

\bibitem{mnih2015human}
V.~Mnih, K.~Kavukcuoglu, D.~Silver, A.~A. Rusu, J.~Veness, M.~G. Bellemare, A.~Graves, M.~Riedmiller, A.~K. Fidjeland, G.~Ostrovski \emph{et~al.}, ``Human-level control through deep reinforcement learning,'' \emph{Nature}, vol. 518, no. 7540, pp. 529--533, Feb. 2015.

\bibitem{schaul2015prioritized}
T.~Schaul, J.~Quan, I.~Antonoglou, and D.~Silver, ``Prioritized experience replay,'' \emph{arXiv preprint arXiv:1511.05952}, Nov. 2015.

\bibitem{4445757}
L.~Busoniu, R.~Babuska, and B.~De~Schutter, ``A comprehensive survey of multiagent reinforcement learning,'' \emph{IEEE Trans. Syst., Man, Cybern. C, Appl. Rev.}, vol.~38, no.~2, pp. 156--172, Mar. 2008.

\bibitem{7989385}
S.~Gu, E.~Holly, T.~Lillicrap, and S.~Levine, ``Deep reinforcement learning for robotic manipulation with asynchronous off-policy updates,'' in \emph{Proc. IEEE Int. Conf. Robot. Autom. (ICRA)}, Singapore, Jul. 2017, pp. 3389--3396.

\bibitem{10233034}
J.~Zhang, L.~Che, and M.~Shahidehpour, ``Distributed training and distributed execution-based stackelberg multi-agent reinforcement learning for {EV} charging scheduling,'' \emph{IEEE Trans. Smart Grid}, vol.~14, no.~6, pp. 4976--4979, Nov. 2023.

\bibitem{10547350}
A.~Kopic, E.~Perenda, and H.~Gacanin, ``A collaborative multi-agent deep reinforcement learning-based wireless power allocation with centralized training and decentralized execution,'' \emph{IEEE Transactions on Communications}, vol.~72, no.~11, pp. 7006--7016, Nov. 2024.

\bibitem{11045994}
J.~You, Z.~Jia, C.~Dong, Q.~Wu, and Z.~Han, ``Joint computation offloading and resource allocation for uncertain maritime mec via cooperation of aavs and vessels,'' \emph{IEEE Transactions on Vehicular Technology}, vol.~74, no.~11, pp. 18\,081--18\,095, Nov. 2025.

\bibitem{Block_Size-1}
A.~Miglani and N.~Kumar, ``{BloomACS}: Bloom filter-based access control scheme in blockchain-enabled {V2G} networks,'' \emph{IEEE Trans. Intell. Transp. Syst.}, vol.~25, no.~9, pp. 10\,636--10\,651, Sep. 2024.

\bibitem{11284890}
J.~You, Z.~Jia, C.~Cui, C.~Dong, Q.~Wu, and Z.~Han, ``Hierarchical task offloading and trajectory optimization in low-altitude intelligent networks via auction and diffusion-based {MARL},'' \emph{IEEE Trans. Cogn. Commun. Netw.}, early access, Dec. 2025.

\end{thebibliography}
\begin{IEEEbiography}[{\includegraphics[width=1in,height=1.25in,clip,keepaspectratio]{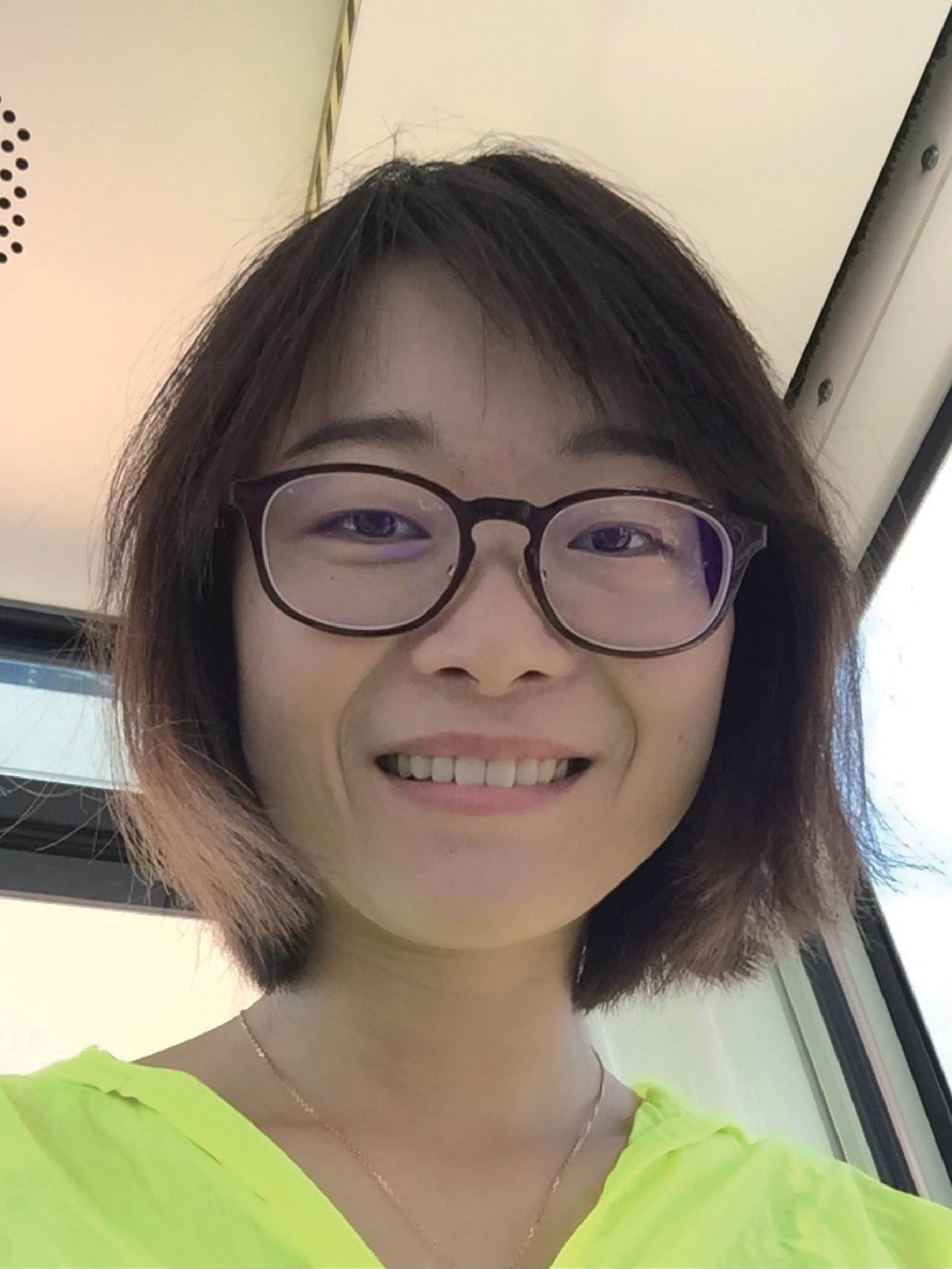}}] {Ziye Jia} (Member, IEEE) received the B.E., M.S., and Ph.D. degrees in communication and information systems from Xidian University, Xi’an, China, in 2012, 2015, and 2021, respectively. From 2018 to 2020, she was a Visiting Ph.D. Student with the Department of Electrical and Computer Engineering, University of Houston. She is currently an Associate Professor with the Key Laboratory of Dynamic Cognitive System of Electromagnetic Spectrum Space, Ministry of Industry and Information Technology, Nanjing University of Aeronautics and Astronautics, Nanjing, China. Her current research interests include space-air-ground networks, aerial access networks, UAV networking, resource optimization and machine learning.She is currently an Associate Editor of the IEEE Internet of Things Journal. She served as the TPC for IEEE GLOBECOM 2025/2024/2023,IEEE WCNC 2025/2024, IEEE WCSP 2025, etc.
\end{IEEEbiography}
\begin{IEEEbiography}[{\includegraphics[width=1in,height=1.25in,clip,keepaspectratio]{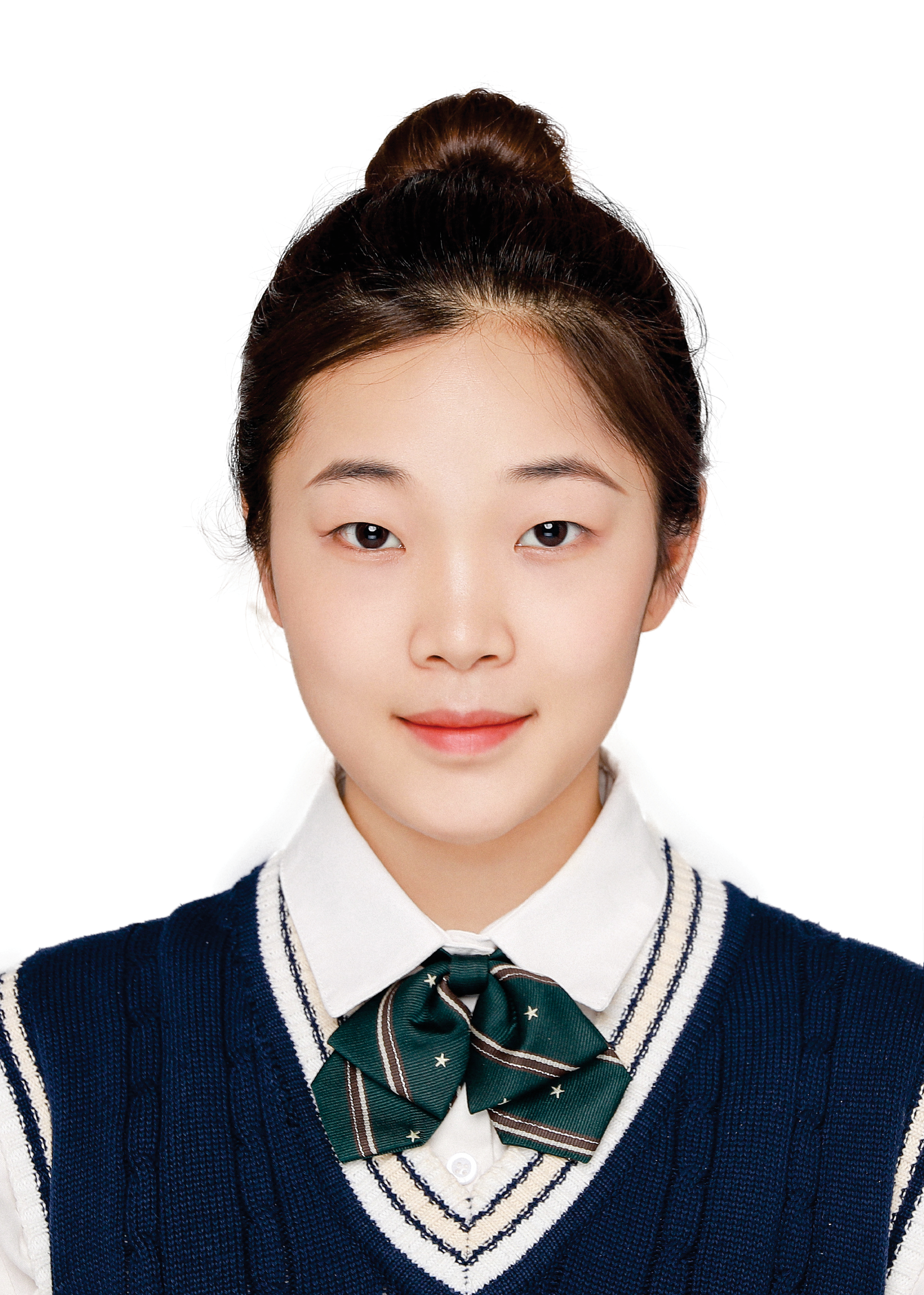}}] {Sijie He} is a postgraduate student with the College of Electronic and Information Engineering, Nanjing
University of Aeronautics and Astronautics, Nanjing,
China. Her current research interests include the low-altitude intelligent networks, trusted routing, blockchain, 
multi-agent deep reinforcement learning, and zero-trust.
\end{IEEEbiography}
\begin{IEEEbiography}[{\includegraphics[width=1in,height=1.25in,clip,keepaspectratio]{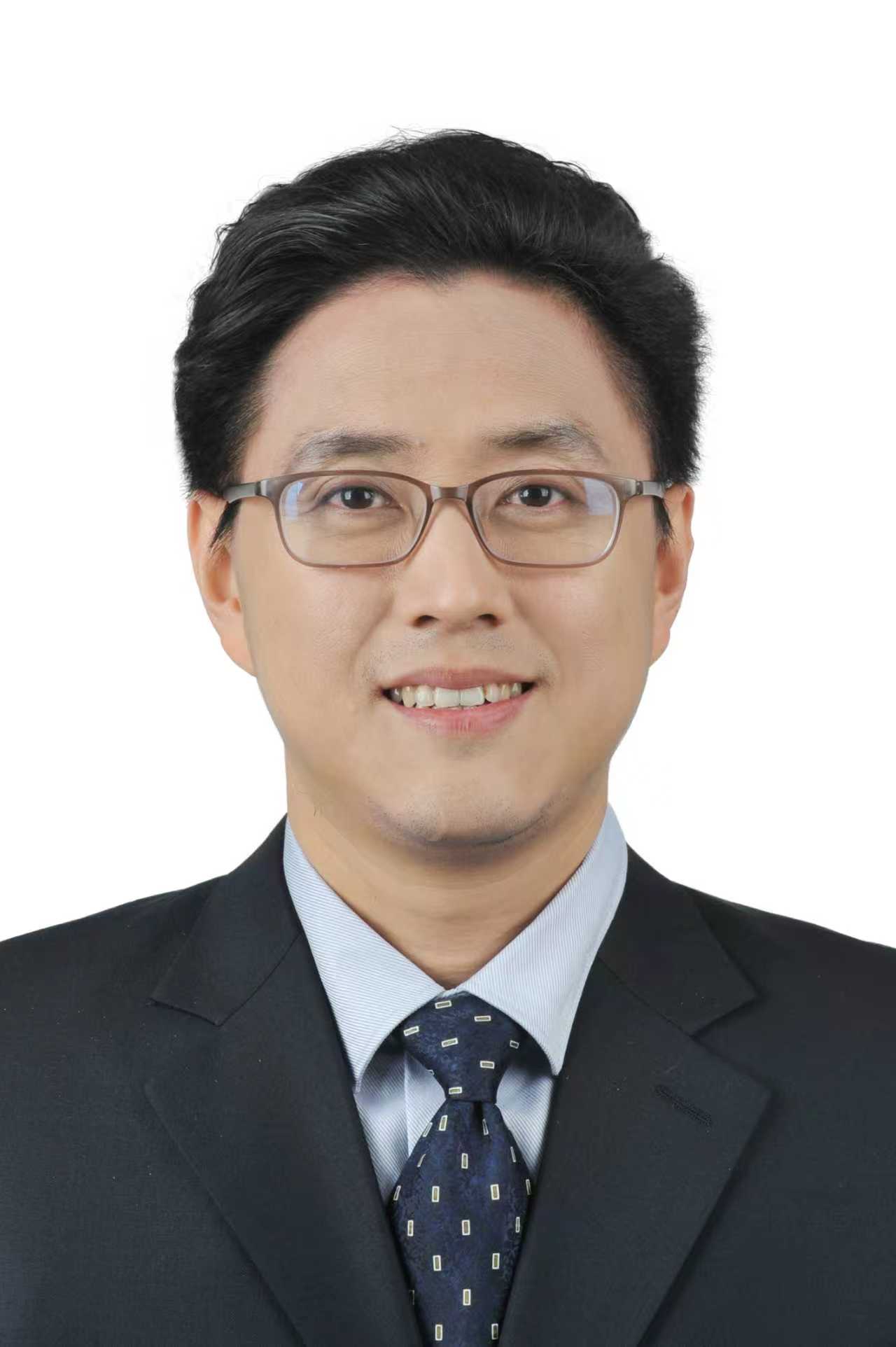}}] {Ligang Yuan,} Associate Professor College of Civil Aviation, Nanjing University of Aeronautics and Astronautics.
His research focuses on theoretical research, key technology development, and applied system design in air traffic flow management, low-altitude airspace management, and system performance evaluation. He has published more than 30 academic papers and filed nearly 30 invention patents. He participated in the formulation of two standards and guidance manuals for the International Civil Aviation Organization (ICAO). He has led the demonstration and construction of the Jiangsu Provincial Low-Altitude Flight Service Center and platform, as well as the low-altitude flight service centers and platforms in Nanjing and Yangzhou.
\end{IEEEbiography}

\begin{IEEEbiography}[{\includegraphics[width=1in,height=1.25in,clip,keepaspectratio]{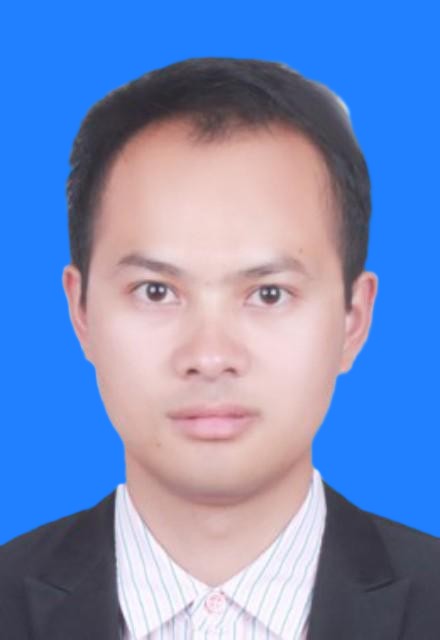}}]{Fuhui Zhou} (Senior Member, IEEE) received the Ph.D. degree from Xidian University, Xi’an, China, in 2016. He is currently a Full Professor with Nanjing University of Aeronautics and Astronautics, Nanjing, China, where he is also with the Key Laboratory of Dynamic Cognitive System of Electromagnetic Spectrum Space. He has published more than 200 papers in internationally renowned journals and conferences in the field of communications. He has been selected for one ESI hot article and 13 ESI highly cited articles. His research interests include cognitive radio, cognitive intelligence, knowledge graph, edge computing, and resource allocation. He received four Best Paper Awards at international conferences, such as IEEE GLOBECOM and IEEE ICC. He was awarded as the 2021 Most Cited Chinese Researchers by Elsevier, the Stanford World's Top 2\% Scientists, the IEEE ComSoc Asia–Pacific Outstanding Young Researcher and Young Elite Scientist Award of China, and the URSI GASS Young Scientist. He serves as an Editor for IEEE Transactions on Communications, IEEE Systems Journal, IEEE Wireless Communications Letters, IEEE Access, and Physical Communication.
\end{IEEEbiography}
\begin{IEEEbiography}[{\includegraphics[width=1in,height=1.25in,clip,keepaspectratio]{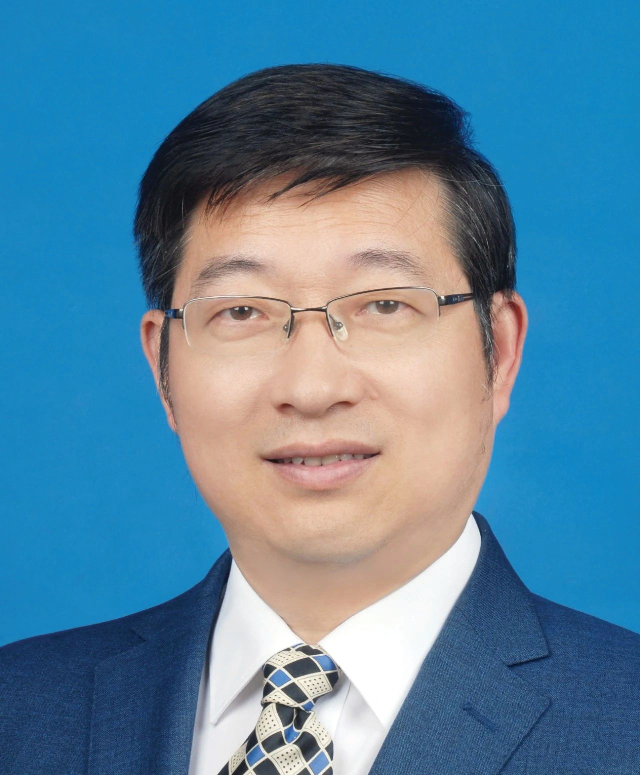}}] {Qihui Wu} (Fellow, IEEE) received the B.S. degree in communications engineering and the M.S. and Ph.D. degrees in communications and information systems from the Institute of Communications Engineering, Nanjing, China, in 1994, 1997, and 2000, respectively. From 2003 to 2005, he was a Post-Doctoral Research Associate with Southeast University, Nanjing. From 2005 to 2007, he was an Associate Professor with the College of Communications Engineering, PLA University of Science and Technology, Nanjing, where he was a Full Professor, from 2008 to 2016. From March 2011 to September 2011, he was an Advanced Visiting Scholar with the Stevens Institute of Technology, Hoboken, NJ, USA. Since May 2016, he has been a Full Professor with the College of Electronic and Information Engineering, Nanjing University of Aeronautics and Astronautics, Nanjing. His current research interests include wireless communications and statistical signal processing, with an emphasis on system design of software defined radio, cognitive radio, and smart radio.
\end{IEEEbiography}
\begin{IEEEbiography}[{\includegraphics[width=1in,height=1.25in,clip,keepaspectratio]{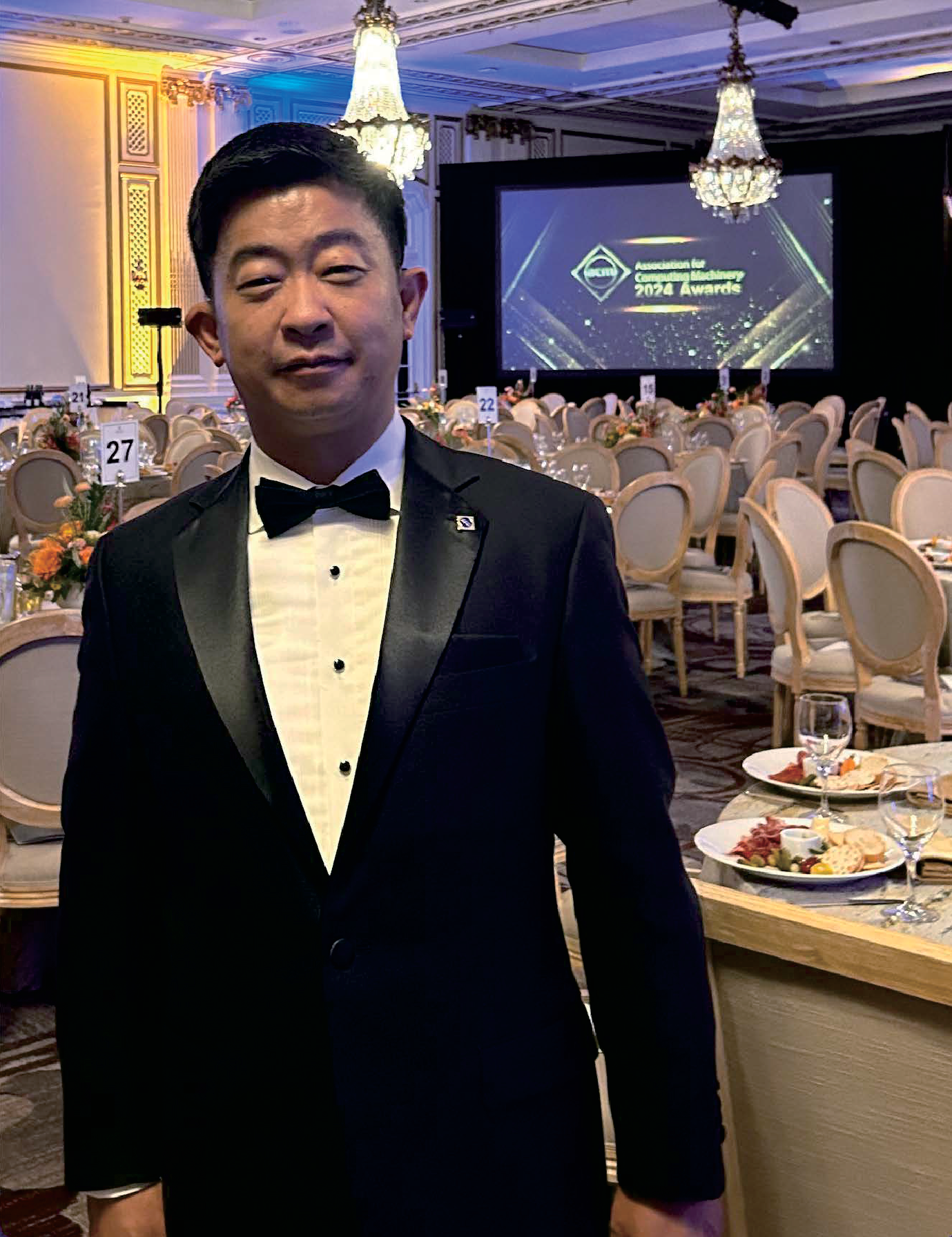}}] {Zhu Han} (Fellow, IEEE) (S'01-M'04-SM'09-F'14) received the B.S. degree in electronic engineering from Tsinghua University, in 1997, and the M.S. and Ph.D. degrees in electrical and computer engineering from the University of Maryland, College Park, in 1999 and 2003, respectively. 

From 2000 to 2002, he was an R\&D Engineer of JDSU, Germantown, Maryland. From 2003 to 2006, he was a Research Associate at the University of Maryland. From 2006 to 2008, he was an assistant professor at Boise State University, Idaho. Currently, he is a John and Rebecca Moores Professor in the Electrical and Computer Engineering Department as well as in the Computer Science Department at the University of Houston, Texas. Dr. Han’s main research targets on the novel game-theory related concepts critical to enabling efficient and distributive use of wireless networks with limited resources. His other research interests include wireless resource allocation and management, wireless communications and networking, quantum computing, data science, smart grid, carbon neutralization, security and privacy.  Dr. Han received an NSF Career Award in 2010, the Fred W. Ellersick Prize of the IEEE Communication Society in 2011, the EURASIP Best Paper Award for the Journal on Advances in Signal Processing in 2015, IEEE Leonard G. Abraham Prize in the field of Communications Systems (best paper award in IEEE JSAC) in 2016, IEEE Vehicular Technology Society 2022 Best Land Transportation Paper Award, and several best paper awards in IEEE conferences. Dr. Han was an IEEE Communications Society Distinguished Lecturer from 2015 to 2018 and ACM Distinguished Speaker from 2022 to 2025, AAAS fellow since 2019, and ACM Fellow since 2024. Dr. Han is also the winner of the 2021 IEEE Kiyo Tomiyasu Award (an IEEE Field Award), for outstanding early to mid-career contributions to technologies holding the promise of innovative applications, with the following citation: ``for contributions to game theory and distributed management of autonomous communication networks." Dr. Han is honored Lifetime Chair Professor of National Yang Ming Chiao Tung University, Taiwan, Eminent Scholar of Kyung Hee University, South Korea and Global Professor of Keio University, Japan. 
\end{IEEEbiography}
\begin{IEEEbiography}[{\includegraphics[width=1in,height=1.25in,clip,keepaspectratio]{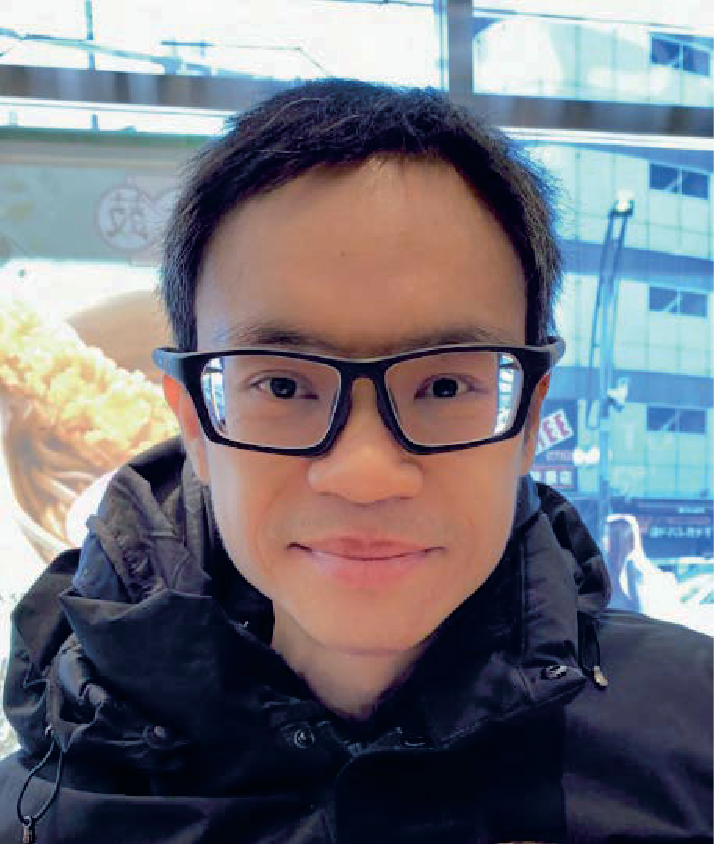}}] {Dusit Niyato} (Fellow, IEEE) is a professor in the College of Computing and Data Science, at Nanyang Technological University, Singapore. He received B.Eng. from King Mongkuts Institute of Technology Ladkrabang (KMITL), Thailand and Ph.D. in Electrical and Computer Engineering from the University of Manitoba, Canada. His research interests are in the areas of mobile generative AI, edge intelligence, quantum computing and networking, and incentive mechanism design.
\end{IEEEbiography}
\end{document}